\documentclass[acmsmall]{acmart}

\newcommand\newnote[1]{\textcolor{black}{#1}}
\AtBeginDocument{%
  \providecommand\BibTeX{{%
    \normalfont B\kern-0.5em{\scshape i\kern-0.25em b}\kern-0.8em\TeX}}}

\setcopyright{acmcopyright}
\copyrightyear{2018}
\acmYear{2018}
\acmDOI{XXXXXXX.XXXXXXX}

\acmJournal{CSUR}
\acmVolume{37}
\acmNumber{4}
\acmArticle{111}
\acmMonth{8}

\usepackage{multirow}
\usepackage{caption}
\usepackage{subcaption}
\usepackage{graphicx}
\usepackage{pifont}
\usepackage{fontawesome}
\usepackage{wasysym}
\usepackage{longtable}
\begin{document}

\title{A Survey of Protocol Fuzzing}

\author{Xiaohan Zhang}
\authornote{Both authors contributed equally to this research.}
\authornote{Work done while visiting Nanyang Technological University and working in Shanghai Jiao Tong University.}
\email{xhzhang1@sjtu.edu.cn}
\orcid{0000-0003-3260-4530}
\affiliation{%
  \department{the State Key Laboratory of Integrated Services Networks, and Engineering Research Center of Big Data Security, Ministry of Education, and the School of Cyber Engineering}
  \institution{Xidian University}
  \streetaddress{2 South Taibai Road}
  \city{Xi An}
  \state{Shaanxi}
  \country{China}
  \postcode{710071}
}
\author{Cen Zhang}
\authornotemark[1]
\affiliation{%
  \institution{Nanyang Technological University}
  \streetaddress{50 Nanyang Ave}
  \country{Singapore}
  \postcode{639798}
}
\email{cen001@e.ntu.edu.sg}
\orcid{0000-0001-5603-1322}

\author{Xinghua Li}
\authornote{Corresponding author.}
\affiliation{%
  \department{the State Key Laboratory of Integrated Services Networks, and Engineering Research Center of Big Data Security, Ministry of Education, and the School of Cyber Engineering}
  \institution{Xidian University}
  \streetaddress{2 South Taibai Road}
  \city{Xi An}
  \state{Shaanxi}
  \country{China}
  \postcode{710071}
}
\email{xhli1@mail.xidian.edu.cn}
\orcid{0000-0002-5583-4155}

\author{Zhengjie Du}
\email{dz1833006@smail.nju.edu.cn}
\orcid{0009-0002-7032-2681}
\author{Bing Mao}
\email{maobing@nju.edu.cn}
\orcid{0000-0002-7066-2144}
\affiliation{%
  \institution{Nanjing University}
  \streetaddress{No. 163 Xianlin Avenue, Qixia District}
  \city{Nanjing City}
  \state{Jiangsu Province}
  \country{China}
  \postcode{210023}
}

\author{Yuekang Li}
\email{yuekang.li@ntu.edu.sg}
\orcid{0000-0003-4382-0757}
\author{Yaowen Zheng}
\email{yaowen.zheng@ntu.edu.sg}
\orcid{0000-0002-8953-0782}
\affiliation{%
  \institution{Nanyang Technological University}
  \streetaddress{50 Nanyang Ave}
  \country{Singapore}
  \postcode{639798}
}

\author{Yeting Li}
\affiliation{%
  \institution{Institute of Information Engineering, Chinese Academy of Sciences}
  \streetaddress{No. 19, Shucun Road, Haidian District}
  \city{Beijing}
  \country{China}
  \postcode{100085}
}
\email{liyeting@iie.ac.cn}
\orcid{0000-0003-0991-4231}

\author{Li Pan}
\affiliation{
    \institution{Shanghai Jiao Tong University}
    \streetaddress{No. 800 Dongchuan Road}
    \city{Shanghai}
    \country{China}
    \postcode{200240}
}
\email{panli@sjtu.edu.cn}
\orcid{0000-0002-0424-9845}

\author{Yang Liu}
\affiliation{%
  \institution{Nanyang Technological University}
  \streetaddress{50 Nanyang Ave}
  \country{Singapore}
  \postcode{639798}
}
\email{yangliu@ntu.edu.sg}
\orcid{0000-0001-7300-9215}

\author{Robert H. Deng}
\affiliation{%
  \institution{Singapore Management University}
  \streetaddress{81 Victoria St}
  \country{Singapore}
  \postcode{188065}
}
\email{robertdeng@smu.edu.sg}
\orcid{0000-0003-3491-8146}

\renewcommand{\shortauthors}{To be filled, et al.}
\newcommand{\cen}[1]{\textcolor{cyan}{{[cen: #1]}}}

\begin{abstract}
Communication protocols form the bedrock of our interconnected world, yet vulnerabilities within their implementations pose significant security threats. 
Recent developments have seen a surge in fuzzing-based research dedicated to uncovering these vulnerabilities within protocol implementations. 
However, there still lacks a systematic overview of protocol fuzzing for answering the essential questions such as what the unique challenges are, how existing works solve them, etc.
To bridge this gap, we conducted a comprehensive investigation of related works from both academia and industry. 
Our study includes a detailed summary of the specific challenges in protocol fuzzing and provides a systematic categorization and overview of existing research efforts. Furthermore, we explore and discuss potential future research directions in protocol fuzzing. 
\end{abstract}



\maketitle

\section{Introduction}

Communication protocols, such as TCP (Transmission Control Protocol) \cite{balakrishnan1995improving}, TLS (Transport Layer Security) \cite{tlsdoc}, Bluetooth \cite{blespec}, etc., serve as the cornerstone of communication by defining the rules for message exchange between parties.
As these protocols underpin publicly accessible services, their security is paramount, and the vulnerabilities contained can lead to severe consequences.
A stark illustration of this is the Heartbleed vulnerability in OpenSSL, an implementation of the TLS protocol.
Upon its disclosure, Heartbleed was found to affect over 17\% of servers worldwide \cite{heartbleed,openssl,heartbleedblog}, demonstrating the extensive impact that a single vulnerability can have.
Moreover, recent statistical analyzes signal an upward trend in high-risk software vulnerabilities within network services \cite{googlevulsheet}, underscoring the increasing risks to network security.
Given this context, the development of automated methods to detect vulnerabilities in network protocol implementations is not just beneficial but essential for the safeguarding of modern network services.

Fuzzing, as a software testing technique, was brought to the forefront by an empirical study conducted by Miller et al. in 1990 \cite{miller1990empirical}.
This method involves the generation of a large number of random mutated testcases aimed at triggering abnormal runtime behaviors within a software program.
Due to its simplicity and scalability, fuzzing has proven to be highly effective in uncovering a wide variety of bugs, leading to its widespread adoption \cite{fuzzingsurvey-roadmap,tse19-fuzzing-survey,tr18-fuzzing-survey,zhang2023automata,zheng2022efficient,zhang2021apicraft}.
However, fuzzing protocol implementations, as opposed to general software such as command-line tools \cite{pldi19-parser,pham2019smart,wang2017skyfire}, introduces additional challenges.
These complexities are largely due to the peculiarities associated with effectively testing the intricate communication logic that protocols entail, ranging from methodological considerations to tool-specific requirements.
In response to these challenges, there is a notable trend toward the creation of advanced fuzzing methods explicitly tailored for protocol testing \cite{sp15-composite,sec15-tlsstatefuzzing,ccs16-tlsattacker,icst20-aflnet,ndss24-chatafl,sec23-bleem,sec22-braktooth,tdsc22-greyhound,sec2022stateful}.
Despite this progress, there still remains a significant gap in research dedicated to systematically examining the distinctive challenges inherent to this field, thoroughly summarizing the existing solutions and discussing future directions.
To fill this gap, we extensively discuss and analyze protocol fuzzing specifics in the following content of this article.

\subsection{Motivation}
The main motivations for this survey are as follows:

\begin{itemize}
    \item 
    Protocols are the essential rules that dictate how our devices and applications communicate, making them both ubiquitous and critically important.
    Because these protocols are everywhere, it is of the utmost importance to ensure they are secure against potential threats.
    Fuzzing plays a key role in finding and fixing security issues within these systems.
    In light of this, building the first end-to-end guide covering the overview and specifics of protocol fuzzing is highly valuable for both researchers and those in the tech industry.
    
    \item Protocol fuzzing presents unique challenges that set it apart from general application fuzzing, grounded in the intricacies of the communication protocols themselves.
    Firstly, there is the need to adhere to strict rules that dictate not just the structure of the messages but also the strict sequence and context in which these messages are sent and received \cite{802.1x,tlsdoc,dtlsdoc}.
    This makes the testing process complex as it requires an in-depth understanding of how these communication protocols operate and change over time.
    Secondly, protocols are built to address various attributes beyond simple message exchange.
    They must account for factors such as timing and how multiple messages or actions can occur simultaneously, which introduces more variables into the mix when testing for security issues \cite{6129367,katz2006parallel,jung2001dns}.
    Third, the widespread use of protocols across different technology levels and systems adds another layer of complexity.
    They are embedded everywhere, from hardware to software applications that we interact with daily, leading to various testing scenarios and discovering potential vulnerabilities at every layer \cite{sec20-frankenstein,sec21-LIGHTBLUE,atc20-sweyntooth,sec22-braktooth,tdsc22-greyhound,acsac21-ics3fuzzer,dac20-icsprotocol}.
    Given these realities, it becomes imperative to establish a comprehensive understanding of protocol-specific challenges.
    
    \item Many protocol fuzzing works have been completed, but so far no systematic review on protocol fuzzing has been conducted.
    Although some survey articles \cite{tr18-fuzzing-survey,tse19-fuzzing-survey,fuzzingsurvey-roadmap} on traditional software fuzzing are available, they cannot provide a systematic overview of the current state and future directions based on existing work solving protocol-specific challenges.
\end{itemize}

\subsection{Survey Dimensions}
This survey aims to provide an overview of the protocol-specific challenges, the corresponding solutions, and the future directions. Specifically, this survey is organized around the following dimensions:

\begin{itemize}
    \item \textbf{Dimension 1:} Examine the differences between traditional fuzzing and protocol fuzzing.
    \item \textbf{Dimension 2:} 
    Review how existing works address the challenges in protocol fuzzing.
    \item \textbf{Dimension 3:} Explore potential future directions in the field.
\end{itemize}

In Section \ref{sec:overview}, we provide an in-depth examination of the distinctive differences between protocols and traditional fuzzing targets to address Dimension 1.  Then, in Sections \ref{sec:input} to \ref{sec:bugdetector}, we detail the techniques used in existing protocol fuzzers to address Dimension 2. Lastly, Dimension 3 is discussed in Section \ref{sec:future}.

\subsection{Collection Strategy}

\begin{table}[]
\centering
\caption{Selected influential conferences and journals}
\label{tab:whitelists}
\resizebox{\columnwidth}{!}{%
\begin{tabular}{|l|l|l|}
\hline
\textbf{Research Area} & \textbf{Type} & \textbf{Name} \\ \hline
\multirow{2}{*}{\textbf{Cyber Security}} & Conferences & ACSAC, CCS, CODASPY, DSN, ICDCS, ICICS, NDSS, SP, USENIX, WiSec, Blackhat*, DEFCON*, RSA* \\ \cline{2-3} 
 & Journals & TDSC, TIFS \\ \hline
\multirow{2}{*}{\textbf{System Architecture}} & Conferences & ASPLOS, ATC, DAC, Eurosys, Mobisys, OSDI \\ \cline{2-3} 
 & Journals & TC \\ \hline
\multirow{2}{*}{\textbf{Communication}} & Conferences & INFOCOM, MobiCOM, NSDI, SIGCOMM \\ \cline{2-3} 
 & Journals & TMC, TNSM, TON  \\ \hline
\multirow{2}{*}{\textbf{Software Engineering}} & Conferences & ASE, FSE, ICSE, ICST, ISSTA \\ \cline{2-3} 
 & Journals & TOSEM, TSE  \\ \hline
\end{tabular}%

}
\begin{minipage}{\linewidth}
  \small *: industrial conferences.
\end{minipage}
\end{table}

\begin{figure}
\centering
\begin{minipage}[b]{0.5\textwidth}
  \centering
  \includegraphics[width=0.8\textwidth]{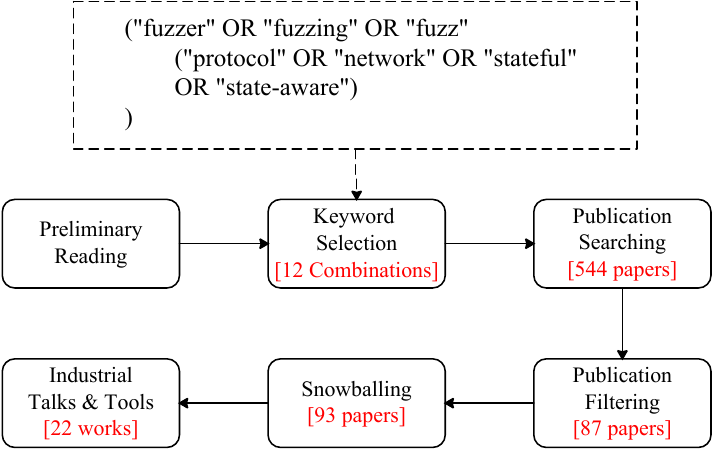}
  \caption{Search criteria.}
  \label{fig:searchcriteria}
  \par\vspace{0pt}
\end{minipage}%
\begin{minipage}[b]{0.5\textwidth}
  \centering
  \includegraphics[width=0.8\textwidth]{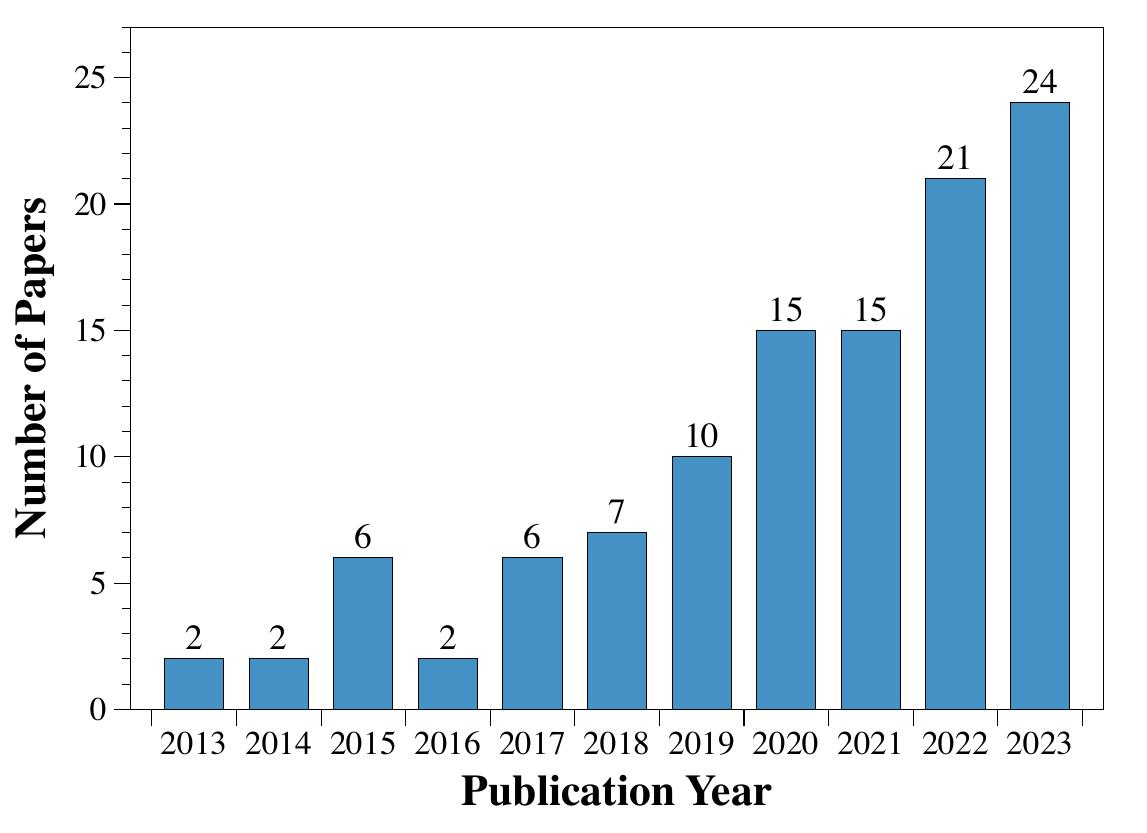}
  \caption{Distribution of papers along publication years.}
  \label{fig:papercount}
  \par\vspace{0pt}
\end{minipage}
\end{figure}

In this survey, we focus on stateful network protocols and the various techniques that are directly related to the fuzz testing of their implementations.
To collect the relevant publications, we followed the procedures depicted in Fig. \ref{fig:searchcriteria}. 
First, we performed a preliminary reading and summarized 12 different keyword combinations that can be used to search for related works. 
Then, searching for these keyword combinations on Google Scholar, we collected \newnote{544} published articles from \newnote{January 2013 to June 2024}. 
After that, we manually filtered out papers irrelevant to protocol fuzzing or not published in influential publications listed in Table \ref{tab:whitelists}.
At that time, the number of articles was reduced to \newnote{87}.
Note that all pre-print papers were kept to remove publication bias \cite{greyliterature}.
And a paper is relevant if its key contribution is in the scope of protocol fuzzing or that paper is a bug detection tool and has picked at least one protocol implementation as its evaluation target.
The latter criterion is based on the heuristic that a bug detection tool probably \newnote{has} proposed protocol-specific techniques if it uses protocol implementations as its evaluation targets.
Next, we performed snowballing and inverse snowballing to obtain a more comprehensive view.
Six more papers were added in this procedure. 
Finally, we applied the above collection process to the released talks of several mainstream industrial security conferences such as BlackHat.
\newnote{22} industrial works were added, including \newnote{20} related talks and three open source protocol fuzzers with more than 50 stars on Github. The ascending trajectory of publications, as illustrated in Fig. \ref{fig:papercount}, underscores the burgeoning research interest in protocol fuzz testing, affirming its emergence as a focal point within the field.

The remainder of the paper is organized as follows.
Section \ref{sec:background} introduces the background knowledge of protocol fuzzing. 
Section \ref{sec:overview} introduces the main differences between general fuzzing targets and protocols and then summarizes the major enhancements of existing protocol fuzzers.
The next three sections detail the existing techniques for each key component of protocol fuzzing.
Section \ref{sec:input} discusses the progress in input generator component.
Section \ref{sec:executor} introduces the techniques for improving the executor component.
Section \ref{sec:bugdetector} manifests the taxonomy of oracles used in the bug detector component. 
Section \ref{sec:future} offers future directions.

\section{Background}
\label{sec:background}

\subsection{Communication Protocols} 
\label{sebsec:protocols}

A communication protocol is a set of rules that enables the exchange of information between two or more entities within a communication system, utilizing any form of physical quantity variation. 
The implementation of a communication protocol generally involves multiple phases \cite{comer2013internetworking}. First, the protocol is conceptually designed, which includes defining the rules, behaviors, and functions it will perform based on the protocol's needs, taking into account factors such as efficiency, reliability, scalability, and security. The outcome of the design phase is a \newnote{set of} \textbf{specifications}. Then, during the development phase, the design of the protocol is translated into concrete \textbf{implementations}. This can be in the form of software, hardware, or a combination of both. Once developed, the protocol undergoes rigorous testing to confirm that it adheres to the protocol specifications and meets the performance and reliability requirements. Among them, fuzzing, which this paper focuses on, is a commonly used technology for testing protocol implementations. Eventually, the protocol implementation will be deployed in a real-world environment. 

In addition to the fundamental task of data exchange, protocols encompass a myriad of other critical communication functionalities, introducing new layers of complexity \cite{marsden1986communication}. This includes tasks such as routing, detection of transmission errors, managing timeouts and retries, confirmations, flow control, and sequence control.
Each of these functionalities embodies a set of strategies and implementations that collectively ensure the efficacy and reliability of communication protocols. The intricate integration of these functions demonstrates the sophisticated nature of protocol design and its pivotal role in modern communication systems.

\subsection{Types of Protocols}
\label{subsec:protocoltypes}

 \begin{figure}[!t] 
 \centering
 \includegraphics[width=0.8\textwidth]{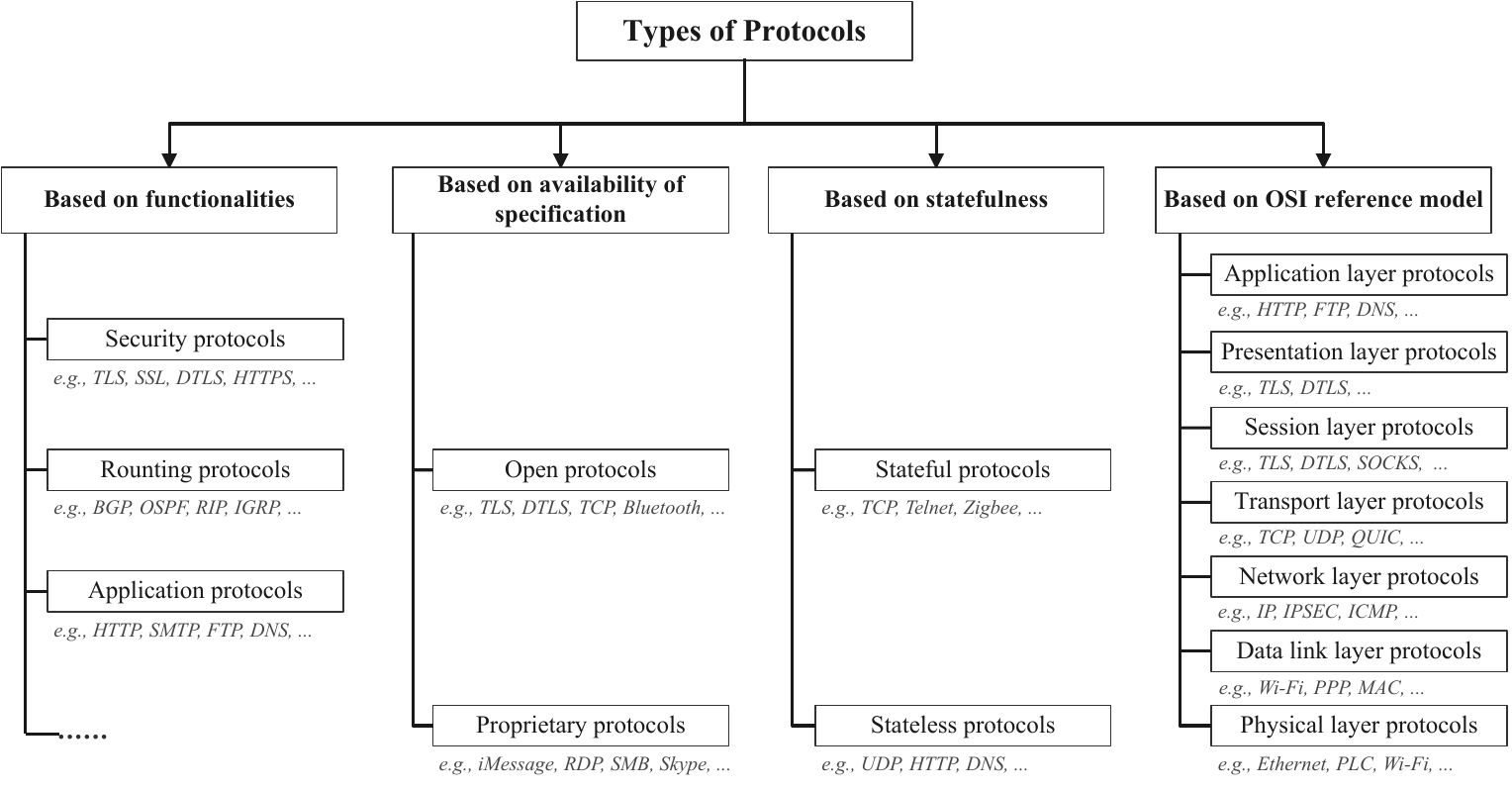} 
 \caption{Types of protocols.}
 \label{fig.types}
 \end{figure}

Protocols can be classified from various perspectives, such as functionality, the accessibility of their specifications, and their alignment with the layers of the OSI network reference model, as shown in Fig. \ref{fig.types}.

From a functional point of view, protocols exhibit a broad spectrum of varieties, each tailored to fulfill unique operational objectives. For example, security protocols are designed primarily to ensure the integrity and confidentiality of data, as exemplified by TLS \cite{tlsdoc} and DTLS (Datagram Transport Layer Security) \cite{dtlsdoc}. Routing protocols, such as BGP (Border Gateway Protocol), are dedicated to efficiently managing the routes of data packets traversing the network. Furthermore, application protocols, such as HTTP (Hypertext Transfer Protocol) for web services and SMTP (Simple Mail Transfer Protocol) for email, are specialized to enable specific functionalities at the application layer.

When considering the availability of protocol specifications, a distinction is drawn between \textit{open protocols} and \textit{proprietary protocols}. \textit{Open protocols}, like TCP, have publicly accessible specifications, allowing widespread scrutiny and implementation. In contrast, \textit{proprietary protocols} such as Microsoft's RDP (Remote Desktop Protocol) \cite{rdpdoc}, are governed by individual entities, with specifications that are not fully public. The availability of protocol specifications is crucial for various stages of fuzzing, such as crafting inputs, constructing state machines, and detecting bugs. It is important to clarify that the classification into \textit{open} and \textit{proprietary} pertains solely to the availability of specifications and is independent of the accessibility of source codes for protocol implementations.

Regarding the statefulness, protocols are bifurcated into stateful and stateless categories. Stateful protocols, such as TLS \cite{ccs16-tlsattacker,sec15-tlsstatefuzzing} and TCP \cite{ndss18-tcpcongestion}, require multiple rounds of interaction. Stateless protocols, such as UDP and HTTP, do not maintain state information across requests.

Based on the OSI network reference model, protocols can be classified into seven distinct layers: physical, data link, network, transport, session, presentation, and application. The protocol layers each solve a distinct class of communication problems. Among them, the lower-level protocols have a higher coupling with the physical hardware. 
It is pertinent to note that not every protocol aligns precisely with a single layer in the OSI model. For example, TLS/DTLS contains the functionality of the session and representation layers; the Wi-Fi protocol contains the main functionality of the physical and data link layers \cite{802.11,chen2005wireless}.
Given varying interpretations of the protocol layering in numerous sources, we categorize these protocols based on their primary functions. 

\section{Protocol Fuzzing Overview}
\label{sec:overview}

\subsection{Differences between protocol fuzzing and traditional fuzzing} 
\label{subsec:protocoldifferences} 

In this subsection, we discuss the unique challenges associated with protocol fuzzing as identified in the literature, addressing Survey Dimension 1.  
Protocol implementations differ from traditional fuzzing targets in two key aspects: first, they exhibit higher communication complexity; second, their testing environments are relatively more constrained.
These differences not only highlight the specificity of protocol fuzzing, but also correspond to a set of inherent challenges. 

\subsubsection{High communication complexity}

The high complexity of communication can be discussed in the following two aspects.

\textbf{Respecting Semantic Constraints in Communication.} 
Protocols serve as the backbone for facilitating communication between different systems by providing a standardized set of rules for message exchange.
This communication is inherently complex, often involving a multi-round process where multiple steps must be sequentially executed for the exchange to be successful.
Such protocols inherently require stateful implementations, with each stage of communication building upon the previous one \cite{802.11,802.1x,tlsdoc,dtlsdoc}.
In testing scenarios, this means that deeper layers of the protocol implementation cannot be tested until the earlier constraints are satisfactorily met -- these are the strict semantic constraints inherent in communication protocols.
Semantic constraints come in two primary forms: intra-message and inter-message constraints.
Intra-message constraints pertain to the structure and content of individual messages, ensuring that data fields are syntactically correct and semantically meaningful within the context of that message. Taking TCP as an example, in a TCP segment, there are several critical fields such as the source port, destination port, sequence number, acknowledgment number, data offset, and control flags (like SYN, ACK) \cite{marsden1986communication}. Each of these fields must adhere to specific formats and rules. 
Inter-message constraints, on the other hand, govern the relationship and sequence of multiple messages, requiring that they adhere to the established protocol sequence and context for the conversation to progress \cite{comer2013internetworking}.
For instance, the establishment of a TCP connection involves a ``three-way handshake" process: the client first sends a SYN message, followed by the server responding with a SYN-ACK message, and finally, the client sends an ACK message to complete the connection.
Violations of either type of constraints during communication can result in the fuzzing becoming non-progressive \cite{icst20-aflnet,ccs16-tlsattacker,atc21-tcpfuzz,ndss18-tcpcongestion,blackhat-us-17-wifuzz}.

\textbf{Testing Different Properties of Communication Process.} 
Besides basic message exchange functionality, protocols need to guarantee a series of additional features that form a more secure or reliable communication such as time requirements, authentication, confidentiality, and concurrency\cite{6129367,katz2006parallel,jung2001dns}.
Effectively testing these attributes in implementations requires a more complex form of testing that goes beyond typical application fuzzing, which mainly focuses on altering structured inputs to find issues \cite{fuzzingsurvey-roadmap,tse19-fuzzing-survey}.
Each attribute may require significant modification or even a redesign of the fuzzing framework, including the development of specialized input generator, feedback mechanisms, and oracles to facilitate effective testing \cite{hussain2021noncompliance,ji2023finding,ccs16-tlsattacker,atc21-tcpfuzz,blackhat-us-17-wifuzz,asiaccs19-mossot,atc20-sweyntooth,icse22-ltlprop,chen2022tyr,ma2023loki}. 
For example, in the context of crafting a fuzzer that aims to detect traffic amplification attacks within protocol implementations \cite{sec22-ampfuzz}, an oracle is needed to identify disproportional request-to-response data volume ratios, indicative of an amplification factor.
Currently, the input generator needs to be adeptly redesigned to generate specific variations of protocol messages that can maximize the potential amplification factor.
Moreover, the amplification factor can be used as a feedback to guide the fuzzer in exploring the input space more effectively.

\subsubsection{Constrained Testing Environment} 

Protocol fuzzing usually faces a constrained testing environment due to the tight coupling between protocols and the hardware.
Firstly, numerous protocols are designed for communications between low-level physical devices or for communications \newnote{occurring} in specialized sectors, such as protocols residing in the lower layers of the OSI reference model, \textit{i.e.}, the physical and data link layers \cite{chen2005wireless,802.11,802.1x,zigbeedoc,blespec}, or protocols designed for specific sectors such as automotive \cite{dac21-PAVFuzz,someipdoc,rtpsdoc}, industrial control system (ICS) \cite{icst19-seqfuzzer,arxiv22-fieldfuzz}, electrical grids, and aviation systems.
In these cases, the testing throughput will be limited by the hardware dependencies, such as the lack of auotmation \cite{dsn22-l2fuzz,defcon30-someip}, the bottleneck for scalable fuzzing \cite{arxiv22-fieldfuzz,sec20-frankenstein}, etc.
In addition, these physical dependencies also limit the application of advanced fuzzing techniques.
This is because many advanced fuzzing techniques require greybox or whitebox testing information from the test target, which cannot be satisfied due to the lack of program analysis frameworks on these specific hardware \cite{tcad22-charon,wisec21-schepers,dac21-PAVFuzz}.

\subsection{Summary of Existing Protocol Fuzzers}

A general fuzzer consists of three basic components, namely input generator, executor, and bug collector. In one fuzzing iteration, \textbf{ input generator} first produces a testing input to \textbf{executor}. Then \textbf{executor} executes the PUT with the given input and collects runtime information for the other two components. Finally, the \textbf{bug collector} checks the runtime information to determine whether the input has triggered a bug. 

The unique characteristics of protocols introduced in Section 3.1 present distinct challenges in the design of these three components. The high communication complexity demands that the input generator not only adheres to semantic constraints, including both intra- and inter-message constraints, but also generates inputs from a multitude of testing perspectives to enhance the discovery of vulnerabilities. 
Similarly, high communication complexity also poses challenges for the bug detector, requiring it to detect whether test cases violate various security properties of the communication process.
Furthermore, the constraints of the testing environment impose limitations on the executor’s scalability and its capability to collect runtime information, further influencing the bug collector's reliance on a limited set of bug detection mechanisms. These challenges underscore the need for specialized adaptations in the fuzzing framework to effectively address the intricacies of protocol testing.

\begin{figure}[t] 
\centering
\includegraphics[width=\textwidth]{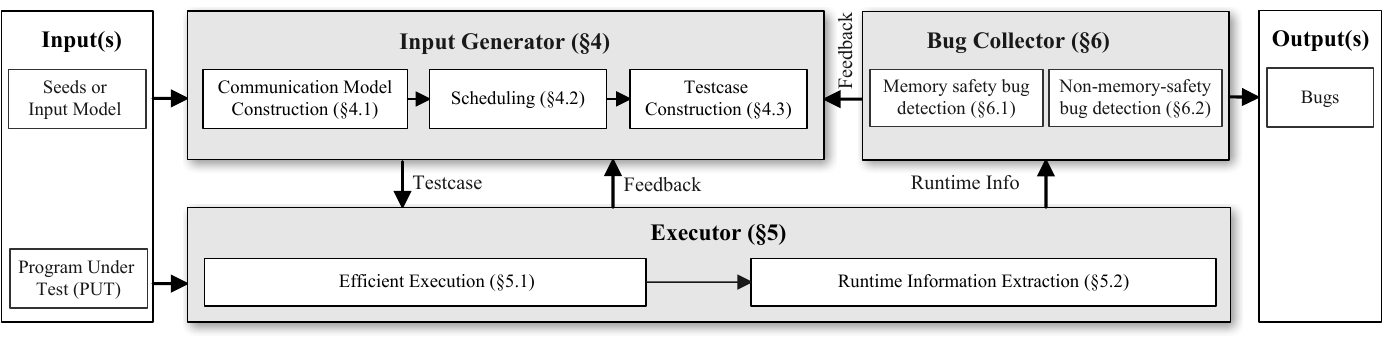} 
\caption{Summarized Workflow of Existing Protocol Fuzzers.}
\label{fig.workflow}
\end{figure}

We have analyzed current protocol fuzzing research and encapsulated these efforts into a technical framework for protocol fuzzers, as depicted in Fig.~\ref{fig.workflow}.
Note that existing protocol fuzzing work still follows the high-level concepts of general fuzzing but proposes specific enhancements in these subcomponents to solve protocol-specific challenges.

\textbf{Input generator.} Ideally, this component is responsible for generating inputs to expose vulnerabilities inside PUTs as effectively as possible.
To realize this, protocol fuzzers usually implement input generators with three main subphases including \textit{communication model construction}, \textit{scheduling} and \textit{testcase construction}.

\textbf{Executor.} In pursuit of an ideal executor for protocol fuzzing, contemporary research has concentrated on two critical aspects: \textit{Efficient Execution} and \textit{Runtime Information Extraction}. The former explores the development of an efficient, automated, and scalable testing environment, enhancing the execution of protocol testing. The latter focuses on creating an analysis environment that extracts essential runtime information, thereby informing and improving the input generation and bug detection processes.

\textbf{Bug collector.} The primary objective of the bug collector component is twofold: to increase the diversity of bug types it can detect and to enhance the accuracy of these detections. The component is finely tuned to meticulously identify a broad spectrum of vulnerabilities, ranging from \textit{memory-safety bugs} like buffer overflows to more subtle \textit{non-memory-safety} bugs such as logic errors and specification \newnote{violations}. 

\section{Input Generator}
\label{sec:input}

In this section, we will introduce in detail how the existing work improves the input generator to solve the unique challenges in protocol fuzzing. 
As shown in Table~\ref{table-inputgen}, we summarize the techniques used in existing work in designing the three key phases in input generator. The covered works are selected from our paper set as long as their main techniques are directly related to the input generator. In addition to the statistics of the mentioned three phases for these works, the table also lists their feedback information. According to Fig. \ref{fig.workflow}, the feedback information can be provided by the executor or the bug collector. We only discuss how the feedback information is used in input generator but leave the details related to the feedback collection for Section~\ref{subsec:extraction}.

\tiny
\begin{longtable}{|l|l|l|l|l|l|l|l|}
\caption{
Protocol fuzzers and their optimization solutions used in input generator.
} \label{table-inputgen} \\

\hline
\textbf{Years} & \textbf{Work} & \textbf{Tax} & \textbf{Target} & \textbf{\begin{tabular}[c]{@{}c@{}}Comm Model \\ Construction\end{tabular}} & \textbf{Scheduling} & \textbf{\begin{tabular}[c]{@{}c@{}}Construction \\ Level\end{tabular}} & \textbf{Feedback} \\ \hline
\endfirsthead

\multicolumn{8}{c}
{{\tablename\ \thetable{} Protocol fuzzers and their optimization solutions used in input generator. (Continued)}} \\
\hline
\textbf{Year} & \textbf{Work} & \textbf{Tax} & \textbf{Target} & \textbf{\begin{tabular}[c]{@{}c@{}}Comm Model \\ Construction\end{tabular}} & \textbf{Scheduling} & \textbf{\begin{tabular}[c]{@{}c@{}}Testcase \\ Construction\end{tabular}} & \textbf{Feedback} \\ \hline
\endhead

\hline
\endfoot

\hline
\endlastfoot

2013 & \textbf{BED}\cite{bed} & \CIRCLE & General & Manual & Sequential & P & State  \\ \hline
2013 & \textbf{Tsankov \textit{et al.}} \cite{issta13-semi-valid} & \CIRCLE & General & Manual & - & P \& S & -  \\ \hline
2014 & \textbf{Peach}\cite{peach} & \CIRCLE & General & Manual & Sequential & P & State  \\ \hline
2015 & \textbf{Pulsar}\cite{securecomm15-pulsar} & \CIRCLE & General & TAPL & SCHS & P & State  \\ \hline 
2015 & \textbf{Beurdouche \textit{et al.}}\cite{sp15-composite} & \CIRCLE & TLS & Manual & Random & S & State  \\ \hline
2015 & \textbf{Ruiter \textit{et al.}}\cite{sec15-tlsstatefuzzing} & \CIRCLE & TLS & TAAL & Random & S & State  \\ \hline
2016 & \textbf{TLS-Attacker}\cite{ccs16-tlsattacker} & \CIRCLE & TLS & Manual & Random & P \& S & - \\ \hline
2016 & \textbf{Driller}\cite{ndss16-driller} & \Circle & General & - & SPMS & P & Code Cov \\ \hline
2017 & \textbf{Fan \textit{et al.}}\cite{icics17-fan} & \CIRCLE & General & TAPL & -  & P \& S & - \\ \hline
2017 & \textbf{WiFuzz}\cite{blackhat-us-17-wifuzz} & \CIRCLE & Wi-Fi & Manual & Sequential & P \& S & - \\ \hline 
2018 & \textbf{TCPWN}\cite{ndss18-tcpcongestion} & \CIRCLE & TCP & Manual & Sequential & P \& S & State  \\ \hline 
2018 & \textbf{IoTFuzzer}\cite{ndss18-iotfuzzer} & \CIRCLE & IoT \cite{mqttdoc,coapdoc} & - & - & P & -  \\ \hline 
2018 & \textbf{Danial \textit{et al.}}\cite{eurospw18-openvpn} & \CIRCLE & OpenVPN\cite{openvpndoc} & Manual & Random & S & - \\ \hline
2018 & \textbf{DELTA}\cite{blackhat-us-18-delta} & \CIRCLE & OpenFlow\cite{openflowdoc} & Manual & -  & P \& S & -  \\ \hline
2019 & \textbf{SeqFuzzer}\cite{icst19-seqfuzzer} & \CIRCLE & ICS & TAPL & - & P & State  \\ \hline 
2019 & \textbf{Polar}\cite{tecs19-polar} & \LEFTcircle & ICS & - & SPMS & P & Code Cov  \\ \hline
2019 & \textbf{IoTHunter} \cite{ccs19-iothunter} & \LEFTcircle & IoT & TAAL & Sequential & P & Code Cov \\ \hline
2019 & \textbf{MoSSOT} \cite{asiaccs19-mossot} & \CIRCLE & SSO\cite{ssodoc} & Manual & Sequential & P \& GUI Ops & - \\ \hline
2019 & \textbf{Chen \textit{et al.}} \cite{feast19-yurong} & \LEFTcircle & General & - & SPHS & P & State \& Code Cov \\ \hline
2019 & \textbf{Fuzzowski}\cite{defcon27-fuzzowski} & \LEFTcircle & General & Manual & - & P & State \& Code Cov \\ \hline
2020 & \textbf{Exploiting Dissent}\cite{tdsc17-differencialTLS} & \CIRCLE & TLS & - & Random & P & - \\ \hline 
2020 & \textbf{DTLS-Fuzzer}\cite{sec20-dtls,icst22-dtlsfuzzer} & \CIRCLE & DTLS & TAAL & Random & S & State  \\ \hline
2020 & \textbf{AFLNET} \cite{icst20-aflnet}& \LEFTcircle & General & TAAL & SRHS & P & State \& Code Cov   \\ \hline 
2020 & \textbf{SweynTooth} \cite{atc20-sweyntooth}& \CIRCLE & BLE\cite{blespec} & Manual & SPHS & P \& S & State \& \# of Bugs \\ \hline
2020 & \textbf{Frankenstein} \cite{sec20-frankenstein}& \LEFTcircle & Bluetooth & Manual & Random & P \& S & State \& Code Cov \\ \hline
2020 & \textbf{Peach*} \cite{dac20-icsprotocol}& \LEFTcircle & ICS & - & - & P & Code Cov  \\ \hline
2020 & \textbf{aBBRate} \cite{raid20-abbrate}& \CIRCLE & TCP & Manual & Sequential & S & State \\ \hline
2020 & \textbf{IJON} \cite{sp20-ijon} & \LEFTcircle & General & PAL & Random & P & State \& Code Cov \\ \hline
2020 & \textbf{FuSeBMC} \cite{arxiv20-FuSeBMC} & \Circle & General & - & Sequential & P & Code Cov \\ \hline
2020 & \textbf{DPIFuzz} \cite{acsac20-dpifuzz} & \CIRCLE & QUIC\cite{quicdoc} & Manual & Random & P \& S & - \\ \hline
2020 & \textbf{Zou \textit{et al.}}\cite{arxiv20-zou} & \CIRCLE & General & - & - & P & - \\ \hline
2021 & \textbf{ICS$^3$Fuzzer} \cite{acsac21-ics3fuzzer}& \CIRCLE & ICS \cite{supervisory1doc,supervisory2doc} & PAL & SCHS  & P \& GUI Ops & State  \\ \hline 
2021 & \textbf{StateAFL} \cite{arxiv21-stateafl}& \LEFTcircle & General & PAL & SPHS \& SRHS & P & \begin{tabular}[l]{@{}l@{}}State \& Code Cov \\ \& \# of Bugs\end{tabular}  \\ \hline
2021 & \textbf{TCP-Fuzz} \cite{atc21-tcpfuzz}& \LEFTcircle & TCP & Manual & Random & P \& S \& Syscall & State Transition \\ \hline 
2021 & \textbf{Snipuzz}\cite{ccs21-snipuzz} & \CIRCLE & IoT & - & - & P & -  \\ \hline 
2021 & \textbf{Z-Fuzzer}\cite{wisec21-zfuzzer} & \LEFTcircle & Zigbee & - & - & P \& Interrupt & Code Cov  \\ \hline
2021 & \textbf{PAVFuzz}\cite{dac21-PAVFuzz} & \LEFTcircle & AV \cite{someipdoc,rtpsdoc} & Manual & Sequential & P & Code Cov \\ \hline 
2021 & \textbf{Aichernig \textit{et al.}}\cite{icst21-aichernig} & \CIRCLE & IoT & TAAL & - & P & -  \\ \hline 
2017 & \textbf{Owfuzz}\cite{blackhat-eu-21-owfuzz} & \CIRCLE & Wi-Fi & Manual & - & P & \multicolumn{1}{l|}{State} \\ \hline
2022 & \textbf{Meng \textit{et al.}}\cite{icse22-ltlprop} & \LEFTcircle & General & Manual & Property-Guided & P & State \\ \hline
2022 & \textbf{Greyhound}\cite{tdsc22-greyhound} & \LEFTcircle & Wi-Fi & Manual & SPHS & P \& S & State \& \# of Bugs \\ \hline
2022 & \textbf{SGFuzz}\cite{sec2022stateful} & \LEFTcircle & General & PAL & SRMS & P & State \& Code Cov  \\ \hline
2022 & \textbf{Braktooth}\cite{sec22-braktooth} & \LEFTcircle & Bluetooth & TAAL & SPHS & P & State Transition \\ \hline 
2022 & \textbf{L2Fuzz}\cite{dsn22-l2fuzz} & \CIRCLE & Bluetooth L2CAP & Manual & Sequential & P & State Cov  \\ \hline 
2022 & \textbf{AmpFuzz}\cite{sec22-ampfuzz} & \LEFTcircle & UDP & - & - & P & BAF  \\ \hline 
2022 & \textbf{FUME}\cite{infocom22-fume} & \CIRCLE & MQTT\cite{mqttdoc} & - & - & P & Response Freshness  \\ \hline 
2022 & \textbf{Garbelini \textit{et al.}}\cite{10001673} & \CIRCLE & 4G/5G & TAPL & - & P \& S & -  \\ \hline 
2023 & \textbf{FeildFuzz}\cite{arxiv22-fieldfuzz} & \CIRCLE & Codesys v3\cite{codesysdoc} & - & - & P & Code Cov  \\ \hline 
2023 & \textbf{BLEEM}\cite{sec23-bleem} & \CIRCLE & General & TAAL & SRHS & P \& S & State  \\ \hline 
2023 & \textbf{Tyr}\cite{chen2022tyr} & \LEFTcircle & Blockchain & Manual & SRHS & P & State \& Code Cov \\ \hline 
2023 & \textbf{CHATAFL}\cite{ndss24-chatafl} & \LEFTcircle & General & LLM & LLM & P & State \& Code Cov  \\ \hline 
2023 & \textbf{EmNetTest}\cite{amusuo2023systematically} & \CIRCLE & General & Manual & Sequential & P \& S & State \\ \hline 
2023 & \textbf{DYFuzzing}\cite{ammann2024dy} & \LEFTcircle & General & Manual & SPMS & P \& S & Code Cov \& \# of Bugs \\ \hline 
2023 & \textbf{FuzzPD}\cite{kim2023fuzz} & \CIRCLE & USBPD & Manual & - & P & State \\ \hline 
2023 & \textbf{Mallory}\cite{ccs23-Mallory} & \LEFTcircle & Distributed Sys & TAAL & SPMS & P \& S & Event Trace \\ \hline 
\end{longtable}
\vspace{-8pt}
\noindent\parbox{\textwidth}{
  \footnotesize
  \Circle: Whitebox Fuzzer; \CIRCLE: Blackbox Fuzzer; \LEFTcircle: Greybox Fuzzer; Tax: Taxonomy; General: The fuzzer is not designed for a specific type of protocol; PAL: Program-Analysis-assisted Learning; TAAL: Traffic-Analysis-based Active Learning; TAPL: Traffic-Analysis-based Passive Learning; SRMS: State Rarity-preferred Monolithic Scheduling; SPHS: State Performance-preferred Hierarchical Scheduling; SCHS: State Complexity-preferred Hierarchical Scheduling; SPMS: State Performance-preferred Monolithic Scheduling; SRHS: State Rarity-preferred Hierarchical Scheduling; -: Not implemented; GUI Ops: GUI Operations; BAF: Bandwidth Amplification Factor; P: Packet level construction; S: Sequence level construction.
}
\normalsize

\subsection{Communication Model Construction}
\label{subsec:statemachine}

\newnote{To enhance traditional fuzzers with semantic constraints, it is essential to develop a communication model of the protocol to guide the fuzzing process. This communication model includes both a state model and a data model. The state model details the transitions between different protocol states, while the data model specifies the format and structure of the messages that drive these state transitions.
Typically, existing works represent the protocol's communication model as a state machine or its variants. A state machine is a data structure that describes the internal state transitions of a protocol implementation. 
State machines can be represented as directed graphs, such as a deterministic finite automaton (DFA) or a Mealy machine \cite{sec15-tlsstatefuzzing,ccs16-tlsattacker,tdsc17-differencialTLS,sec20-dtls,atc20-sweyntooth,sec20-frankenstein,wisec21-zfuzzer,tdsc22-greyhound}. 
In these graphs, nodes represent the internal states of an entity, while edges represent state transitions caused by receiving or sending certain types of messages.}
By referring to the communication model, protocol fuzzers can be aware of the current target state, and can generate testcases according to the data model of message types that are acceptable in the current state, thereby improving the effectiveness of testcases. Note that a protocol implementation may have multiple communication models, as it may behave differently depending on its working mode or configurations.
For example, a Wi-Fi device can be configured to run in AP-mode, STA-mode, or P2P-mode \cite{blackhat-as-20-xie,blackhat-eu-21-owfuzz} and a SIP implementation can be configured as a client, server, or proxy \cite{defcon21-Ozavci}, each of them reacts differently to requests,  thus having a different communication model.
Most existing work treats these implementations running in different configurations as different targets: They construct one communication model for one given configuration.

\begin{figure}[h] 
\centering
\includegraphics[width=0.8\textwidth]{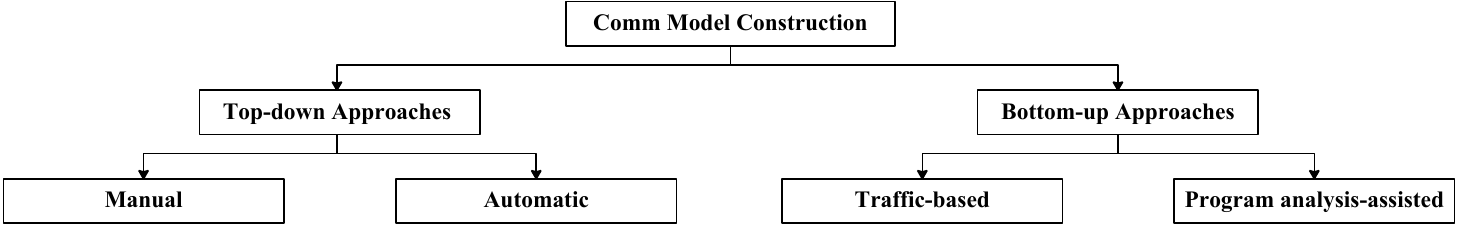}
\caption{Taxonomy of communication model construction techniques.}
\label{fig.section3}
\end{figure}

In this section, we provide a taxonomy of the existing work based on its communication model construction methods.
As shown in Fig.~\ref{fig.section3}, they are divided into two categories: \textit{(i) top-down approaches}, and \textit{(ii) bottom-up approaches}.

\subsubsection{Top-Down Approaches} Top-down approaches construct the protocol communication model by learning from the textual description of the protocol, such as specifications or documents. 
Top-down approaches require protocol specifications as input, thus \newnote{are} mostly used by open protocols.  
Benefiting from global protocol knowledge of the specifications and the precisely defined states and transitions, communication models constructed by top-down approaches are relatively complete and accurate.
It is worth noting that the constructed communication model may still differ from the implementation's. 
This is because developers may customize or extend the design described in the specification according to the practical situation. 
This difference may affect the final fuzzing performance.
Methodologically, there are \textbf{manual} and \textbf{automatic} ways to construct a communication model from the protocol documents. 

\textbf{Manual Construction} (labeled as ``\textit{manual}'' in Table \ref{table-inputgen} column 4): 
Most existing works construct a communication model manually with considerable domain expertise \cite{sp15-composite,dsn15-snake,ccs16-tlsattacker,sp17-iotcube,ndss18-tcpcongestion,asiaccs19-mossot,atc20-sweyntooth,sec20-frankenstein,raid20-abbrate,atc21-tcpfuzz,tdsc22-greyhound,dsn22-l2fuzz,defcon27-fuzzowski,9868872,amusuo2023systematically,ammann2024dy}. 
For example, Garbelini \textit{et al.} \cite{atc20-sweyntooth} and GREYHOUND \cite{tdsc22-greyhound} construct a holistic state machine of Bluetooth Low Energy (BLE) and Wi-Fi by referring to the core design of the protocol specifications \cite{802.11,chen2005wireless,802.1x,blespec} to guide fuzzing. 
Though manually constructing a state machine is an error-prone and labor-intensive task, the benefit here is that human experts can flexibly customize (tailor or extend) the state machine to maximize its effect toward the work's goal.
For example, to detect state machine bugs caused by incorrect multiplexing between different protocol modes, \textit{e.g.}, different protocol versions, extensions, authentication modes, or key-exchange methods, Beurdouche \textit{et al.} \cite{sp15-composite} manually construct a composite state machine including all valid state transitions across protocol modes. 
That composite state machine is then used to generate deviant traces as testcases to discover invalid state transitions. Similarly, FuzzPD \cite{kim2023fuzz} is meticulously designed to accommodate the unique dual-role characteristic inherent in the USB power delivery protocol (USBPD), where each device simultaneously functions as both a power source and a power sink. By integrating the state machines of these two roles, FuzzBD is able to support seamless switching of power roles on the fly during the fuzzing process. 
Unlike the above-discussed works, some work chooses to learn partial information of a state machine for guidance.
Zou \textit{et al.} propose TCP-Fuzz, a novel approach that incorporates 15 dependency rules manually extracted from RFC documents. These rules encompass various dependencies, including packet-to-packet, syscall-to-packet, and syscall-to-syscall interactions. Using these rules, TCP-Fuzz adeptly generates testcases by simultaneously producing the interdependent packets and syscalls.
Another example is L2Fuzz \cite{dsn22-l2fuzz}. The authors construct a map delineating the valid commands pertinent to each of the 19 states identified in the protocol. This mapping facilitates the generation of testcases that are specifically tailored to produce commands acceptable in the current state, thereby enhancing the relevance and effectiveness of the testing process. 
Some works also address the problem that the communication models between the specification and implementation are not completely equal.
Using heterogeneous Single-Sign-On (SSO) platforms as an example, MoSSOT \cite{asiaccs19-mossot} constructs a state machine of a regular SSO process first and then analyzes the practical SSO network traffics of different SSO platforms to learn the implementation details, such as key parameters in each action. These implementation details refine the state transition conditions in the state machines of different SSO platforms. 

\textbf{Automatic Construction:} To automate the error-prone and labor-intensive process of manual communication model construction, some works automatically retrieve the semantic constraint from the protocol specification \cite{icse19-restler,sp22-auto-fsm-extraction}. 
For example, RESTler \cite{icse19-restler} learns the message dependency relationships based on the return types from Swagger specification, which is a structural specification format describing the RESTful API endpoints, methods, parameters and return types. 
Pacheco \textit{et al.} propose to use Natural Language Processing (NLP) to extract a finite state machine (FSM) from the protocol specifications \cite{sp22-auto-fsm-extraction}. 
Note that these two papers are not listed in Table \ref{table-inputgen} since these works are not building stateful protocol fuzzers. 
\newnote{With the advancement of large language model (LLM) technology, numerous studies have begun to employ LLM techniques to automate the learning and comprehension of protocol specifications, such as mGPTFuzz\cite{sec24-mgptfuzz}, CHATAFL\cite{ndss24-chatafl} and LLMIF\cite{sp24-llmif}. These approaches aim to provide guidance during fuzzing processes, particularly in inferring the current protocol state and generating appropriate test messages. The integration of LLMs offers a promising avenue to improve the efficiency and effectiveness of protocol fuzzing by leveraging their sophisticated language understanding capabilities to interpret complex protocol specifications and guide the testing strategy accordingly.}

\subsubsection{Bottom-Up Approaches} 

Bottom-Up approaches provide another solution for communication model reconstruction.
These approaches utilize the observable information of a protocol implementation to reconstruct the communication model.
Since they do not rely on textual documentation or specifications, they are suitable for proprietary protocols. 
Unlike top-down approaches, which have clear definitions of protocol states in the documents, the definitions of states in bottom-up approaches are purpose-specific and may vary among use cases, methods, and implementations. 
For example, AFLNET \cite{icst20-aflnet} determines a protocol state according to the status code of the PUT's response. 
Another example is StateAFL \cite{arxiv21-stateafl}, which groups the memory layout of long-lived memory as different states.
From the learning source's point of view, these methods can be divided into two categories, namely \textbf{ traffic analysis-based approaches} and \textbf{ program analysis-assisted approaches}. 

\textbf{Traffic-Analysis-Based Approaches:} The traffic-analysis-based methods focus on reconstructing the protocol communication model purely from the observed network traffic traces \cite{sec15-tlsstatefuzzing,icst19-seqfuzzer,tecs19-polar,sec20-dtls,icst20-aflnet,sec21-mpinspector}.
This kind of approach is easy to operate and works well in cases that the program execution cannot be traced, \textit{e.g.}, cannot obtain the firmware containing the target program. 
The traffic analysis-based communication model construction approaches can be \textit{passive} or \textit{active}. 

\begin{itemize}
    \item \textit{Passive learning} (labeled as ``\textit{TAPL}'' in Table \ref{table-inputgen} column 4) methods mainly rely on a set of pre-collected network traces of the PUT with other entities to infer the communication model \cite{securecomm15-pulsar,icics17-fan,icst19-seqfuzzer,iot22-cgfuzzer}. 
    The learning algorithms proposed by existing works can be divided into two categories: statistics-based and neural network-based algorithms.
    For the former category, Pulsar \cite{securecomm15-pulsar} builds a second-order Markov model by computing the probability of the occurrence of the adjacent messages in the network trace corpus and then minimizes this Markov model into a DFA. After receiving a message, Pulsar matches it with one of the states in the inferred DFA to select a valid response template for building a new testcase.   
    For the latter category, Fan \textit{et al.} \cite{icics17-fan} and SeqFuzzer\cite{icst19-seqfuzzer} use LSTM to learn the grammar and temporal features of stateful protocols. Specifically, they employ \newnote{Long short-term memory (LSTM)} as the encoder and decoder of the sequence-to-sequence (seq2seq) model \cite{sutskever2014sequence}. Seq2seq model is an encoder-decoder model structure that can handle input and output sequences of different lengths. The encoder LSTM model learns the features of the protocol via captured network traces, while the decoder LSTM model is used to generate fuzzing inputs. 
    \textit{Passive} network trace-based state machine learning methods are easy to operate and fast-running. However, the quality of the constructed state machine depends on the coverage of captured traffics. 
    In practice, it is hard to capture a comprehensive set of message types and sequences, causing the constructed communication model to lack parts of uncaptured states or state transitions. 
     
    \item \textit{Active learning} (labeled as ``\textit{TAAL}'' in Table \ref{table-inputgen} column 4) methods involve learning the communication model during the fuzzing process \cite{sec15-tlsstatefuzzing,esorics18-extendstatefuzzing,eurospw18-openvpn,sec20-dtls,icst21-aichernig,icst22-dtlsfuzzer,sagonas2023edhoc,icdcs23-dp-reverser,10001673,icst20-aflnet,sec23-bleem,sec22-braktooth}. 
    These approaches can be categorized on the basis of whether the global state set is predefined. 
    \textbf{The first category does not define the global state set in advance}, \textit{i.e.}, meaning it does not predetermine the number and nature of possible states in the state machine. This approach employs automata active learning algorithms to discern the state machine of the target. The learning algorithms are based on user-defined input/output alphabets and mappers between alphabets and concrete messages. Starting from an empty state machine, these algorithms iteratively propose and refine the model by interacting with the target protocol implementation, ceasing only when no counterexamples to the learned state machine are found. 
    Most works in this category \cite{sec15-tlsstatefuzzing,esorics18-extendstatefuzzing,eurospw18-openvpn,sec20-dtls,icst21-aichernig,icst22-dtlsfuzzer,sagonas2023edhoc} utilize Angluin's $L^*$ algorithm, defining input alphabets based on protocol specifications and translating these into actual messages using message templates. 
    \newnote{DYNPRE \cite{ndss24-dynpre} incorporates an adaptive message rewriting technique to manage session-specific identifiers while interacting with the server. It uses byte-level mutations and server feedback to deduce the meaning and format of messages, identify different message types, and create a precise and minimal protocol state machine based on observed message patterns.}
Conversely, \textbf{the second category pre-defines the state set} through a rule-based method to circumvent the complexities of automata learning algorithms, and learns the transitions between states by mutating known messages. For example, AFLNET \cite{icst20-aflnet} uses response message status codes to infer the current protocol state, mutating real message sequences to uncover transitions. Bleem \cite{sec23-bleem} utilizes the Scapy library to parse messages and abstract them into various message types by retaining all fields of the enumeration type. This strategy is based on empirical observations from more than 50 protocols supported by Scapy, where different enumeration field values typically signify distinct packet or frame types. Bleem then uses these abstracted message traces to construct a guiding graph for fuzzing. Another example is Braktooth \cite{sec22-braktooth}, which defines eight rules that map messages to states based on message characteristics. It operates as a proxy between the PUT and a standard protocol stack, mutating communications to explore additional state transitions. Similarly, Garbelini \textit{et al.}\cite{10001673} establish mapping rules to identify states and learn the state machine using capture traces (\textit{i.e.}, pcap files). 
\end{itemize}

\textbf{Program-Analysis-Assisted Approaches} (labeled as ``\textit{PAL}'' in Table \ref{table-inputgen} column 4): Compared with traffic-analysis-based approaches, program-analysis-assisted approaches additionally use internal execution information to construct the communication model. 
In general, internal execution information includes the results of static and dynamic program analysis, which requires not only the access to the program but also the availability of analysis frameworks such as program instrumentation tools.
Based on the type of internal execution information used, existing work can be divided into \textit{execution-trace-based} and \textit{state-variable-based}.
\textit{Execution-Trace-Based} approaches recognize different internal execution states according to the execution trace of the target.
For example, ICS$^3$Fuzzer\cite{acsac21-ics3fuzzer} dynamically instruments the target supervisory software to collect the trace.
By comparing the identity of the execution traces, ICS$^3$Fuzzer can distinguish whether the PUT is in a different state. 
The \textit{state-variable-based} approaches detect protocol state transitions by tracking the value changes of state variables during input processing \cite{arxiv21-stateafl,sec22-stateinspector,sec2022stateful,qin2023nsfuzz}.
These approaches are based on the simple observation that most protocol implementations use certain variables to store the current state. 
Therefore, they identify these variables as state variables and use their values to distinguish between different states. 
For example, StateAFL \cite{arxiv21-stateafl} identifies possible state variables by identifying long-lived data structures in memory snapshots.
Similarly, STATEINSPECTOR \cite{sec22-stateinspector} identifies state variables by locating memory regions in heap memory that kept the same values in the execution of each message sequence.
Differently, SGFuzz \cite{sec2022stateful} identifies state variables through regular expressions by automatically extracting all \textit{enum} type variables that are assigned at least once. The insight behind this approach is based on the investigation that most protocol implementations use \textit{enum}-type state variables. 
\newnote{STATELIFTER \cite{sec23-statelifter} views the loops in protocol parsers as the core of the state machine. Through static analysis, it traverses the loop structures in the code and maps the different paths that each loop iteration may execute to distinct states. The dependencies between loop iterations are treated as state transitions within the state machine.
ParDiff \cite{zheng2024pardiff} uses static symbolic analysis to extract finite state machines (FSMs) from protocol implementations by identifying and converting the relevant message parsing constraints into ordered state transitions.}

\newnote{Some works have also used program analysis-assisted approaches to construct the data models of protocol implementations \cite{ccs07-polyglot,sp09-prospex,tdsc23-spenny,ccs23-netlifter}. Polyglot \cite{ccs07-polyglot} leverages dynamic binary analysis to extract protocol information by closely monitoring how a program processes network data, revealing intricate protocol semantics and structures without source code dependency. Recent work Netlifter\cite{ccs23-netlifter} uses static analysis to derive precise protocol specifications directly from source code, employing Abstract Format Graphs (AFG) to capture and visualize complex data structure relationships and dependencies with high accuracy. Spenny\cite{tdsc23-spenny} integrates dynamic analysis with symbolic execution to precisely reverse-engineer industrial control system protocols.}

\subsection{Task Scheduling}
\label{subsec:schedule}

In the realm of recent protocol fuzzing research, the scheduling phase has been distinctly categorized based on the methodology employed to handle state-related complexities. This classification leads to two primary categories: \textit{\textbf{Hierarchical Approaches}} and \textit{\textbf{Monolithic Approaches}}. 

\textbf{Hierarchical Approaches} decompose the scheduling process into two discrete phases: (1) Inter-state scheduling: This phase involves selecting a state to fuzz using a state scheduling algorithm based on the priority or relevance of the state. (2)Intra-state scheduling: Once the target state is selected, a general scheduling algorithm is applied to optimize fuzzing within that state.
By separating these phases, hierarchical approaches allow for more nuanced control over the fuzzing process. For example, if we want to test a protocol after a handshake is completed, the inter-state scheduling phase will first prioritize this state. Then, the intra-state scheduling phase will generate specific test cases to be executed within the post-handshake state using traditional scheduling methods (e.g., seed scheduling, byte scheduling, mutation strategy scheduling). In this paradigm, the heuristics used by the scheduling process mainly fall into three categories, namely \textbf{\textit{rarity-preferred}} (SRHS in Table \ref{table-inputgen} column 5), \textbf{\textit{performance-preferred}} (SPHS in Table \ref{table-inputgen} column 5), and \textbf{\textit{complexity-preferred}} (SCHS in Table \ref{table-inputgen} column 5), as detailed in Table \ref{tab:scheduleinfo}.
\textbf{\textit{Rarity-preferred}} heuristics allocate more resources to seldomly exercised states, hypothesizing that these states harbor more undiscovered adjacent states or code logics \cite{icst20-aflnet,arxiv21-stateafl,nsdi19-act,borcherding2023bandit}. 
\textbf{\textit{Performance-preferred}} heuristics prioritize states demonstrating higher code coverage or bug discovery rates \cite{feast19-yurong,icst20-aflnet,arxiv21-stateafl,atc20-sweyntooth,tdsc22-greyhound}. 
Furthermore, some works utilize \textbf{\textit{complexity-preferred}} heuristics, favoring states with greater complexity (\textit{i.e.}, connected to more basic blocks) or deeper states (\textit{i.e.}, further from the initial state) \cite{securecomm15-pulsar,acsac21-ics3fuzzer}.
For example, ICS$^3$Fuzzer \cite{acsac21-ics3fuzzer} inclines to choose the deeper states and those states that exercise more basic blocks. 
As a generation-based fuzzer, Pulsar \cite{securecomm15-pulsar} calculates the weight of all states that can be reached from the current state and then selects the state that has the maximum weight to be tested next.
In detail, the weight of a state is calculated as the sum of all mutable fields in a fixed number of transitions.  
However, since all these state selection algorithms are implemented and evaluated separately on different platforms and targets, it is difficult to make a fair comparison and achieve conclusive findings.
Liu \textit{et al.} \cite{dongge-state-selection} evaluate the three existing state selection algorithms of AFLNet \cite{icst20-aflnet}, including a rarity-preferred algorithm, an algorithm that randomly selects states, and a sequential state selection algorithm. They find that these algorithms achieved very similar results in terms of code coverage. They attribute the reasons to the coarse-grained state abstraction of AFLNET and the inaccurate estimation of the state productivity. 
Therefore, they propose the AFLNETLEGION algorithm \cite{dongge-state-selection} to address these issues, which is based on a variant of the Monte Carlo tree search algorithm \cite{ase20-legion}. 

In contrast, \textbf{monolithic approaches} employ a single, unified scheduling phases where state-related information is integrated directly into the scheduling algorithm. This means that the scheduler considers the performance of seeds in various states as part of its decision-making process, rather than treating state transitions and test case generation as separate steps.
For example, SGFuzz \cite{sec2022stateful} divides the states into rare states and normal states according to the exercised times. When assigning energy to seeds, it calculates the proportion of the rare states that are exercised by each seed and adds this proportion as one of the parameters on the basis of the original power scheduling algorithm. In a similar way, SGFuzz assigns more energy to the seeds containing state transitions that correspond to the expected protocol behaviors.
This is because SGFuzz expects that these valid state transitions are easier to be mutated to other invalid state transitions, thus incurring error handling logic. 
Similarly, LTL-Fuzzer \cite{icse22-ltlprop} also schedules the entire seed, prioritizing seeds that are closer to the target code locations during execution. 

In addition to state-related information, many works utilize other categories of information for scheduling purposes. However, the scheduling algorithms based on these categories of information are generally universal and have been well discussed in the literature \cite{tse19-fuzzing-survey,fuzzingsurvey-roadmap}. Therefore, we did not discuss them in detail.

\begin{table}[t]
\centering
\caption{Categories of scheduling related information}
\label{tab:scheduleinfo}
\resizebox{\textwidth}{!}{%
\renewcommand{\arraystretch}{1.2}
\begin{tabular}{|l|l|l|l|}
\hline
\textbf{Scheduling Type} & \textbf{Infomation} & Hierarchical & Monolithic \\ \hline
\textit{\textbf{Rarity-preferred}} & State exercised times & \cite{icst20-aflnet,arxiv21-stateafl,nsdi19-act} & \cite{sec2022stateful} \\ \hline
\textit{\textbf{Performance-preferred}} & Contribution to new code coverage, Contribution to new state coverage, Contribution to new bugs & \cite{feast19-yurong,icst20-aflnet,arxiv21-stateafl,atc20-sweyntooth,tdsc22-greyhound} & - \\ \hline
\textit{\textbf{Complexity-preferred}} & Count of connected basic blocks, Depth of state, Mutation opportunities & \cite{securecomm15-pulsar,acsac21-ics3fuzzer} & - \\ \hline
\textit{\textbf{Others}} & Distance from the key statement & - & \cite{icse22-ltlprop} \\ \hline
\end{tabular}%
}
\end{table}

\subsection{Testcase Construction}
\label{subsec:input}

The construction strategy used in protocol fuzzing can be categorized into \textit{packet-level} and \textit{sequence-level}.

\subsubsection{Packet-Level Construction Strategy (labeled as \textit{``Packet"} in Table \ref{table-inputgen} column 6)}
Packet-Level construction strategies of protocol fuzzers basically inherit the common strategies of general fuzzers. For example, elements such as relation, fixup and transform in Peach Fuzzer \cite{peach} are often used to describe the relationships between protocol fields such as length, checksums, and encoding transformations. General mutation methods like bit flip and set zero are also commonly used in generating packet-level testcases.
In this paragraph, more consideration is given to the construction strategies that take advantage of protocol-specific characteristics to \textit{ reduce input space} or \textit{ improve the effectiveness of bug triggering}. 
For the former purpose (\textit{i.e., reducing input space}), SPIDER\cite{arxiv22-SPIDER} leverages the domain-specific insight that most Openflow messages trigger new system events in existing SDN controllers that affect state computation and resource footprints.
Based on the insight, SPIDER can directly generate event sequences rather than generating Openflow messages, significantly reducing the input space.
In addition, L2Fuzz \cite{dsn22-l2fuzz} divides the L2CAP packet format into the field that can be mutated and keeps the other fields unchanged to generate testcases that are less likely to be rejected. IPSpex \cite{zheng2022ipspex} combines network traffic and execution traces of network packet construction to extract the message field semantics of ICS protocols.
Strategies targeting the later purpose (\textit{i.e., improving the effectiveness of triggering bugs}) are mainly heuristics summarized from practices.
For example, EmNetTest \cite{amusuo2023systematically} systematically generates validly constructed packets with invalid header fields or truncated headers. The insight behind this strategy is gained from a comprehensive study of 61 reported vulnerabilities in Embedded Network Stacks (ENS). 
Similar strategies are mentioned in many industry conference works. BadMesher \cite{blackhat-eu-21-badmesher} adopts several domain-specific strategies, such as setting the length field to margin values, and randomly deleting some fields, to improve the effectiveness of triggering bugs in Wi-Fi mesh devices.
Yen \textit{et al.} \cite{blackhat-eu-21-dds} find that some strategies such as mutating the ID field to a nonexisting ID, changing the port number or length field to a boundary value (\textit{e.g.}, 0xFF/0x00), and changing IP to some random addresses, can be quite effective in fuzzing Data Distribution Service (DDS) protocol. 
Similarly, BrokenMesh \cite{blackhat-us-22-brokenmesh} adopts some strategies such as mutating the packet count or the length field in fuzzing the Bluetooth Mesh protocol. 
\newnote{TaintBFuzz \cite{esorics23-taintbfuzz} employs static taint analysis to identify the connections between Zigbee protocol fields and the variables used to make path decisions in Zigbee implementations, thereby prioritizing mutations that are anticipated to reveal additional code paths and improve the efficacy of testing.}
\newnote{MPFuzz \cite{emsoft24-mpfuzz} uses a global synchronization mechanism to share key field information across different fuzzing instances and performs targeted mutations based on the semantic characteristics of these fields, making it easier to discover potential vulnerabilities.}

\subsubsection{Sequence-Level Construction Strategy (labeled as \textit{``Sequence"} in Table \ref{table-inputgen} column 6)}  
Protocol fuzzers may adopt some sequence-level construction strategies.
These strategies proactively construct message sequences that deviate from the regular protocol state machine, expecting to trigger more non-memory-safety bugs of the PUT.
\textbf{Generation-based fuzzers} and \textbf{mutation-based fuzzers} operate differently in sequence-level construction. 

\begin{enumerate}
    \item \textbf{Generation-Based Fuzzers}: These fuzzers construct message sequences leveraging established protocol knowledge, such as standard state machines and inter-message dependency relationships. Notable examples include works \cite{sp15-composite,ccs16-tlsattacker,blackhat-us-17-wifuzz,acsac20-dpifuzz} that generate aberrant message traces by applying strategies like the addition or removal of random protocol messages to valid sequences derived from standard state machines. Projects like Sweyntooth \cite{atc20-sweyntooth}, Greyhound \cite{tdsc22-greyhound}, and Braktooth \cite{sec22-braktooth} meticulously monitor state transitions of PUT and strategically inject valid packets at incorrect states to elicit anomalies, in accordance with the state machine model. Recent research by Fiterau-Brostean \textit{et al.}\cite{fiterau2023automata} proposes a novel method for detecting state machine bugs by inputting a catalog of finite automatons which indicate certain types of state machine bugs, as well as a model of the PUT's. It can then analyze the models and produces testcases that expose the bug. 
    \item \textbf{Mutation-Based Fuzzers}: These fuzzers predominantly employ simple yet effective strategies to mutate the message sequences of seeds. This includes techniques such as packet shuffling, random insertion, or deletion. For instance, AFLNET \cite{icst20-aflnet} constructs message sequences by maintaining a pool of messages from network traces that can be integrated into or substituted for existing seeds. AFLNET further employs a blend of byte-level and sequence-level operators, including replacement, insertion, duplication, and deletion of messages, to craft the message sequences. Similarly, DYFuzzing \cite{amusuo2023systematically} mutates seeds and applies Dolev-Yao (DY) attacker strategies.
    Frankenstein \cite{sec20-frankenstein} reorganizes known message sequences to enhance code coverage. He \textit{et al.} \cite{9868872} propose a unique fuzzer for the 5G non-access-stratum (NAS) protocol, which extracts packets from captures into a structured message table. This fuzzer then applies different mutation techniques tailored to the types of key fields within the protocol messages, significantly enhancing the intelligence and precision of the message mutation process. For example, length fields are mutated combining boundary and intermediate values, including 0, maximum, minimum, and random intermediate values.
    It is important to note that mutation-based fuzzers must judiciously manage the correlation of specific fields in message sequences, such as session numbers, counters, or timestamps. Indiscriminate mutations in these fields could render the input ineffective and lead to early rejection. To address this challenge, AFLNET \cite{icst20-aflnet} modifies the code of the PUT to use a fixed session number \footnote{https://github.com/aflnet/aflnet/blob/master/README.md\#step-0-server-and-client-compilation--setup}, thus ensuring the effectiveness of the fuzzing process. 
\end{enumerate}
\section{Executor}
\label{sec:executor}

\begin{figure}[!tp] 
\centering
\includegraphics[width=0.8\textwidth]{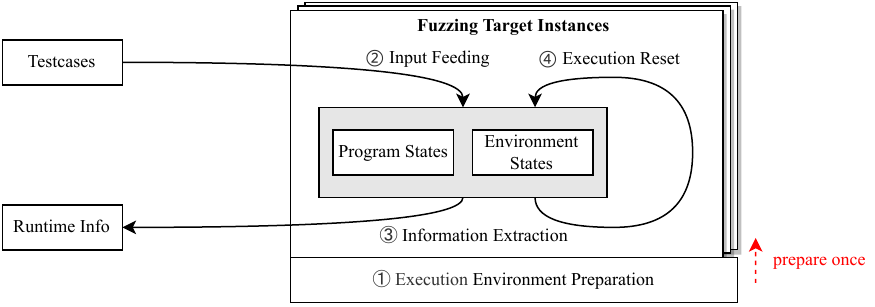}
\caption{Detailed workflow of executors in protocol fuzzing.}
\label{fig-executor}
\end{figure}

\begin{table}[!t]
\centering
\caption{Protocol fuzzers and their optimization of executor.  }
\label{table-executor}
\centering
\resizebox{\textwidth}{!}{%
\begin{tabular}{|l|l|c|l|llll|ll|}
\hline
\multicolumn{1}{|c|}{\multirow{2}{*}{\textbf{Year}}} & \multicolumn{1}{c|}{\multirow{2}{*}{\textbf{Work}}} & \multicolumn{1}{c|}{\multirow{2}{*}{\textbf{Tax}}} & \multicolumn{1}{c|}{\multirow{2}{*}{\textbf{Target}}} & \multicolumn{4}{c|}{\textbf{Efficient Execution}} & \multicolumn{2}{c|}{\textbf{Runtime Info Extraction}} \\ \cline{5-10} 
\multicolumn{1}{|c|}{} & \multicolumn{1}{c|}{} & \multicolumn{1}{c|}{} & \multicolumn{1}{c|}{} & \multicolumn{1}{c|}{\textbf{Env Prep}} & \multicolumn{1}{c|}{\textbf{Input Feeding}} & \multicolumn{1}{c|}{\textbf{\begin{tabular}[c]{@{}c@{}} Execution \\ State Reset\end{tabular}}} & \multicolumn{1}{c|}{\textbf{\begin{tabular}[c]{@{}c@{}} Execution \\ Env Reset\end{tabular}}} &  \multicolumn{1}{c|}{\textbf{Runtime Info}} & \multicolumn{1}{c|}{\textbf{\begin{tabular}[c]{@{}c@{}} Monitoring \\ Method\end{tabular}}} \\ \hline
2014 & \textbf{Gorenc \textit{et al.}}\cite{defcon22-sms} & \CIRCLE & SMS/MMS\cite{mmsdoc} & \multicolumn{1}{l|}{HIL} & \multicolumn{1}{l|}{OTA} & \multicolumn{1}{l|}{VMSR} & DR & \multicolumn{1}{l|}{-} & - \\ \hline
2017 & \textbf{WiFuzz}\cite{blackhat-us-17-wifuzz} & \CIRCLE & Wi-Fi & \multicolumn{1}{l|}{HIL} & \multicolumn{1}{l|}{OTA} & \multicolumn{1}{l|}{RM} & - & \multicolumn{1}{l|}{State} & Resp \\ \hline %
2018 & \textbf{TCPWN}\cite{ndss18-tcpcongestion} & \CIRCLE & TCP & \multicolumn{1}{l|}{CUVM} & \multicolumn{1}{l|}{Socket (MiTM)} & \multicolumn{1}{l|}{-} & - & \multicolumn{1}{l|}{State} & Resp \\ \hline
2019 & \textbf{SeqFuzzer}\cite{icst19-seqfuzzer} & \CIRCLE & ICS & \multicolumn{1}{l|}{HIL} & \multicolumn{1}{l|}{Socket (P2P)} & \multicolumn{1}{l|}{-} & - & \multicolumn{1}{l|}{-} & - \\ \hline
2019 & \textbf{MoSSOT}\cite{asiaccs19-mossot} & \CIRCLE & SSO & \multicolumn{1}{l|}{CUVM} & \multicolumn{1}{l|}{Socket (MiTM)} & \multicolumn{1}{l|}{VMSR} & VMSR & \multicolumn{1}{l|}{State} & Resp \\ \hline
2019 & \textbf{Chen \textit{et al.}}\cite{feast19-yurong} & \LEFTcircle & General & \multicolumn{1}{l|}{-} & \multicolumn{1}{l|}{File} & \multicolumn{1}{l|}{PSR} & - & \multicolumn{1}{l|}{State \& Code Cov} & SSI \\ \hline
2019 & \textbf{Fw-Fuzz}\cite{gao2022fw} & \LEFTcircle & General & \multicolumn{1}{l|}{-} & \multicolumn{1}{l|}{Socket (P2P)} & \multicolumn{1}{l|}{ProcR} & - & \multicolumn{1}{l|}{Code Cov} & SDI \\ \hline
2019 & \textbf{Park \textit{et al.}}\cite{blackhat-eu-19-park} & \LEFTcircle & RDP\cite{rdpdoc} & \multicolumn{1}{l|}{-} & \multicolumn{1}{l|}{Virtual Channels} & \multicolumn{1}{l|}{-} & - & \multicolumn{1}{l|}{Code Cov} & SDI \\ \hline 
2020 & \textbf{Exploiting Dissent}\cite{tdsc17-differencialTLS} & \CIRCLE & TLS & \multicolumn{1}{l|}{-} & \multicolumn{1}{l|}{Socket (P2P)} & \multicolumn{1}{l|}{ProcR} & - & \multicolumn{1}{l|}{State} & Resp \\ \hline
2020 & \textbf{DTLS-Fuzzer}\cite{sec20-dtls} & \CIRCLE & DTLS & \multicolumn{1}{l|}{-} & \multicolumn{1}{l|}{Socket (P2P)} & \multicolumn{1}{l|}{MR} & - & \multicolumn{1}{l|}{State} & Resp \\ \hline  
2020 & \textbf{AFLNET}\cite{icst20-aflnet} & \LEFTcircle & General & \multicolumn{1}{l|}{-} & \multicolumn{1}{l|}{Socket (P2P)} & \multicolumn{1}{l|}{MR} & UPSR & \multicolumn{1}{l|}{State \& Code Cov} & Resp \& SSI \\ \hline
2020 & \textbf{SweynTooth}\cite{atc20-sweyntooth} & \CIRCLE & BLE & \multicolumn{1}{l|}{HIL} & \multicolumn{1}{l|}{OTA} & \multicolumn{1}{l|}{ProcR} & - & \multicolumn{1}{l|}{State} & Resp \\ \hline 
2020 & \textbf{Frankenstein}\cite{sec20-frankenstein} & \LEFTcircle & Bluetooth & \multicolumn{1}{l|}{SE} & \multicolumn{1}{l|}{Shared memory} & \multicolumn{1}{l|}{VMSR} & - & \multicolumn{1}{l|}{Code Cov} & SDI \\ \hline
2020 & \textbf{Peach*}\cite{dac20-icsprotocol} & \LEFTcircle & ICS & \multicolumn{1}{l|}{-} & \multicolumn{1}{l|}{Socket (P2P)} & \multicolumn{1}{l|}{ProcR} & UPSR & \multicolumn{1}{l|}{Code Cov} & SSI \\ \hline 
2020 & \textbf{aBBRate}\cite{raid20-abbrate} & \CIRCLE & TCP & \multicolumn{1}{l|}{CUVM} & \multicolumn{1}{l|}{Socket (MiTM)} & \multicolumn{1}{l|}{-} & - & \multicolumn{1}{l|}{State} & Resp \\ \hline
2020 & \textbf{BaseSAFE}\cite{wisec20-BaseSAFE} & \LEFTcircle & LTE & \multicolumn{1}{l|}{SE} & \multicolumn{1}{l|}{Shared-Memory} & \multicolumn{1}{l|}{PSR} & - & \multicolumn{1}{l|}{Code Cov} & SDI \\ \hline
2020 & \textbf{ToothPicker}\cite{woot20-ToothPicker} & \LEFTcircle & Bluetooth & \multicolumn{1}{l|}{HIL} & \multicolumn{1}{l|}{FHPI} & \multicolumn{1}{l|}{ThrdR}  & - & \multicolumn{1}{l|}{Code Cov} & SDI \\ \hline
2021 & \textbf{ICS$^3$Fuzzer}\cite{acsac21-ics3fuzzer} & \CIRCLE & ICS & \multicolumn{1}{l|}{-} & \multicolumn{1}{l|}{Socket (P2P)} & \multicolumn{1}{l|}{ProcR} & - & \multicolumn{1}{l|}{Exec Traces} & SDI \\ \hline
2021 & \textbf{StateAFL}\cite{arxiv21-stateafl} & \LEFTcircle & General & \multicolumn{1}{l|}{-} & \multicolumn{1}{l|}{Socket (P2P)} & \multicolumn{1}{l|}{PSR} & UPSR & \multicolumn{1}{l|}{State \& Code Cov} & SSI \\ \hline
2021 & \textbf{TCP-Fuzz}\cite{atc21-tcpfuzz} & \LEFTcircle & TCP & \multicolumn{1}{l|}{-} & \multicolumn{1}{l|}{Socket (P2P)} & \multicolumn{1}{l|}{-} & - & \multicolumn{1}{l|}{Branch Cov} & SSI \\ \hline
2021 & \textbf{Snipuzz}\cite{ccs21-snipuzz} & \CIRCLE & IoT & \multicolumn{1}{l|}{HIL} & \multicolumn{1}{l|}{Socket (P2P)} & \multicolumn{1}{l|}{MR \& PhyR} & - & \multicolumn{1}{l|}{Code Cov} & Resp \\ \hline 
2021 & \textbf{Z-Fuzzer}\cite{wisec21-zfuzzer} & \LEFTcircle & Zigbee & \multicolumn{1}{l|}{SE} & \multicolumn{1}{l|}{Socket (P2P)} & \multicolumn{1}{l|}{ProcR} & - & \multicolumn{1}{l|}{Code Cov} & SDI \\ \hline
2021 & \textbf{PAVFuzz}\cite{dac21-PAVFuzz} & \LEFTcircle & AV & \multicolumn{1}{l|}{-} & \multicolumn{1}{l|}{Socket (P2P)} & \multicolumn{1}{l|}{ProcR} & UPSR & \multicolumn{1}{l|}{Code Cov} & SSI \\ \hline
2021 & \textbf{Schepers \textit{et al.}}\cite{wisec21-schepers} & \CIRCLE & Wi-Fi & \multicolumn{1}{l|}{-} & \multicolumn{1}{l|}{Virtual Interface} & \multicolumn{1}{l|}{-} & - & \multicolumn{1}{l|}{Code Cov} & SSI \\ \hline
2021 & \textbf{Wu \textit{et al.}}\cite{blackhat-as-21-wu} & \CIRCLE & EV Fast Charging & \multicolumn{1}{l|}{HIL} & \multicolumn{1}{l|}{CAN Bus (MiTM)} & \multicolumn{1}{l|}{-} & - & \multicolumn{1}{l|}{-} & - \\ \hline 
2022 & \textbf{Meng \textit{et al.}}\cite{icse22-ltlprop} & \LEFTcircle & General & \multicolumn{1}{l|}{-} & \multicolumn{1}{l|}{Socket (P2P)} & \multicolumn{1}{l|}{PSR} & - & \multicolumn{1}{l|}{Property-Guided} & SSI \\ \hline
2022 & \textbf{Greyhound}\cite{tdsc22-greyhound} & \LEFTcircle & Wi-Fi & \multicolumn{1}{l|}{HIL} & \multicolumn{1}{l|}{OTA} & \multicolumn{1}{l|}{ProcR} & - & \multicolumn{1}{l|}{State} & Resp \\ \hline 
2022 & \textbf{SGFuzz}\cite{sec2022stateful} & \LEFTcircle & General & \multicolumn{1}{l|}{-} & \multicolumn{1}{l|}{Shared-Memory} & \multicolumn{1}{l|}{-} & - & \multicolumn{1}{l|}{State \& Code Cov} & SSI \\ \hline 
2022 & \textbf{Braktooth}\cite{sec22-braktooth} & \LEFTcircle & Bluetooth & \multicolumn{1}{l|}{HIL} & \multicolumn{1}{l|}{OTA} & \multicolumn{1}{l|}{ProcR} & - & \multicolumn{1}{l|}{State} & Resp \\ \hline 
2022 & \textbf{SNPSFuzzer}\cite{arxiv2022-snpsfuzzer} & \LEFTcircle & General & \multicolumn{1}{l|}{-} & \multicolumn{1}{l|}{Socket (P2P)} & \multicolumn{1}{l|}{PSR} & UPSR & \multicolumn{1}{l|}{Code Cov} & SSI \\ \hline
2022 & \textbf{Nyx-net}\cite{eurosys22-nyxnet} & \LEFTcircle & General & \multicolumn{1}{l|}{CUVM} & \multicolumn{1}{l|}{File} & \multicolumn{1}{l|}{VMSR} & VMSR & \multicolumn{1}{l|}{Code Cov} & HA/SSI \\ \hline
2022 & \textbf{SnapFuzz}\cite{arxiv2022-snapfuzz} & \LEFTcircle & General & \multicolumn{1}{l|}{-} & \multicolumn{1}{l|}{UDS} & \multicolumn{1}{l|}{PSR} & IMFR & \multicolumn{1}{l|}{Code Cov} & SSI \\ \hline
2022 & \textbf{AmpFuzz}\cite{sec22-ampfuzz} & \LEFTcircle & UDP & \multicolumn{1}{l|}{-} & \multicolumn{1}{l|}{Socket (P2P)} & \multicolumn{1}{l|}{-} & - & \multicolumn{1}{l|}{BAF \& Code Cov}  & Resp \& SSI \\ \hline
2022 & \textbf{L2Fuzz}\cite{dsn22-l2fuzz} & \CIRCLE & Bluetooth L2CAP & \multicolumn{1}{l|}{HIL} & \multicolumn{1}{l|}{OTA} & \multicolumn{1}{l|}{-} & - & \multicolumn{1}{l|}{State} & Resp \\ \hline 
2022 & \textbf{Song \textit{et al.}}\cite{defcon30-someip} & \CIRCLE & SOME/IP\cite{someipdoc} & \multicolumn{1}{l|}{HIL} & \multicolumn{1}{l|}{CAN Bus} & \multicolumn{1}{l|}{-} & - & \multicolumn{1}{l|}{State} & Resp \\ \hline 
2022 & \textbf{Charon}\cite{tcad22-charon} & \LEFTcircle & ICS & \multicolumn{1}{l|}{-} & \multicolumn{1}{l|}{Socket (MiTM)} & \multicolumn{1}{l|}{-} & - & \multicolumn{1}{l|}{State \& Code Cov} & Resp \& SSI  \\ \hline 
2023 & \textbf{FieldFuzz}\cite{arxiv22-fieldfuzz} & \CIRCLE & Codesys v3 & \multicolumn{1}{l|}{CUVM} & \multicolumn{1}{l|}{Socket} & \multicolumn{1}{l|}{RM} & - & \multicolumn{1}{l|}{Code Cov} & Resp \\ \hline
2023 & \textbf{BLEEM}\cite{sec23-bleem} & \CIRCLE & General & \multicolumn{1}{l|}{-} & \multicolumn{1}{l|}{Socket (MiTM)} & \multicolumn{1}{l|}{RM} & - & \multicolumn{1}{l|}{State} & Resp  \\ \hline 
2023 & \textbf{NS-Fuzz}\cite{qin2023nsfuzz} & \LEFTcircle & General & \multicolumn{1}{l|}{-} & \multicolumn{1}{l|}{Socket (P2P)} & \multicolumn{1}{l|}{PSR} & UPSR & \multicolumn{1}{l|}{State \& Code Cov} & SSI  \\ \hline 
2023 & \textbf{HNPFuzzer}\cite{fu2023framework} & \LEFTcircle & General & \multicolumn{1}{l|}{-} & \multicolumn{1}{l|}{Shared-Memory} & \multicolumn{1}{l|}{PSR} & UPSR & \multicolumn{1}{l|}{State \& Code Cov} & Resp \& SSI  \\ \hline 
\end{tabular}
}
\begin{minipage}{\linewidth}
  \footnotesize \Circle: Whitebox Fuzzer; \CIRCLE: Blackbox Fuzzer; \LEFTcircle: Greybox Fuzzer; Tax: Taxonomy; HIL: Hardware-In-the-Loop; General: The fuzzer is not designed for a specific type of protocol; CUVM: Commonly Used Virtual Machine; SE: Specialized Emulation; OTA: Over-the-air; UDS: Unix Domain Socket; MiTM: Man-in-The-Middle-based packet injection;  VMSR: Virtual Machine-level Snapshot and Recovery mechanism; ProcR: Process Restart; PhyR: Physical Reset; ThrdR: Thread Restart; DR: Database Reset; PSR: Process-level Snapshot and Recovery mechanism; UPSR: User-Provided Script Reset; HA: Hardware-Assisted mechanism; EOB: Externally-Observable-Behavior-based method; SDI: Software Dynamic Instrumentation; SSI: Software Static Instrumentation; BAF: Bandwidth Amplification Factor; -: Not implemented; RM: Reset Message; IMFR: In-Memory Filesystem Reset; Resp: Responses.
\end{minipage}
\end{table}

In this section, we will introduce the key improvements of protocol fuzzers in the executor in detail. As shown in Fig. \ref{fig-executor}, an executor in protocol fuzzing normally includes four key processes.
First, the executor needs to prepare an executable execution environment for PUT (\ding{172}. Execution Environment Preparation), and then send input to PUT through the input feeding mechanism (\ding{173}. Input Feeding), extract runtime information during the input processing (\ding{174}. Information Extraction), and reset the execution state and the environment state to a specific state after the execution of the current iteration is complete (\ding{175}. Execution Reset).

In Table. \ref{table-executor}, we summarize the key techniques and improvements in efficient execution (including \ding{172}, \ding{173}, \ding{175}) and runtime information extraction (\ding{174}) of the existing protocol fuzzing works. The works in the table are selected from the collection of papers because they are directly related to the executor.

\subsection{Efficient Execution}
\label{subsec:executionacc}

In protocol fuzzing, there are commonly two directions to improve the fuzzing efficiency: 1) establishing an execution environment that enables parallel testing for \textbf{scalability improvement} (\ding{172} in Fig. \ref{fig-executor}); 
2) \textbf{reducing the execution cost} of each testing iteration (\ding{173} and \ding{175} in Fig. \ref{fig-executor}).

\subsubsection{Scalability Improvement} 

Scalable fuzzing, in this context, refers to the capacity to create multiple testing environments for parallel fuzzing. This is crucial in protocol fuzzing, where many fuzzing targets are closely bounded to hardware. The traditional approach for parallel testing of purchasing multiple physical devices can be economically burdensome and inefficient. 
For protocol fuzzing, since many fuzzing targets depend on specialized execution environments, concurrent testing of these targets can only be carried out by purchasing multiple physical devices, leading to high economic costs and waste.

Emulation emerges as a key solution for scalable fuzzing. It offers a virtual execution environment for the PUT, reducing the dependency on specialized hardware and facilitating the creation of numerous parallel testing instances. This capability significantly enhances scalability, allowing for extensive fuzzing operations across multiple environments. 
Some of the protocol fuzzers leverage the existing emulation solutions to scale the fuzzing process (labeled as \textit{``CUVM"} in Table. \ref{table-executor} column 4)\cite{ndss18-tcpcongestion,raid20-abbrate,nsdi19-act,asiaccs19-mossot,eurosys22-nyxnet}.

However, two difficulties hinder the usage of emulation in protocol fuzzing. 
The first is the availability of the protocol implementation binary, as many firmware images are not publicly available. Second, compared to the diversity of hardware, existing emulators can only support a small fraction of them. 
These difficulties lead to a lot of work still performing fuzzing in a hardware-in-the-loop way (labeled as \textit{``HIL"} in Table. \ref{table-executor} column 4) \cite{tecs19-polar,atc20-sweyntooth,ccs21-snipuzz,tdsc22-greyhound,defcon22-sms,dsn22-l2fuzz}. 

Some works address these issues according to the characteristics of different devices (labeled as \textit{``SE"} in Table. \ref{table-executor} column 4). 
For the first challenge, existing work obtains the target binaries by intercepting Over-The-Air (OTA) firmware updates, or extracting using vendor-specific command or debugging ports. 
For example, Frankenstein \cite{sec20-frankenstein} leverages the \textit{Patchram} mechanism, a Broadcom vendor-specific command that can be used to temporarily patch breakpoints to the ROM, to take the memory snapshot of a physical Bluetooth chip and emulate it in an unmodified version of QEMU. 
To address the second challenge, the existing work often uses an approach called rehosting to partially emulate the functionality of the physical hardware \cite{wisec20-BaseSAFE,9678653}. 
For example, BaseSafe \cite{wisec20-BaseSAFE} selectively rehosts several signaling message parser functions utilizing the Unicorn engine, which is a popular CPU emulator \cite{quynh2015unicorn}. 

\subsubsection{Execution Cost Reduction}

Another direction to improve fuzzing efficiency is to optimize the intermediate execution steps in each iteration. 
In the following, we present the progress of existing work towards this direction, which mainly focuses on two subprocedures: \textbf{ input feeding} (\ding{173} in Fig. \ref{fig-executor}) and \textbf{execution reset} (\ding{175} in Fig. \ref{fig-executor}), specifically.

\textbf{Input Feeding.} Input feeding mechanism acts as a pipeline between the input generator and the PUT to pass the testcase to the PUT for parsing and execution. 
According to the Inter-Process Communication (IPC) mechanisms that the communication between Fuzzer and PUT relies on, existing approaches can be roughly divided into four categories, namely \textit{OTA-based}, \textit{socket-based}, \textit{shared-memory-based}, and \textit{file-based} approaches.
\textit{OTA-} and \textit{socket-based} approaches are mostly used when the fuzzer and PUT cannot be deployed on the same physical device. The latter two approaches can be used to speed up input feeding when the PUT and the fuzzer can be deployed on the same device.

\begin{itemize}
    \item \textit{OTA-Based Input Feeding} (labeled as \textit{``OTA"} in Table. \ref{table-executor} column 5). In general, OTA-based input feeding mechanisms are mostly used in the scenario of fuzzing the implementations of protocols that are typically closed in nature and tightly integrated with hardware components, such as Wi-Fi \cite{tdsc22-greyhound,blackhat-us-17-wifuzz,defcon26-RFfuzzer}, Bluetooth (including classic Bluetooth and BLE) \cite{sec22-braktooth,dsn22-l2fuzz,atc20-sweyntooth}, LTE \cite{wisec20-BaseSAFE}, \newnote{Zigbee \cite{ma2023no}}, 4G/5G \cite{10001673,9868872} and SMS/MMS protocols \cite{defcon22-sms}. 
    In this approach, PUT and the fuzzer need to be deployed in adjacent physical spaces and communicate with each other on specific frequency bands. 
    Thus, OTA-based fuzzers require the use of radio-frequency transceiver devices with receive and transmit functions, such as a software-defined radio (SDR) to handle signals over a wide tuning range.
    OTA-based fuzzing provides the capability to test the entire protocol stack including the physical layer.
    However, OTA-based approaches are the slowest among the above-mentioned approaches. 
    Therefore, many wireless protocol fuzzers try to use other input feeding mechanisms to have better performance, which will be introduced in the following.

    \item \textit{Socket-Based Input Feeding} (labeled as \textit{``Socket"} in Table. \ref{table-executor} column 5).  Socket-based input feeding mechanisms are mostly used in protocol implementations based on TCP/IP infrastructures. 
    In common cases under these approaches, the fuzzer and the PUT communicate with each other through IP addresses, via socket mechanisms including TCP socket and UDP socket.
    The socket-based approaches include two deployment modes, one is point-to-point (P2P) communication between the fuzzer and PUT \cite{ndss18-iotfuzzer,tdsc17-differencialTLS,sec20-dtls,icst20-aflnet,arxiv21-stateafl,atc21-tcpfuzz,icse22-ltlprop,arxiv2022-snpsfuzzer,arxiv2022-snapfuzz}. The fuzzer can play the role of a client or server depending on the role of the PUT. 
    The other deployment mode is Man-in-the-middle (MiTM), in which the fuzzer acts as a proxy between two communication parties and performs mutation or injection to the normal communication traffic \cite{ndss18-tcpcongestion,raid20-abbrate}. The MiTM-based input feeding is mostly used in the scenario where the protocol involves certain contextual information (checksum, packet sequences, etc.) that cannot keep valid by mutating static seeds. 
    However, both modes need to address two challenges. 
    First, socket communication is quite heavy and involves lots of context switches. 
    Existing works improve the efficiency of the socket-based input feeding mechanisms by avoiding the use of these expensive network functions. 
    For example, 
    SnapFuzz \cite{arxiv2022-snapfuzz} replaces the original internet socket with UNIX domain socket \cite{coffield1987tutorial}, a lightweight IPC mechanism that does not have the routing, checksum calculation operations that IP sockets have. 
    Second, it's hard for the fuzzer to determine whether the PUT 
    has already finished processing the previous message and is ready to receive the next message. The PUT may reject the messages coming too early when the target is not ready, thus causing the fuzzer to desynchronize from its state machine. 
    To solve this issue, Fiterau-Brostean \textit{et al.} \cite{sec20-dtls} and AFLNET \cite{icst20-aflnet} set static time intervals to wait for the PUT to initialize, process requests, and send responses. 
    However, static timers are too coarse-grained and can waste a lot of time waiting for the timeout, thus slowing down the fuzzing process. 
    SnapFuzz \cite{arxiv2022-snapfuzz} and AMPFuzz \cite{sec22-ampfuzz} develop a more fine-grained method to inspect the state of the socket. Specifically, they use the function call to related network system calls such as $recv()$, $recvfrom()$ as a sign of ready to receive the next message. They monitor all these function calls through binary rewriting and compile-time code instrumentation, and then notify the fuzzer to send the next iteration of input.

    \item \textit{File-Based Input Feeding} (labeled as \textit{``File"} in Table. \ref{table-executor} column 5). File-based input feeding leverages static or dynamic instrumentation techniques to replace heavy network operations with file operations to achieve a performance boost. 
    For example, Yurong \textit{et al.} \cite{feast19-yurong} transform socket communication to file operations using preloading customized libraries \cite{preeny} under the circumstance that the source code of the PUT is not available. Similarly, Nyx-net \cite{eurosys22-nyxnet} injects a library into the target to hook the network functions of the target connection to obtain their associated file descriptors and injected fuzzing input to the right place.      
    \item \textit{Shared-Memory-Based Input Feeding} (labeled as \textit{``Shared-Memory"} in Table. \ref{table-executor} column 5). Shared-Memory-Based input feeding writes the fuzzing input to the address of shared memory and hooks the related functions to read the testcase from shared memory \cite{wisec20-BaseSAFE,sec20-frankenstein,fu2023framework,bars2023fuzztruction}. 
    For example, BaseSafe \cite{wisec20-BaseSAFE} executes each generated testcase in a forked copy of the target process, and the input for each run is copied to the appropriate address in the corresponding child process. Similarly, Frankenstein \cite{sec20-frankenstein} creates a virtual modem to inject custom packets. The fuzzed input is written to the receive buffer in RAM that is mapped to the hardware receive buffer using direct memory access (DMA). Also, HNPFuzzer \cite{fu2023framework} emulates network functions based on shared memory to reduce the time consumption due to message transmission between fuzzer and PUT. 
    \item \textit{Others.} There are also works that rely on specialized communication channels to feed fuzzing inputs. For example, to fuzz the client of the Remote Desktop Protocol (RDP), Park \textit{et al.} leverage the virtual channel, an abstraction layer in RDP that is used to transport data, to actively send fuzzing input from the server to the client \cite{blackhat-eu-19-park}. Song \textit{et al.} use a media converter to convert traffic between Automotive Ethernet and standard Gigabit Ethernet, and fuzz the SOME/IP protocol stack of the electronic control unit (ECU) \cite{defcon30-someip}, which is a control communication protocol between ECUs. 
\end{itemize}

\textbf{Execution Reset.} 
After each iteration of execution, it is necessary to reset the PUT to a specified state and wait for the next iteration of fuzzing. 
This is because each testcase may affect both the internal execution states of the PUT (\textit{e.g.}, global variables) or influence the execution environment (\textit{e.g.}, file system, databases). 
Execution without reset makes the PUT behave more non-deterministicly, making it harder to reproduce the bugs. 
For example, when fuzzing an FTP server, a testcase may cause a file to be created under the shared folder. If the shared folder is not reset, the FTP server will report an error if the following testcase tries to create a file with the same name, which means that the same testcase results in different behavior of PUT.

Based on the analysis of the reset of execution methods of existing work, the process mainly involves three key sub-phases, which are \textit{1. reset time selection}, \textit{2. execution state reset} and  \textit{3. execution environment reset}. 
Firstly, the executor assesses whether the current iteration has been concluded during the \textit{reset time selection} phase. This determination precedes any actions taken to reset the execution. Once it is confirmed that the current execution cycle is complete, the process continues with the \textit{execution state reset} and \textit{execution environment reset}, which reset the runtime state of the PUT and the associated external execution environments, respectively.
Below we summarize the progress of existing works in these three key phases separately.

\textit{1. Reset Strategy Selection.} 
A reset strategy is mainly used to determine the appropriate time or execution point for performing a reset, which has a significant influence on the performance of fuzzing.
An early reset may cause the target to terminate when it is still performing some tasks that may be vulnerable, and a late reset may affect the efficiency of the test. 
A common approach is to set a fixed time interval before resetting the execution. For example, AFLNET \cite{icst20-aflnet} allows the user to manually configure the time delays before restarting the PUT.
    However, this approach is relatively coarse-grained and it is hard to determine an appropriate time interval.
    In order to precisely control when to reset the execution, some works use program analysis to find the location that indicates the end of the execution of an iteration, and instrument the target program to terminate at these code locations \cite{arxiv2022-snapfuzz,eurosys22-nyxnet,sec22-ampfuzz,qin2023nsfuzz}.
    For example, AMPFuzz \cite{sec22-ampfuzz} performs static analysis and injects termination calls into code branches that do not contain message-sending APIs.
    In addition, some works choose not to perform an execution reset after each fuzzing iteration for performance boost.
    For example, Charon \cite{tcad22-charon} leverage a program status inferring module to infer the time point at which the PUT has finished processing the packet, thereby detecting the coverage of specific inputs and avoiding the need to repeatedly restart the PUT to collect feedback. Similarly, SGFuzz \cite{sec2022stateful} does not restart the PUT in every iteration. Instead, it performs a post-analysis to attribute the relationship between inputs and the target program's behavior. Specifically, it collects all the inputs on which the PUT has been executed and minimizes the input list to a minimal message sequence that can trigger the bug. 
    
\textit{2. Execution State Reset.} 
    Execution state reset is responsible for resetting the context of the running PUT process to a specified state, including the data in registers and memory, etc.
    The existing execution state reset mechanisms can be divided into three categories, namely \textit{message-based reset}, \textit{process restart}, and \textit{snapshot \& recovery}.
    
    \textit{Message-based reset} (labeled as \textit{``MR"} in Table. \ref{table-executor} column 6) operates by sending a specific type of message that forces the PUT to terminate the ongoing session and revert to its initial state \cite{blackhat-us-17-wifuzz,sec20-dtls,arxiv22-fieldfuzz}. For instance, when fuzzing the Wi-Fi Access Point (AP), WiFuzz uses a deauthentication message to reset its state \cite{blackhat-us-17-wifuzz}. 
    Message-based reset is easy-to-use, but it only supports a limited set of protocols as not every protocol is designed with a reset message. Furthermore, while it can reset the explicit protocol state of the PUT, it cannot reset the implicit state of the test target, such as global variables and memory that has been allocated but not freed.

    Another commonly used approach to reset the execution state is to \textit{kill the target process and restart} (labeled as \textit{``ProcR"} in Table. \ref{table-executor} column 6) \cite{ndss18-iotfuzzer,dac20-icsprotocol,acsac21-ics3fuzzer}. However, it is a relatively heavy operation for fuzzing, as the restart of the program involves multiple expensive pre-processing steps, such as loading the program into memory, dynamic linking, etc., resulting in inefficiencies.

    The \textit{snapshot \& recovery} mechanism has been integrated into fuzzing. This approach involves checkpointing PUT at a specific runtime state and then resetting it back to that checkpoint after each fuzzing iteration. This method effectively bypasses the repeated execution of resource-intensive initialization operations, thereby enhancing fuzzing efficiency. 
    Protocol fuzzing, in particular, derives significant benefits from snapshot technology. Protocols are predominantly stateful, which implies that the input often comprises multiple prefix messages that guide the PUT to a designated state before introducing the crafted message. It is common for testcases to share the same prefix message sequences, especially when a specific state requires repeated exploration. Implementing snapshot technology in the protocol fuzzing process eliminates redundant executions associated with parsing these shared packet sequences, thereby markedly boosting fuzzing efficiency. 
    The snapshot methodologies used in the current protocol fuzzing research can be broadly categorized into two types: \textbf{\textit{process-level snapshots}} and \textbf{\textit{virtual machine-level snapshots}}.
    
    \begin{itemize}
        \item \textbf{\textit{Process-Level snapshot mechanisms}} (labeled as \textit{``PSR"} in Table. \ref{table-executor} column 6) rely on system call capabilities provided by the operating system to realize their functionality. Generally, based on the APIs used, existing methods can be categorized into two types: \textit{fork-based} and \textit{ptrace-based}. 
        \textit{Fork-based} snapshot mechanisms are widely used in several well-known general-purpose fuzzers, including AFL\cite{afl}. Specifically, AFL inserts a piece of fork-server code into the PUT program binary, which is executed before the $main()$ function. Following a signal from the AFL fuzzing side, the fork-server generates a child process via the $fork()$ function, and this child process continues with the $main()$ function. 
        Since the fork-server has already loaded all kinds of resources, each child process only needs to execute the main function's code, thereby bypassing the costly preprocessing steps and enhancing efficiency.        
        This mechanism has been adopted by many protocol fuzzers for state resetting \cite{tecs19-polar,icst20-aflnet,arxiv21-stateafl,wisec20-BaseSAFE,arxiv2022-snpsfuzzer,arxiv2022-snapfuzz,qin2023nsfuzz,fu2023framework}.         
        Furthermore, some works have extended the original fork-server mechanism in AFL to allow conditional multiple initializations at different code points, enabling the fuzzer to conveniently switch between various states of the protocol and thus boosting the fuzzing process \cite{feast19-yurong,arxiv2022-snapfuzz}.
        \textit{Ptrace-based} snapshot mechanisms, such as CRIU and DMTCP, leverage the debugging API $ptrace()$ to collect all the process context information and save it as image files \cite{arxiv2022-snpsfuzzer}. In the restoration process, these snapshot mechanisms read the dumped image files and recreate the process using syscalls such as $fork()$ or $clone()$. Unlike the fork-based snapshot, which requires predetermined snapshot conditions (\textit{i.e.}, the location of the fork-server call) before execution, ptrace-based snapshots can checkpoint at any state during runtime.               
        \item \textit{\textbf{Virtual-Machine-Level snapshot mechanisms}} (labeled as \textit{``VMSR"} in Table. \ref{table-executor} column 6) utilize the capabilities of virtual machine hypervisors to capture snapshots of the entire virtual machine at a specific time point, typically facilitated through a hypercall \cite{eurosys22-nyxnet,asiaccs19-mossot}. When hypercalls are invoked, the program running within the virtual machine exits the VM context and transfers control to the hypervisor. 
		Although the hypervisor-based approach is user-friendly, requiring no instrumentation, it is somewhat less efficient and more space-consuming because of its large granularity. 
		To enhance the practicality of using virtual machine-level snapshots in protocol fuzzing, Nyx-net \cite{eurosys22-nyxnet} implements an incremental snapshot approach to reduce the overhead associated with creating and removing snapshots. Specifically, Nyx-net establishes a root snapshot in pristine state, and each execution iteration commences from this root snapshot. In subsequent fuzzing iterations, Nyx-net generates incremental snapshots based on the root snapshot following the execution of an input message. Consequently, Nyx-net has a great performance boost on the testcases that share the same prefix message sequences. 
\end{itemize}

\textit{3. Execution Environment Reset.} The reset of the execution environment primarily involves resetting the filesystem or database that may be affected by the PUT. 
Many fuzzers require users to provide a cleanup script to revert all changes \cite{icst20-aflnet,qin2023nsfuzz,fu2023framework,arxiv2022-snpsfuzzer,arxiv21-stateafl}, necessitating substantial manual effort to analyze the PUT's potential impact on the external environment. 
To address this issue, Snapfuzz \cite{arxiv2022-snapfuzz} leverages a custom in-memory filesystem, where modifications are automatically discarded after the completion of a fuzzing iteration. Furthermore, the hypervisor-based snapshot mechanism. 
In addition, the hypervisor-based snapshot mechanism (VMSR in Table. \ref{table-executor} column 7) \cite{eurosys22-nyxnet}, which captures the state of the entire virtual machine, can reset both the execution state and the environment simultaneously.

\subsection{Runtime Information Extraction}
\label{subsec:extraction}
 
In general, the runtime information extraction methods used in existing works can be divided into three categories according to their generality.

\textbf{Hardware-Assisted methods} (labeled as ``\textit{HA}'' in Table \ref{table-executor} column 9) capitalize on the unique capabilities inherent to certain specialized hardware devices to glean runtime information. 
A prime example of this method is demonstrated by Nyx-net \cite{eurosys22-nyxnet}, which employs Intel Processor Trace (Intel PT). This feature, unique to certain high-end Intel CPUs, allows for the detailed recording of software execution aspects, such as control flow paths, thus enabling the comprehensive collection of in-depth coverage information.

\textbf{Software-Based methods} leverage the capability of the software execution environment, \textit{e.g.}, compiler, operating system, virtual machine hypervisor, etc., to obtain the runtime information. Instrumentation is the most commonly used method for realizing runtime information extraction, which inserts information collection function calls into the program at specific code point. 
Program instrumentation can be static (labeled as ``\textit{SSI}'' in Table \ref{table-executor} column 9) or dynamic (labeled as ``\textit{SDI}'' in Table \ref{table-executor} column 9) \cite{acsac21-ics3fuzzer,blackhat-eu-19-park,woot20-ToothPicker}. The former happens before the PUT runs and can be performed at compile time \cite{tecs19-polar,icst20-aflnet,dac20-icsprotocol,arxiv21-stateafl,atc21-tcpfuzz,icse22-ltlprop,sec2022stateful,arxiv2022-snpsfuzzer,eurosys22-nyxnet,wisec21-zfuzzer} or by directly rewriting the binary \cite{arxiv2022-snapfuzz}. 
The latter happens while the PUT is running, leveraging tools such as DynamoRIO \cite{blackhat-eu-19-park} or Frida \cite{woot20-ToothPicker} to inject hooking functions at specific code points to collect runtime information. 

\textbf{Externally-Observable-Behavior-Based methods} are the most general class of methods, as it does not rely on any support of the execution environment and can be used in a blackbox manner. 
There are various externally observable behaviors, such as the output of the program (labeled as ``\textit{Resp}'' in Table \ref{table-executor} column 9) and the side-channel information such as power consumption and response time.
The heuristics behind these observable behavior-based methods are that the differences in these behaviors can represent the PUT is under different states or having gone through different execution paths. 
Specifically, AFLNET \cite{icst20-aflnet} and Fieldfuzz \cite{arxiv22-fieldfuzz} identify different protocol states according to the status code in the response messages. 
Snipuzz \cite{ccs21-snipuzz} and FUME \cite{infocom22-fume} adopt the heuristic that different response messages mean different execution paths. Thus, they keep the input that can cause a different response as a seed for subsequent mutation testing, expecting to increase the coverage.
\newnote{Aafer \textit{et al.} \cite{sec21-androidtv} and Logos \cite{issta24-logos} use the execution logs as a feedback to refine the input generation grammars, as developers usually add log statements to indicate the detailed information about input validation.}
Observing side channel information such as system status, power consumption, and response time, Flowfuzz \cite{blackhat-us-17-flowfuzz} determines whether hardware switches have gone through different execution paths.

\section{Bug Collector}
\label{sec:bugdetector}

\begin{figure}[!htp] 
\centering
\includegraphics[width=0.8\textwidth]{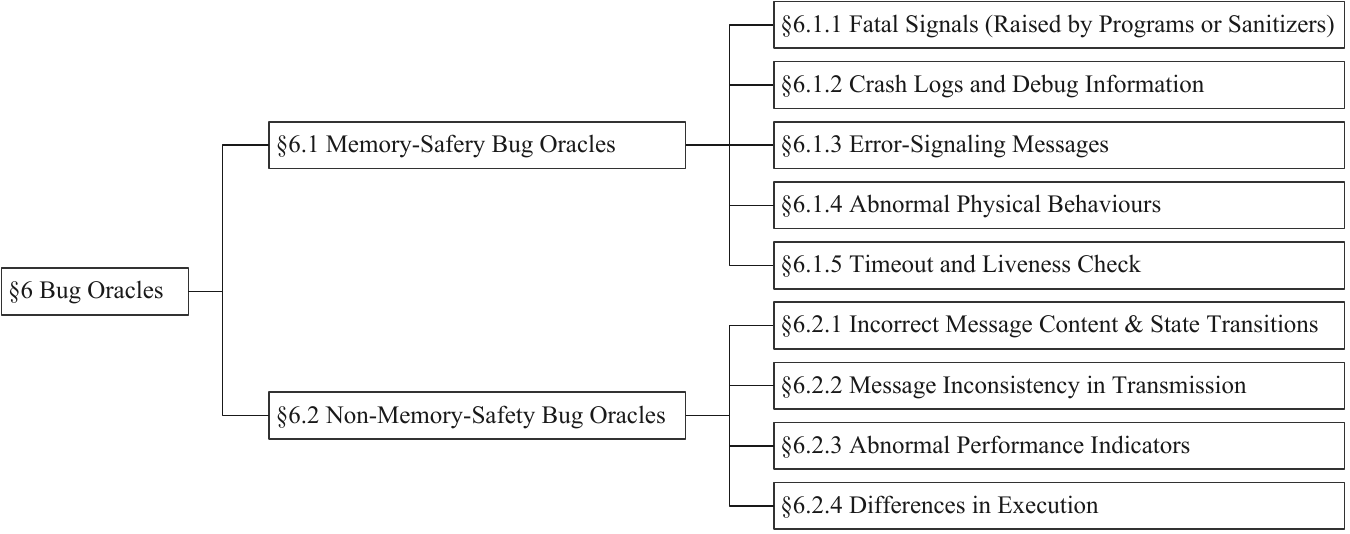}
\caption{Taxonomy of bug oracles in protocol fuzzing.}
\label{fig.oracles}
\end{figure}

To address the challenges in protocol fuzzing, existing work designs both memory-safety bug oracles and non-memory safety bug oracles according to various information sources, as shown in Fig. \ref{fig.oracles}.

\subsection{Memory-Safety Bug Oracles}
\label{subsec:memorybug}

\subsubsection{Fatal Signals (Raised by Programs or Sanitizers).} have been extensively utilized as a pivotal mechanism for bug detection in a plethora of contemporary studies \cite{ccs16-tlsattacker,tecs19-polar,sec20-frankenstein,dac20-icsprotocol,arxiv21-stateafl,atc21-tcpfuzz,wisec21-zfuzzer,sec2022stateful,poncelet2022so}. Predominantly, memory-safety bugs manifest through the overwriting of data or code pointers with invalid values, leading to critical process disruptions such as segmentation faults or process terminations, thereby generating fatal signals like SIGSEGV, SIGABRT, and others. Fuzzers can detect faults by checking whether the PUT process is dying from these signals.
Addressing the subset of memory-safety bugs that do not immediately precipitate program crashes, fuzzers employ sanitizers. Sanitizers are bug detection tools that are specifically engineered to identify and highlight unsafe or undesirable memory access patterns. Upon identification of such anomalies, sanitizers are designed to terminate PUT, thus signaling the presence of a potential bug \cite{serebryany2012addresssanitizer,song2019sok,stepanov2015memorysanitizer,serebryany2009threadsanitizer}.
Sanitizers can be enabled at compile-time\cite{ccs16-tlsattacker,tecs19-polar,sec20-frankenstein,dac20-icsprotocol,arxiv21-stateafl,atc21-tcpfuzz,wisec21-zfuzzer,sec2022stateful} or enabled dynamically at runtime \cite{wisec20-BaseSAFE}.

\begin{table}[!t]
\centering
\caption{Protocol fuzzers and their oracles}
\label{table-oracles}
\centering
\resizebox{\textwidth}{!}{%
\footnotesize
\begin{tabular}{|l|l|c|l|ll|}
\hline
\multicolumn{1}{|c|}{\multirow{2}{*}{\textbf{Year}}} & \multicolumn{1}{c|}{\multirow{2}{*}{\textbf{Work}}} & 
\multicolumn{1}{c|}{\multirow{2}{*}{\textbf{Tax}}} & 
\multicolumn{1}{c|}{\multirow{2}{*}{\textbf{Target}}} & \multicolumn{2}{c|}{\textbf{Bug Detector}} \\ \cline{5-6} 
\multicolumn{1}{|c|}{} & \multicolumn{1}{c|}{} & \multicolumn{1}{c|}{} & \multicolumn{1}{c|}{} & \multicolumn{1}{c|}{\textbf{Memory-Safety Bug Oracles}} & \multicolumn{1}{c|}{\textbf{Non-Memory-Safety Bug Oracles}} \\ \hline
2013 & \textbf{Tsankov \textit{et al.}}\cite{issta13-semi-valid} & \CIRCLE & General & \multicolumn{1}{l|}{Sanitizer} & - \\ \hline
2015 & \textbf{Doona}\cite{doona} & \CIRCLE & General & \multicolumn{1}{l|}{Fatal Signals} & - \\ \hline
2015 & \textbf{Pulsar}\cite{securecomm15-pulsar} & \CIRCLE & General & \multicolumn{1}{l|}{Timeout} & - \\ \hline
2015 & \textbf{Ruiter \textit{et al.}}\cite{sec15-tlsstatefuzzing} & \LEFTcircle & TLS & \multicolumn{1}{l|}{-} & Manual \\ \hline
2015 & \textbf{Beurdouche \textit{et al.}}\cite{sp15-composite} & \CIRCLE & TLS & \multicolumn{1}{l|}{Timeout} & Incorrect State Transitions \\ \hline
2016 & \textbf{TLS-Attacker}\cite{ccs16-tlsattacker} & \CIRCLE & TLS & \multicolumn{1}{l|}{Sanitizer} & Incorrect State Transitions\\ \hline
2017 & \textbf{WiFuzz}\cite{blackhat-us-17-wifuzz} & \CIRCLE & Wi-Fi & \multicolumn{1}{l|}{Timeout} & Incorrect Content \& State Transitions\\ \hline
2018 & \textbf{TCPWN}\cite{ndss18-tcpcongestion} & \CIRCLE & TCP & \multicolumn{1}{l|}{-} & Abnormal Performance Indicators\\ \hline
2018 & \textbf{IoTFuzzer}\cite{ndss18-iotfuzzer} & \CIRCLE & IoT & \multicolumn{1}{l|}{Liveness Check} & - \\ \hline
2019 & \textbf{SeqFuzzer}\cite{icst19-seqfuzzer} & \CIRCLE & ICS & \multicolumn{1}{l|}{-} & Incorrect Content \& State Transitions\\ \hline
2019 & \textbf{ACT}\cite{nsdi19-act} & \LEFTcircle & TCP & \multicolumn{1}{l|}{-} & Abnormal Performance Indicators \\ \hline
2019 & \textbf{MoSSOT}\cite{asiaccs19-mossot} & \CIRCLE & SSO & \multicolumn{1}{l|}{Fatal Signals} & Incorrect State Transitions \\ \hline 
2020 & \textbf{Exploiting Dissent}\cite{tdsc17-differencialTLS} & \CIRCLE & TLS & \multicolumn{1}{l|}{-} & DE \\ \hline
2020 & \textbf{SweynTooth}\cite{atc20-sweyntooth} & \CIRCLE & BLE & \multicolumn{1}{l|}{Crash Logs / Timeout} & Incorrect State Transition \\ \hline
2020 & \textbf{aBBRate}\cite{raid20-abbrate} & \CIRCLE & TCP & \multicolumn{1}{l|}{-} & Abnormal Performance Indicators \\ \hline
2020 & \textbf{DPIFuzz}\cite{acsac20-dpifuzz} & \CIRCLE & QUIC & \multicolumn{1}{l|}{Sanitizer} & DE \\ \hline
2020 & \textbf{BaseSAFE}\cite{wisec20-BaseSAFE} & \LEFTcircle & LTE & \multicolumn{1}{l|}{Sanitizer} & - \\ \hline
2021 & \textbf{ICS$^3$Fuzzer}\cite{acsac21-ics3fuzzer} & \CIRCLE & ICS & \multicolumn{1}{l|}{Crash Logs} & - \\ \hline
2021 & \textbf{TCP-Fuzz}\cite{atc21-tcpfuzz} & \LEFTcircle & TCP & \multicolumn{1}{l|}{Fatal Signals} & Inconsistency in Transmission \& DE  \\ \hline
2021 & \textbf{Snipuzz}\cite{ccs21-snipuzz} & \CIRCLE & IoT & \multicolumn{1}{l|}{Liveness Check} & - \\ \hline
2021 & \textbf{PAVFuzz}\cite{dac21-PAVFuzz} & \LEFTcircle & AV & \multicolumn{1}{l|}{Fatal Signals} & - \\ \hline
2021 & \textbf{Aichernig \textit{et al.}}\cite{icst21-aichernig} & \CIRCLE & MQTT & \multicolumn{1}{l|}{-} & DE \\ \hline
2021 & \textbf{Roitburd \textit{et al.}}\cite{cns21-Roitburd} & \CIRCLE & AnyConnect\cite{ciscodoc} & \multicolumn{1}{l|}{Liveness Check} & - \\ \hline
2021 & \textbf{Owfuzz}\cite{blackhat-eu-21-owfuzz} & \CIRCLE & Wi-Fi & \multicolumn{1}{l|}{Liveness Check \& ESM} & - \\ \hline
2021 & \textbf{BadMesher}\cite{blackhat-eu-21-badmesher} & \CIRCLE & Wi-Fi Mesh & \multicolumn{1}{l|}{Fatal Signals \& Liveness Check} & -\\ \hline
2022 & \textbf{Meng \textit{et al.}}\cite{icse22-ltlprop} & \LEFTcircle & General & \multicolumn{1}{l|}{Fatal Signals} & Incorrect State Transitions \\ \hline
2022 & \textbf{Greyhound}\cite{tdsc22-greyhound} & \LEFTcircle & Wi-Fi & \multicolumn{1}{l|}{Fatal Signals / Timeout} & Incorrect State Transitions \\ \hline
2022 & \textbf{Braktooth}\cite{sec22-braktooth} & \LEFTcircle & General & \multicolumn{1}{l|}{Crash Logs \& APB} & - \\ \hline
2022 & \textbf{L2Fuzz}\cite{dsn22-l2fuzz} & \CIRCLE & Bluetooth L2CAP & \multicolumn{1}{l|}{Crash Logs \& Liveness Check \& ESM} & Incorrect State Transitions \\ \hline 
2022 & \textbf{AmpFuzz}\cite{sec22-ampfuzz} & \LEFTcircle & UDP & \multicolumn{1}{l|}{-} &  Abnormal Performance Indicators \\ \hline
2022 & \textbf{BrokenMesh}\cite{blackhat-us-22-brokenmesh} & \CIRCLE & BLE Mesh & \multicolumn{1}{l|}{Timeout} & - \\ \hline 
2022 & \textbf{PCFuzzer}\cite{liu2022fuzzing} & \CIRCLE & PLC & \multicolumn{1}{l|}{Liveness Check \& APB} & - \\ \hline 
2023 & \textbf{FieldFuzz}\cite{arxiv22-fieldfuzz} & \CIRCLE & Codesys v3 & \multicolumn{1}{l|}{ESM \& Crash Logs \& Timeout} & - \\ \hline
2023 & \textbf{Tyr}\cite{chen2022tyr} & \LEFTcircle & Blockchain & \multicolumn{1}{l|}{-} & Incorrect State Transitions \\ \hline 
2023 & \textbf{LOKI}\cite{ma2023loki} & \LEFTcircle & Blockchain & \multicolumn{1}{l|}{Fatal Signals} & Incorrect State Transitions \\ \hline 
2023 & \textbf{Wang \textit{et al.}}\cite{wang2023nlp} & \CIRCLE & General & \multicolumn{1}{l|}{Crash Logs} & - \\ \hline 
2023 & \textbf{DYFuzzing}\cite{ammann2024dy} & \LEFTcircle & General & \multicolumn{1}{l|}{Sanitizer} & Incorrect Content \& State Transitions \\ \hline 
2023 & \textbf{ResolFuzz}\cite{bushart2023resolfuzz} & \LEFTcircle & DNS & \multicolumn{1}{l|}{-} & DE \\ \hline 

\end{tabular}
}
\begin{minipage}{\linewidth}
  \footnotesize \Circle: Whitebox Fuzzer; \CIRCLE: Blackbox Fuzzer; \LEFTcircle: Greybox Fuzzer; Tax: Taxonomy; General: The fuzzer is not designed for a specific type of protocol; -: Not detectable; ESM: Error-Signaling Messages; APB: Abnormal Physical Behaviors; DE: Differences in Execution.
\end{minipage}
\end{table}

\subsubsection{Crash Logs and Debug Information.} Some works determine whether the PUTs are crashed by analyzing system logs or debug information \cite{atc20-sweyntooth,acsac21-ics3fuzzer,blackhat-eu-21-badmesher,arxiv22-fieldfuzz,sec22-braktooth,dsn22-l2fuzz,wang2023nlp}. These system logs and debug information can be obtained through various channels. 
Specifically, ICS$^3$ Fuzzer uses the Windows EventLog Service to detect crash events on Windows systems \cite{acsac21-ics3fuzzer}. 
Swentooth \cite{atc20-sweyntooth} and Braktooth \cite{sec22-braktooth} propose to collect startup messages or crash messages in the system logs using the debug ports exposed by respective Bluetooth development boards. The startup message is an indicator of program crashes, as Bluetooth devices have a watchdog program to reset the Bluetooth SoC when finding it is unresponsive. Wang \textit{et al.} \cite{wang2023nlp} leverage NLP technology to process the logs and detect unintended behavior of PUT.
Unlike other sources, L2Fuzz \cite{dsn22-l2fuzz} and FieldFuzz \cite{arxiv22-fieldfuzz} identify crashes by checking whether a crash dump was generated.

\subsubsection{Error-Signaling Messages (labeled as ``ESM'' in Table \ref{table-oracles} column 5).} Many protocols use special responses or status codes to indicate internal errors, and therefore can be used for bug detection. For example, L2Fuzz detects Bluetooth L2CAP vulnerabilities by checking whether the received packet contains an error signaling message such as Connection Failed, Connection Aborted, Connection Reset, and Connection Refused \cite{dsn22-l2fuzz}. These error messages indicate that the PUT may be crashed. OWFuzz uses the Deauth / Disassoc frames, the management frames of the Wi-Fi protocol to terminate the communication, as an indicator of anomaly during the fuzzing of the Wi-Fi protocol stacks \cite{blackhat-eu-21-owfuzz}.

\subsubsection{Abnormal Physical Behaviors (labeled as ``APB'' in Table \ref{table-oracles} column 5).} The abnormal physical behavior of the target device, \textit{e.g.}, startup sound, can also be used as a bug oracle. 
For example, when fuzzing the Bluetooth sound device, Braktooth uses the event of repeated startup sound as a bug oracle \cite{sec22-braktooth}. 
This is because Bluetooth devices will be restarted by the watchdog program when an error occurs, and a startup sound will be played during the booting process. Differently, PCFuzzer \cite{liu2022fuzzing} leverages an oscilloscope to collect the physical signal of the output module to monitor the target's status.

\subsubsection{Timeout and Liveness Checks.} 
\label{subsubsec:timeout}

Timeout and liveness checks identify crashes or infinite loops by checking the target's unresponsiveness.
A common way to check the unresponsiveness of a target is to set a static timeout for a response \cite{securecomm15-pulsar,ndss18-iotfuzzer,blackhat-eu-21-owfuzz,blackhat-eu-21-badmesher,ccs21-snipuzz,tdsc22-greyhound,defcon30-someip}. If the response message from the target is not received by the time, it is determined that the target process is dead or enters an infinite loop. This method is suitable for environments with limited debugging techniques, such as unable to obtain process signals or debug logs.
However, setting a fixed timeout is a relatively coarse-grained method, which may introduce false positives due to network fluctuations or excessive load on the target. 
Some works propose several active liveness checks to mitigate the false positive issues. For example, Snipuzz \cite{ccs21-snipuzz} resends input sequences for multiple times to reduce false positives. IoTFuzzer \cite{ndss18-iotfuzzer}, OWFuzz \cite{blackhat-eu-21-owfuzz}, and BadMesher \cite{blackhat-eu-21-badmesher} use heartbeat messages (\textit{e.g.}, ICMP messages) to infer the status of the PUT. 

\subsection{Non-Memory-Safety Bug Oracles}
\label{subsec:statebugdet}
Non-Memory-Safety bugs are bugs that are caused by non-memory access reasons and violate some expected properties, such as logical bugs, RFC violations, or performance-influential bugs.
Non-Memory-Safety bugs are challenging to be identified because they have no uniform observable behavior. 
Detecting non-memory-safety bugs usually requires the user to define the oracle according to the properties destroyed by the target. 
Depending on the properties checked, these oracles can be roughly divided into four categories, namely \textit{incorrect message content \& state transitions}, \textit{inconsistency in transmission}, \textit{abnormal performance indicators}, and \textit{differences in execution}. 
We will describe these techniques in detail in the following subsections. 
It should be noted that although there are various ways to identify possible non-memory-safety bugs, most of these methods can only report suspicious behaviors of the PUT, which still require experts' manual verification to determine impact and exploitability.

\subsubsection{Incorrect Message Content \& State Transitions}
\label{oracle:errorcontent}

\textit{Incorrect Message Content} checks whether the content of the responses violates some semantic constraints.
\textit{Incorrect State Transitions} verifies whether the state transitions are valid or allowed. 
In most cases, these rules are extracted from protocol specifications or designed with expert knowledge. These rules can be in different forms, such as canonical state machines \cite{sp15-composite,ccs16-tlsattacker}, linear-temporal properties \cite{icse22-ltlprop}, constraints of response messages \cite{atc20-sweyntooth,tdsc22-greyhound,ma2023loki}, etc.
For example, Beurdouche \textit{et al.} \cite{sp15-composite} manually construct a standard state machine from the specification and then use this state machine as a ground truth to identify deviant behaviors of the PUT. Utilizing this method, a logical bug was identified in a TLS implementation JSSE\cite{sp15-composite}. This flaw permitted attackers to bypass all messages pertaining to key exchange and authentication, subsequently enabling them to initiate unencrypted communication. 

Given a linear-time temporal logic property that a protocol implementation needs to satisfy, LTL-Fuzzer \cite{icse22-ltlprop} leverages directed greybox fuzzing to direct the fuzzing towards specific locations that can affect the property and checks whether the property is held during each execution iteration. 
In addition, Sweyntooth \cite{atc20-sweyntooth} and Greyhound \cite{tdsc22-greyhound} check whether the received response packet is in the expected type set of the current protocol state. 
Any mismatched message types are labeled anomaly. 
Loki \cite{ma2023loki} extracts rules from the PBFT consensus protocol paper \cite{castro1999practical}, which are used as oracles to detect non-memory-safety bugs in blockchain implementations. For example, Loki identified a bug in Hyperledger Fabric \cite{androulaki2018hyperledger} that can be used to confirm illegal transactions.

\subsubsection{Message Inconsistency in Transmission}

Some works check whether there are non-memory-safety bugs that can lead to integrity break of the protocol. Specifically, since correct data transfer is one of the basic properties of TCP protocol, TCP-Fuzz \cite{atc21-tcpfuzz} designed a data checker on both the sender and the receiver side to check the violation of this property. Whenever a message is sent or received, the data checker checks if the sent message and the received message are identical. 

\subsubsection{Abnormal Performance Indicators}
Some works aim to find network attack strategies that can affect the performance of the PUT, and these works judge the effectiveness of attack strategies by monitoring whether some performance indicators of PUT are beyond the normal range. 
For example, to find the amplification DDoS attack strategies in UDP services, AMPFuzz \cite{sec22-ampfuzz} uses the bandwidth amplification factor (BAF) \cite{rossow2014amplification} of each pair of requests and responses, which is the ratio of the sum of the lengths of all response messages to the length of the attack request, as an indicator to find the message that can maximize the consumption of throughput. 
TCPWN \cite{ndss18-tcpcongestion} and ABBrate \cite{raid20-abbrate} aim to find attack strategies against implementations of TCP congestion control that can increase or decrease the congestion window in a model-guided approach. To detect whether inputs indeed influence the congestion control mechanism, TCPWN obtains the window size from system logs and compares it with an expected baseline.

\subsubsection{Differences in Execution (labeled as ``DE'' in Table \ref{table-oracles} column 6)}

Differential testing involves comparing the execution behaviors of different implementations of the same protocol to investigate potential security impacts. This method is scalable due to its independence from code instrumentation. For example, TCP-Fuzz \cite{atc21-tcpfuzz} compares the outputs of multiple TCP implementations to identify discrepancies. Yang \textit{et al.} \cite{osdi21-consensus} employ differential testing to uncover consensus bugs in Ethereum that could lead to fork attacks. They generate a sequence of transactions as inputs and observe the responses of two Ethereum clients, specifically implemented in Golang and Rust. \newnote{Similarly, IcyChecker \cite{ye2023detecting} identifies blockchain state inconsistency bugs by generating mutated DApp transaction sequences and verifying the consistency of the final states.} \newnote{ParDiff \cite{zheng2024pardiff} utilizes the bisimulation algorithm to compare the FSMs of different protocol implementations and identifies discrepancies by analyzing the differences in state transition conditions.}
However, a significant challenge in this domain is ascertaining which implementation diverges from the protocol's expected behavior, and determining whether observed behavioral differences stem from errors or under-specifications in the protocol's RFC. As such, most of the work that adopts differential testing \cite{atc21-tcpfuzz,osdi21-consensus,bushart2023resolfuzz,kakarla2023oracle} integrates a subsequent manual inspection phase to differentiate actual vulnerabilities from innocuous discrepancies. 

To augment the bug-finding efficiency, some studies compare the PUT with an already well-tested or formally verified implementation, referred to as a `reference stack' \cite{atc21-tcpfuzz}. For example, TCP-Fuzz \cite{atc21-tcpfuzz} employs classical and extensively tested kernel-level TCP stacks, such as Linux TCP or FreeBSD TCP, as a reference to test newer TCP stacks. In such scenarios, if inconsistencies are reported, it strongly suggests the presence of bugs in the newer protocol implementations. This methodology not only identifies discrepancies, but also provides a framework for evaluating the correctness of various protocol implementations.

\section{Directions of Future Research}
\label{sec:future}

So far, we have discussed state-of-the-art protocol fuzzers. In this section, we will address Survey Dimension 3 by discussing the research trends and current challenges of fuzzing techniques based on what we have surveyed.

\subsection{Towards Perfect Communication Model Construction}
The current methods for constructing communication models are far from perfect, often resulting in incomplete or inaccurate knowledge acquisition, or requiring extensive manual effort.
Specifically, as introduced in Section \ref{sec:input}, existing methodologies for constructing communication models can be broadly categorized into top-down and bottom-up approaches. Bottom-up methods are proposed to learn communication models specific to particular protocol implementations \cite{sec15-tlsstatefuzzing,icst19-seqfuzzer,tecs19-polar,sec20-dtls,icst20-aflnet,sec21-mpinspector,acsac21-ics3fuzzer,arxiv21-stateafl,sec22-stateinspector,sec2022stateful,qin2023nsfuzz}, rather than the canonical communication model of the protocols themselves. 
However, for top-down approaches, most existing works still rely heavily on manual processes to construct state machines from protocol specifications \cite{sp15-composite,dsn15-snake,ccs16-tlsattacker,sp17-iotcube,ndss18-tcpcongestion,asiaccs19-mossot,atc20-sweyntooth,sec20-frankenstein,raid20-abbrate,atc21-tcpfuzz,tdsc22-greyhound,dsn22-l2fuzz,defcon27-fuzzowski,9868872}. This manual construction is not only labor intensive, but is also prone to errors.

Existing research \cite{aaai19-specification-learning,sp22-auto-fsm-extraction} has begun to explore the automatic extraction of partial FSMs from protocol specifications using NLP techniques. This exploration has preliminarily validated the feasibility of automating the extraction of protocol communication models. 
However, this method currently falls short of extracting canonical communication models from protocol specifications due to ambiguities and unspecified behaviors in specifications, thus precluding a complete one-to-one translation between the text and the communication model. 

To resolve this, machine learning model-based approaches can be explored for better model construction.
Considering the recent remarkable progress of LLM (Large Language Model) \cite{kakarla2023oracle,sp24-llmif,nikbakht2024tspec,issta24-llmdriver}, one promising direction is to develop LLM-based solutions for more precise model construction. Another possible direction is to combine the other information sources (such as the code of protocol implementations, code commit or comment information during development, program analysis results, etc.) to help better understand the content of the specification. 

\subsection{Towards Multi-Dimension Testing Perspectives}

Existing research focuses more on changing the content of packets or the order of packet sequences. 
This approach, while effective to some extent, overlooked the fact that protocols have multidimensional testing perspectives, e.g., variables such as message latency \cite{6129367}, cache state \cite{jung2001dns}, configurations \cite{10.1145/3580597,5438043}, and concurrency level \cite{katz2006parallel}, as highlighted in Section \ref{subsec:protocoldifferences}. These attributes play a crucial role in deciding the behavior of the target system.
To effectively test these attributes within protocol implementations, it is necessary to create detailed models that accurately represent each attribute, including message latency, cache state, configuration parameters, and concurrency levels.
Additionally, specific oracles and mutators can be designed to assess the correctness of the protocol behavior in various scenarios that encompass these multidimensional aspects.
This direction is interesting and can help establish a more comprehensive evaluation of the protocol's resilience and robustness.

\subsection{Fuzzing Characterized Protocol Targets}

A significant and underexplored future research direction is the fuzzing of characterized protocol targets. Current research has not fully covered various protocols, especially for those with distinct characteristics and importance. The following three areas are particularly noteworthy:

\textbf{1. Domain-Specific Protocols.} Proprietary domain protocols, such as those used in satellite communication \cite{sun2005satellite}, unmanned aerial vehicle (UAV) communication \cite{gupta2015survey}, and Robot Operating System (ROS) \cite{ohkawa2019high}, typically possess a high knowledge threshold and a relatively closed nature. These protocols play a crucial role in many infrastructures, making their security research paramount. Presently, fuzzing research for these protocols is relatively scarce, presenting an opportunity for the academic community to improve testing effectiveness and security through the development of new fuzzing techniques and tools.

\textbf{2. Hardware-Implemented Protocols.} Another direction is to design fuzzers for testing protocols implemented on hardware such as FPGAs \cite{trimberger2014fpga}.
These hardware implementations often exhibit different error characteristics compared to those at the software level, necessitating the development of new approaches to more effectively identify and exploit potential vulnerabilities.

\textbf{3. Multi-Party Protocols.} Another possible direction of protocol fuzzing is to support multi-party protocols.
In general, protocols have many communication modes, such as peer-to-peer mode \cite{ndss18-tcpcongestion,tdsc17-differencialTLS,raid20-abbrate}, server-client (master-slave) mode \cite{blackhat-us-17-wifuzz,ndss18-iotfuzzer,acsac21-ics3fuzzer}, and multi-party mode \cite{sudhodanan2016attack}.
Existing protocol fuzzers focus more on the first two modes by acting as a client/server to test the other \cite{blackhat-us-17-wifuzz,ndss18-iotfuzzer,acsac21-ics3fuzzer}, or playing a role as a peer node to test the PUT \cite{ndss18-tcpcongestion,tdsc17-differencialTLS,raid20-abbrate}.
The multi-party protocols have not been studied. For example, a node can play the role of a computing node, consensus node, or management node in a blockchain network \cite{androulaki2018hyperledger}, each of which is responsible for a different task. 
The correct execution of a smart contract protocol requires the cooperation of these roles.
How to efficiently test these multi-party protocols is an interesting but challenging question.

\subsection{Combining with Other Vulnerability-Finding Techniques} 
Beyond fuzzing, there exists a wide range of vulnerability-finding techniques, such as symbolic execution \cite{sensys08-kleenet,sp17-symcerts,icst22-symbolic,icc18-symbolic,issta17-symbolic,tse14-SymbexNet} and model checking \cite{nsdi04,ase06-javaprovers,10.1007/978-3-540-30579-8_24,2009-ASPIER,dsn17-modelchecking}. Although the combination of these techniques with fuzzing has been explored in general contexts \cite{ndss16-driller,yun2018qsym}, their applications in protocol fuzzing remain relatively underexplored \cite{tse14-SymbexNet}.
This presents a promising future research direction, especially considering the fact that combined approaches still face unique testing challenges for complex communications defined in protocols.
Intuitively, future research can improve existing vulnerability-finding techniques to better solve protocol-specific challenges.
In addition, many protocols are accompanied by high-quality learning sources, such as detailed specifications.
Future research can explore ways to utilize these valuable sources effectively to inform and enhance combined approaches.

\subsection{Shift-Left Protocol Fuzzing} 

Although there are certain research efforts focusing on the integration of general-purpose fuzzing techniques into the development cycle -- such as with tools like libFuzzer, OSS-Fuzz, and research into fuzzing within CI/CD integration testing \cite{9233212} -- few studies have specifically dedicated themselves to bridging the gaps between protocol fuzzing and the development process. 
Protocol fuzzing is distinct from general software fuzzing; it involves rigorously testing the various protocols that allow for communication and data exchange between different software systems and components. Protocol targets generally have a more complex development workflow than general software targets. This complexity arises from their need to precisely follow set standards and specifications to ensure interoperability across diverse systems, leading to unique challenges in integration and testing.
These challenges necessitate a tailored approach to fuzzing that understands and adapts to the intricacies of protocol development.
Therefore, a shift-left approach to protocol fuzzing is needed, which would integrate protocol-specific fuzzing techniques earlier in the software development lifecycle.
This can involve the exploration of designing techniques from the developer's perspective, and HCI (Human-Computer Interaction) \cite{carroll1997human} techniques can also be considered if necessary.
By doing so, it can surface vulnerabilities and issues at an earlier stage where they can be addressed more easily and cost-effectively, ensuring a more robust and secure software ecosystem for protocol implementations.

\begin{acks}
This research is supported by the National Natural Science Foundation of China (Grant Nos. 62125205 and U23A20303), the `111 Center' (B16037), the Key Research and Development Program of Shaanxi under Grant 2023KXJ190, and the Fundamental Research Funds for the Central Universities, No.YJSJ24010.
The Nanyang Technological University (NTU)-DESAY SV Research Program under Grant 2018-0980,
the National Research Foundation Singapore under its AI Singapore Programme (AISG2-RP-2020-019), the National Research Foundation, Singapore, and the Cyber Security Agency under its National Cybersecurity R\&D Programme (NCRP25-P04-TAICeN). Any opinions, findings, conclusions or recommendations expressed in this material are those of the authors and do not reflect the views of National Research Foundation, Singapore, and the Cyber Security Agency of Singapore. 
\end{acks}

\bibliographystyle{ACM-Reference-Format}
\bibliography{main}


\begin{thebibliography}{202}


\ifx \showCODEN    \undefined \def \showCODEN     #1{\unskip}     \fi
\ifx \showDOI      \undefined \def \showDOI       #1{#1}\fi
\ifx \showISBNx    \undefined \def \showISBNx     #1{\unskip}     \fi
\ifx \showISBNxiii \undefined \def \showISBNxiii  #1{\unskip}     \fi
\ifx \showISSN     \undefined \def \showISSN      #1{\unskip}     \fi
\ifx \showLCCN     \undefined \def \showLCCN      #1{\unskip}     \fi
\ifx \shownote     \undefined \def \shownote      #1{#1}          \fi
\ifx \showarticletitle \undefined \def \showarticletitle #1{#1}   \fi
\ifx \showURL      \undefined \def \showURL       {\relax}        \fi
\providecommand\bibfield[2]{#2}
\providecommand\bibinfo[2]{#2}
\providecommand\natexlab[1]{#1}
\providecommand\showeprint[2][]{arXiv:#2}

\bibitem[802(2020)]%
        {802.1x}
 \bibinfo{year}{2020}\natexlab{}.
\newblock \showarticletitle{IEEE Standard for Local and Metropolitan Area Networks--Port-Based Network Access Control}.
\newblock \bibinfo{journal}{\emph{IEEE Std 802.1X-2020 (Revision of IEEE Std 802.1X-2010 Incorporating IEEE Std 802.1Xbx-2014 and IEEE Std 802.1Xck-2018)}} (\bibinfo{year}{2020}), \bibinfo{pages}{1--289}.
\newblock


\bibitem[802(2021)]%
        {802.11}
 \bibinfo{year}{2021}\natexlab{}.
\newblock \showarticletitle{IEEE Standard for Information Technology--Telecommunications and Information Exchange between Systems - Local and Metropolitan Area Networks--Specific Requirements - Part 11: Wireless LAN Medium Access Control (MAC) and Physical Layer (PHY) Specifications}.
\newblock \bibinfo{journal}{\emph{IEEE Std 802.11-2020 (Revision of IEEE Std 802.11-2016)}} (\bibinfo{year}{2021}), \bibinfo{pages}{1--4379}.
\newblock


\bibitem[Aafer et~al\mbox{.}(2021)]%
        {sec21-androidtv}
\bibfield{author}{\bibinfo{person}{Yousra Aafer}, \bibinfo{person}{Wei You}, \bibinfo{person}{Yi Sun}, \bibinfo{person}{Yu Shi}, \bibinfo{person}{Xiangyu Zhang}, {and} \bibinfo{person}{Heng Yin}.} \bibinfo{year}{2021}\natexlab{}.
\newblock \showarticletitle{Android SmartTVs Vulnerability Discovery via Log-Guided Fuzzing}. In \bibinfo{booktitle}{\emph{30th USENIX Security Symposium (USENIX Security 21)}}. \bibinfo{pages}{2759--2776}.
\newblock


\bibitem[Aichernig et~al\mbox{.}(2021)]%
        {icst21-aichernig}
\bibfield{author}{\bibinfo{person}{Bernhard~K. Aichernig}, \bibinfo{person}{Edi Muškardin}, {and} \bibinfo{person}{Andrea Pferscher}.} \bibinfo{year}{2021}\natexlab{}.
\newblock \showarticletitle{Learning-Based Fuzzing of IoT Message Brokers}. In \bibinfo{booktitle}{\emph{2021 14th IEEE Conference on Software Testing, Verification and Validation (ICST)}}. \bibinfo{pages}{47--58}.
\newblock


\bibitem[Alliance(2015)]%
        {zigbeedoc}
\bibfield{author}{\bibinfo{person}{ZigBee Alliance}.} \bibinfo{year}{2015}\natexlab{}.
\newblock \bibinfo{title}{ZigBee Specification}.
\newblock
\newblock


\bibitem[Alshmrany and Cordeiro(2020)]%
        {arxiv20-FuSeBMC}
\bibfield{author}{\bibinfo{person}{Kaled~M. Alshmrany} {and} \bibinfo{person}{Lucas~C. Cordeiro}.} \bibinfo{year}{2020}\natexlab{}.
\newblock \showarticletitle{Finding Security Vulnerabilities in Network Protocol Implementations}.
\newblock  (\bibinfo{year}{2020}).
\newblock
\showeprint[arXiv]{2001.09592}


\bibitem[Ammann et~al\mbox{.}(2024)]%
        {ammann2024dy}
\bibfield{author}{\bibinfo{person}{Max Ammann}, \bibinfo{person}{Lucca Hirschi}, {and} \bibinfo{person}{Steve Kremer}.} \bibinfo{year}{2024}\natexlab{}.
\newblock \showarticletitle{DY Fuzzing: Formal Dolev-Yao Models Meet Cryptographic Protocol Fuzz Testing}. In \bibinfo{booktitle}{\emph{45th IEEE Symposium on Security and Privacy (SP)}}. \bibinfo{pages}{99--99}.
\newblock


\bibitem[Amusuo et~al\mbox{.}(2023)]%
        {amusuo2023systematically}
\bibfield{author}{\bibinfo{person}{Paschal~C Amusuo}, \bibinfo{person}{Ricardo Andr{\'e}s~Calvo M{\'e}ndez}, \bibinfo{person}{Zhongwei Xu}, \bibinfo{person}{Aravind Machinery}, {and} \bibinfo{person}{James~C Davis}.} \bibinfo{year}{2023}\natexlab{}.
\newblock \showarticletitle{Systematically Detecting Packet Validation Vulnerabilities in Embedded Network Stacks}. In \bibinfo{booktitle}{\emph{2023 38th IEEE/ACM International Conference on Automated Software Engineering (ASE)}}. \bibinfo{pages}{926--938}.
\newblock


\bibitem[Andronidis and Cadar(2022)]%
        {arxiv2022-snapfuzz}
\bibfield{author}{\bibinfo{person}{Anastasios Andronidis} {and} \bibinfo{person}{Cristian Cadar}.} \bibinfo{year}{2022}\natexlab{}.
\newblock \showarticletitle{SnapFuzz: An Efficient Fuzzing Framework for Network Applications}.
\newblock  (\bibinfo{year}{2022}).
\newblock
\showeprint[arXiv]{2201.04048}


\bibitem[Androulaki et~al\mbox{.}(2018)]%
        {androulaki2018hyperledger}
\bibfield{author}{\bibinfo{person}{Elli Androulaki}, \bibinfo{person}{Artem Barger}, \bibinfo{person}{Vita Bortnikov}, \bibinfo{person}{Christian Cachin}, \bibinfo{person}{Konstantinos Christidis}, \bibinfo{person}{Angelo De~Caro}, \bibinfo{person}{David Enyeart}, \bibinfo{person}{Christopher Ferris}, \bibinfo{person}{Gennady Laventman}, \bibinfo{person}{Yacov Manevich}, {et~al\mbox{.}}} \bibinfo{year}{2018}\natexlab{}.
\newblock \showarticletitle{Hyperledger fabric: a distributed operating system for permissioned blockchains}. In \bibinfo{booktitle}{\emph{thirteenth EuroSys conference}}. \bibinfo{pages}{1--15}.
\newblock


\bibitem[Asadian et~al\mbox{.}(2022)]%
        {icst22-symbolic}
\bibfield{author}{\bibinfo{person}{Hooman Asadian}, \bibinfo{person}{Paul Fiterău-Broştean}, \bibinfo{person}{Bengt Jonsson}, {and} \bibinfo{person}{Konstantinos Sagonas}.} \bibinfo{year}{2022}\natexlab{}.
\newblock \showarticletitle{Applying Symbolic Execution to Test Implementations of a Network Protocol Against its Specification}. In \bibinfo{booktitle}{\emph{2022 IEEE Conference on Software Testing, Verification and Validation (ICST)}}. \bibinfo{pages}{70--81}.
\newblock


\bibitem[Aschermann et~al\mbox{.}(2020)]%
        {sp20-ijon}
\bibfield{author}{\bibinfo{person}{Cornelius Aschermann}, \bibinfo{person}{Sergej Schumilo}, \bibinfo{person}{Ali Abbasi}, {and} \bibinfo{person}{Thorsten Holz}.} \bibinfo{year}{2020}\natexlab{}.
\newblock \showarticletitle{Ijon: Exploring Deep State Spaces via Fuzzing}. In \bibinfo{booktitle}{\emph{2020 IEEE Symposium on Security and Privacy (SP)}}. \bibinfo{pages}{1597--1612}.
\newblock


\bibitem[Atlidakis et~al\mbox{.}(2019)]%
        {icse19-restler}
\bibfield{author}{\bibinfo{person}{Vaggelis Atlidakis}, \bibinfo{person}{Patrice Godefroid}, {and} \bibinfo{person}{Marina Polishchuk}.} \bibinfo{year}{2019}\natexlab{}.
\newblock \showarticletitle{RESTler: stateful {REST} {API} fuzzing}. In \bibinfo{booktitle}{\emph{41st International Conference on Software Engineering ({ICSE})}}. \bibinfo{pages}{748--758}.
\newblock


\bibitem[AUTOSAR(2016)]%
        {someipdoc}
\bibfield{author}{\bibinfo{person}{AUTOSAR}.} \bibinfo{year}{2016}\natexlab{}.
\newblock \bibinfo{title}{SOME/IP Protocol Specification}.
\newblock
\newblock


\bibitem[Ba et~al\mbox{.}(2022)]%
        {sec2022stateful}
\bibfield{author}{\bibinfo{person}{Jinsheng Ba}, \bibinfo{person}{Marcel B{\"o}hme}, \bibinfo{person}{Zahra Mirzamomen}, {and} \bibinfo{person}{Abhik Roychoudhury}.} \bibinfo{year}{2022}\natexlab{}.
\newblock \showarticletitle{Stateful Greybox Fuzzing}. In \bibinfo{booktitle}{\emph{31st USENIX Security Symposium (USENIX Security)}}. \bibinfo{pages}{3255--3272}.
\newblock


\bibitem[Balakrishnan et~al\mbox{.}(1995)]%
        {balakrishnan1995improving}
\bibfield{author}{\bibinfo{person}{Hari Balakrishnan}, \bibinfo{person}{Srinivasan Seshan}, \bibinfo{person}{Elan Amir}, {and} \bibinfo{person}{Randy~H Katz}.} \bibinfo{year}{1995}\natexlab{}.
\newblock \showarticletitle{Improving TCP/IP performance over wireless networks}. In \bibinfo{booktitle}{\emph{1st annual international conference on Mobile computing and networking}}. \bibinfo{pages}{2--11}.
\newblock


\bibitem[Bars et~al\mbox{.}(2023)]%
        {bars2023fuzztruction}
\bibfield{author}{\bibinfo{person}{Nils Bars}, \bibinfo{person}{Moritz Schloegel}, \bibinfo{person}{Tobias Scharnowski}, \bibinfo{person}{Nico Schiller}, {and} \bibinfo{person}{Thorsten Holz}.} \bibinfo{year}{2023}\natexlab{}.
\newblock \showarticletitle{Fuzztruction: Using Fault Injection-based Fuzzing to Leverage Implicit Domain Knowledge}. In \bibinfo{booktitle}{\emph{32nd USENIX Security Symposium (USENIX Security 23)}}. \bibinfo{pages}{1847--1864}.
\newblock


\bibitem[Beurdouche et~al\mbox{.}(2015)]%
        {sp15-composite}
\bibfield{author}{\bibinfo{person}{Benjamin Beurdouche}, \bibinfo{person}{Karthikeyan Bhargavan}, \bibinfo{person}{Antoine Delignat-Lavaud}, \bibinfo{person}{Cédric Fournet}, \bibinfo{person}{Markulf Kohlweiss}, \bibinfo{person}{Alfredo Pironti}, \bibinfo{person}{Pierre-Yves Strub}, {and} \bibinfo{person}{Jean~Karim Zinzindohoue}.} \bibinfo{year}{2015}\natexlab{}.
\newblock \showarticletitle{A Messy State of the Union: Taming the Composite State Machines of TLS}. In \bibinfo{booktitle}{\emph{2015 IEEE Symposium on Security and Privacy (SP)}}. \bibinfo{pages}{535--552}.
\newblock


\bibitem[Borcherding et~al\mbox{.}(2023)]%
        {borcherding2023bandit}
\bibfield{author}{\bibinfo{person}{Anne Borcherding}, \bibinfo{person}{Mark Giraud}, \bibinfo{person}{Ian Fitzgerald}, {and} \bibinfo{person}{J{\"u}rgen Beyerer}.} \bibinfo{year}{2023}\natexlab{}.
\newblock \showarticletitle{The Bandit’s States: Modeling State Selection for Stateful Network Fuzzing as Multi-armed Bandit Problem}. In \bibinfo{booktitle}{\emph{2023 IEEE European Symposium on Security and Privacy Workshops (EuroS\&PW)}}. \bibinfo{pages}{345--350}.
\newblock


\bibitem[Bushart and Rossow(2023)]%
        {bushart2023resolfuzz}
\bibfield{author}{\bibinfo{person}{Jonas Bushart} {and} \bibinfo{person}{Christian Rossow}.} \bibinfo{year}{2023}\natexlab{}.
\newblock \showarticletitle{ResolFuzz: Differential Fuzzing of DNS Resolvers}. In \bibinfo{booktitle}{\emph{28th European Symposium on Research in Computer Security (ESORICS)}}. \bibinfo{pages}{62--80}.
\newblock


\bibitem[Bytes et~al\mbox{.}(2022)]%
        {arxiv22-fieldfuzz}
\bibfield{author}{\bibinfo{person}{Andrei Bytes}, \bibinfo{person}{Prashant Hari~Narayan Rajput}, \bibinfo{person}{Michail Maniatakos}, {and} \bibinfo{person}{Jianying Zhou}.} \bibinfo{year}{2022}\natexlab{}.
\newblock \bibinfo{title}{FieldFuzz: Enabling vulnerability discovery in Industrial Control Systems supply chain using stateful system-level fuzzing}.
\newblock
\newblock
\showeprint[arXiv]{2204.13499}


\bibitem[Caballero et~al\mbox{.}(2007)]%
        {ccs07-polyglot}
\bibfield{author}{\bibinfo{person}{Juan Caballero}, \bibinfo{person}{Heng Yin}, \bibinfo{person}{Zhenkai Liang}, {and} \bibinfo{person}{Dawn Song}.} \bibinfo{year}{2007}\natexlab{}.
\newblock \showarticletitle{Polyglot: automatic extraction of protocol message format using dynamic binary analysis}. In \bibinfo{booktitle}{\emph{14th ACM Conference on Computer and Communications Security (CCS)}}. \bibinfo{pages}{317–329}.
\newblock


\bibitem[Cao(2021)]%
        {blackhat-eu-21-owfuzz}
\bibfield{author}{\bibinfo{person}{Hongjian Cao}.} \bibinfo{year}{2021}\natexlab{}.
\newblock \bibinfo{title}{Owfuzz: WiFi Nightmare}.
\newblock
\newblock
\urldef\tempurl%
\url{https://www.blackhat.com/eu-21/briefings/schedule/#owfuzz-wifi-nightmare-24338}
\showURL{%
\tempurl}


\bibitem[Carroll(1997)]%
        {carroll1997human}
\bibfield{author}{\bibinfo{person}{John~M Carroll}.} \bibinfo{year}{1997}\natexlab{}.
\newblock \showarticletitle{Human-computer interaction: psychology as a science of design}.
\newblock \bibinfo{journal}{\emph{Annual review of psychology}} \bibinfo{volume}{48}, \bibinfo{number}{1} (\bibinfo{year}{1997}), \bibinfo{pages}{61--83}.
\newblock


\bibitem[Castro et~al\mbox{.}(1999)]%
        {castro1999practical}
\bibfield{author}{\bibinfo{person}{Miguel Castro}, \bibinfo{person}{Barbara Liskov}, {et~al\mbox{.}}} \bibinfo{year}{1999}\natexlab{}.
\newblock \showarticletitle{Practical byzantine fault tolerance}. In \bibinfo{booktitle}{\emph{3rd Symposium on Operating Systems Design and Implementation (OSDI)}}, Vol.~\bibinfo{volume}{99}. \bibinfo{pages}{173--186}.
\newblock


\bibitem[Chaki and Datta(2009)]%
        {2009-ASPIER}
\bibfield{author}{\bibinfo{person}{Sagar Chaki} {and} \bibinfo{person}{Anupam Datta}.} \bibinfo{year}{2009}\natexlab{}.
\newblock \showarticletitle{ASPIER: An Automated Framework for Verifying Security Protocol Implementations}. In \bibinfo{booktitle}{\emph{2009 22nd IEEE Computer Security Foundations Symposium}}. \bibinfo{pages}{172--185}.
\newblock


\bibitem[Chau et~al\mbox{.}(2017)]%
        {sp17-symcerts}
\bibfield{author}{\bibinfo{person}{Sze~Yiu Chau}, \bibinfo{person}{Omar Chowdhury}, \bibinfo{person}{Endadul Hoque}, \bibinfo{person}{Huangyi Ge}, \bibinfo{person}{Aniket Kate}, \bibinfo{person}{Cristina Nita-Rotaru}, {and} \bibinfo{person}{Ninghui Li}.} \bibinfo{year}{2017}\natexlab{}.
\newblock \showarticletitle{SymCerts: Practical Symbolic Execution for Exposing Noncompliance in X.509 Certificate Validation Implementations}. In \bibinfo{booktitle}{\emph{2017 IEEE Symposium on Security and Privacy (SP)}}. \bibinfo{pages}{503--520}.
\newblock


\bibitem[Chen et~al\mbox{.}(2018)]%
        {ndss18-iotfuzzer}
\bibfield{author}{\bibinfo{person}{Jiongyi Chen}, \bibinfo{person}{Wenrui Diao}, \bibinfo{person}{Qingchuan Zhao}, \bibinfo{person}{Chaoshun Zuo}, \bibinfo{person}{Zhiqiang Lin}, \bibinfo{person}{XiaoFeng Wang}, \bibinfo{person}{Wing~Cheong Lau}, \bibinfo{person}{Menghan Sun}, \bibinfo{person}{Ronghai Yang}, {and} \bibinfo{person}{Kehuan Zhang}.} \bibinfo{year}{2018}\natexlab{}.
\newblock \showarticletitle{IoTFuzzer: Discovering Memory Corruptions in IoT Through App-based Fuzzing}. In \bibinfo{booktitle}{\emph{25th Annual Network and Distributed System Security Symposium ({NDSS})}}.
\newblock


\bibitem[Chen et~al\mbox{.}(2005)]%
        {chen2005wireless}
\bibfield{author}{\bibinfo{person}{Jyh-Cheng Chen}, \bibinfo{person}{Ming-Chia Jiang}, {and} \bibinfo{person}{Yi-wen Liu}.} \bibinfo{year}{2005}\natexlab{}.
\newblock \showarticletitle{Wireless LAN security and IEEE 802.11 i}.
\newblock \bibinfo{journal}{\emph{IEEE Wireless Communications}} \bibinfo{volume}{12}, \bibinfo{number}{1} (\bibinfo{year}{2005}), \bibinfo{pages}{27--36}.
\newblock


\bibitem[Chen et~al\mbox{.}(2019)]%
        {feast19-yurong}
\bibfield{author}{\bibinfo{person}{Yurong Chen}, \bibinfo{person}{Tian lan}, {and} \bibinfo{person}{Guru Venkataramani}.} \bibinfo{year}{2019}\natexlab{}.
\newblock \showarticletitle{Exploring Effective Fuzzing Strategies to Analyze Communication Protocols}. In \bibinfo{booktitle}{\emph{3rd ACM Workshop on Forming an Ecosystem Around Software Transformation (FEAST)}}. \bibinfo{pages}{17–23}.
\newblock


\bibitem[Chen et~al\mbox{.}(2023)]%
        {chen2022tyr}
\bibfield{author}{\bibinfo{person}{Yuanliang Chen}, \bibinfo{person}{Fuchen Ma}, \bibinfo{person}{Yuanhang Zhou}, \bibinfo{person}{Yu Jiang}, \bibinfo{person}{Ting Chen}, {and} \bibinfo{person}{Jiaguang Sun}.} \bibinfo{year}{2023}\natexlab{}.
\newblock \showarticletitle{Tyr: Finding Consensus Failure Bugs in Blockchain System with Behaviour Divergent Model}. In \bibinfo{booktitle}{\emph{2023 IEEE Symposium on Security and Privacy (SP)}}. \bibinfo{pages}{2517--2532}.
\newblock


\bibitem[Cisco(2022)]%
        {ciscodoc}
\bibfield{author}{\bibinfo{person}{Cisco}.} \bibinfo{year}{2022}\natexlab{}.
\newblock \bibinfo{title}{Cisco Secure Client Data Sheet}.
\newblock
\newblock
\urldef\tempurl%
\url{https://www.cisco.com/c/en/us/products/collateral/security/anyconnect-secure-mobility-client/secure-mobility-client-ds.html}
\showURL{%
\tempurl}


\bibitem[Coffield and Shepherd(1987)]%
        {coffield1987tutorial}
\bibfield{author}{\bibinfo{person}{David Coffield} {and} \bibinfo{person}{Doug Shepherd}.} \bibinfo{year}{1987}\natexlab{}.
\newblock \showarticletitle{Tutorial guide to Unix sockets for network communications}.
\newblock \bibinfo{journal}{\emph{Computer Communications}} \bibinfo{volume}{10}, \bibinfo{number}{1} (\bibinfo{year}{1987}), \bibinfo{pages}{21--29}.
\newblock


\bibitem[Comer(2013)]%
        {comer2013internetworking}
\bibfield{author}{\bibinfo{person}{Douglas~E. Comer}.} \bibinfo{year}{2013}\natexlab{}.
\newblock \bibinfo{booktitle}{\emph{Internetworking with TCP/IP} (\bibinfo{edition}{6th} ed.)}.
\newblock \bibinfo{publisher}{Addison-Wesley Professional}.
\newblock
\showISBNx{013608530X}


\bibitem[Comparetti et~al\mbox{.}(2009)]%
        {sp09-prospex}
\bibfield{author}{\bibinfo{person}{Paolo~Milani Comparetti}, \bibinfo{person}{Gilbert Wondracek}, \bibinfo{person}{Christopher Kruegel}, {and} \bibinfo{person}{Engin Kirda}.} \bibinfo{year}{2009}\natexlab{}.
\newblock \showarticletitle{Prospex: Protocol Specification Extraction}. In \bibinfo{booktitle}{\emph{2009 30th IEEE Symposium on Security and Privacy}}. \bibinfo{pages}{110--125}.
\newblock


\bibitem[Corporation(2020)]%
        {supervisory1doc}
\bibfield{author}{\bibinfo{person}{Mitsubishi~Electric Corporation}.} \bibinfo{year}{2020}\natexlab{}.
\newblock \bibinfo{title}{GX Works2 - Programmable Controllers MELSEC}.
\newblock
\newblock
\urldef\tempurl%
\url{https://www.mitsubishielectric.com/fa/products/cnt/plceng/smerit/gx_works2/index.html}
\showURL{%
\tempurl}


\bibitem[Dai et~al\mbox{.}(2010)]%
        {5438043}
\bibfield{author}{\bibinfo{person}{Huning Dai}, \bibinfo{person}{Christian Murphy}, {and} \bibinfo{person}{Gail Kaiser}.} \bibinfo{year}{2010}\natexlab{}.
\newblock \showarticletitle{Configuration Fuzzing for Software Vulnerability Detection}. In \bibinfo{booktitle}{\emph{2010 International Conference on Availability, Reliability and Security}}. \bibinfo{pages}{525--530}.
\newblock


\bibitem[Daniel et~al\mbox{.}(2018)]%
        {eurospw18-openvpn}
\bibfield{author}{\bibinfo{person}{Lesly-Ann Daniel}, \bibinfo{person}{Erik Poll}, {and} \bibinfo{person}{Joeri de Ruiter}.} \bibinfo{year}{2018}\natexlab{}.
\newblock \showarticletitle{Inferring OpenVPN State Machines Using Protocol State Fuzzing}. In \bibinfo{booktitle}{\emph{2018 IEEE European Symposium on Security and Privacy Workshops (EuroS\&PW)}}. \bibinfo{pages}{11--19}.
\newblock


\bibitem[de~Ruiter and Poll(2015)]%
        {sec15-tlsstatefuzzing}
\bibfield{author}{\bibinfo{person}{Joeri de Ruiter} {and} \bibinfo{person}{Erik Poll}.} \bibinfo{year}{2015}\natexlab{}.
\newblock \showarticletitle{Protocol State Fuzzing of {TLS} Implementations}. In \bibinfo{booktitle}{\emph{24th USENIX Security Symposium (USENIX Security)}}. \bibinfo{pages}{193--206}.
\newblock


\bibitem[Dierks(2008)]%
        {tlsdoc}
\bibfield{author}{\bibinfo{person}{T. Dierks}.} \bibinfo{year}{2008}\natexlab{}.
\newblock \bibinfo{title}{RFC5246: The Transport Layer Security (TLS) Protocol Version 1.2}.
\newblock
\newblock
\urldef\tempurl%
\url{https://www.rfc-editor.org/rfc/rfc5246}
\showURL{%
\tempurl}


\bibitem[Eddington(2014)]%
        {peach}
\bibfield{author}{\bibinfo{person}{M. Eddington}.} \bibinfo{year}{2014}\natexlab{}.
\newblock \bibinfo{title}{Peach fuzzing platform}.
\newblock
\newblock
\urldef\tempurl%
\url{Available: http://community.peachfuzzer.com/WhatIsPeach.html}
\showURL{%
\tempurl}


\bibitem[Electric(2009)]%
        {supervisory2doc}
\bibfield{author}{\bibinfo{person}{Schneider Electric}.} \bibinfo{year}{2009}\natexlab{}.
\newblock \bibinfo{title}{TwidoSuite Programming Software}.
\newblock
\newblock
\urldef\tempurl%
\url{https://www.se.com/ww/en/download/document/TwidoSuite_V0220_11_SP/}
\showURL{%
\tempurl}


\bibitem[ETSI(2002)]%
        {mmsdoc}
\bibfield{author}{\bibinfo{person}{ETSI}.} \bibinfo{year}{2002}\natexlab{}.
\newblock \bibinfo{title}{Universal Mobile Telecommunications System (UMTS); Multimedia Messaging Service (MMS); Stage 1 (3GPP TS 22.140 version 5.3.0 Release 5)}.
\newblock
\newblock
\urldef\tempurl%
\url{https://www.etsi.org/deliver/etsi_ts/122100_122199/122140/05.03.00_60/ts_122140v050300p.pdf}
\showURL{%
\tempurl}


\bibitem[Fan and Chang(2018)]%
        {icics17-fan}
\bibfield{author}{\bibinfo{person}{Rong Fan} {and} \bibinfo{person}{Yaoyao Chang}.} \bibinfo{year}{2018}\natexlab{}.
\newblock \showarticletitle{Machine Learning for Black-Box Fuzzing of Network Protocols}. In \bibinfo{booktitle}{\emph{Information and Communications Security}}. \bibinfo{pages}{621--632}.
\newblock


\bibitem[Fang et~al\mbox{.}(2021)]%
        {acsac21-ics3fuzzer}
\bibfield{author}{\bibinfo{person}{Dongliang Fang}, \bibinfo{person}{Zhanwei Song}, \bibinfo{person}{Le Guan}, \bibinfo{person}{Puzhuo Liu}, \bibinfo{person}{Anni Peng}, \bibinfo{person}{Kai Cheng}, \bibinfo{person}{Yaowen Zheng}, \bibinfo{person}{Peng Liu}, \bibinfo{person}{Hongsong Zhu}, {and} \bibinfo{person}{Limin Sun}.} \bibinfo{year}{2021}\natexlab{}.
\newblock \showarticletitle{ICS3Fuzzer: {A} Framework for Discovering Protocol Implementation Bugs in {ICS} Supervisory Software by Fuzzing}. In \bibinfo{booktitle}{\emph{Annual Computer Security Applications Conference (ACSAC)}}. \bibinfo{pages}{849--860}.
\newblock


\bibitem[Feng et~al\mbox{.}(2021)]%
        {ccs21-snipuzz}
\bibfield{author}{\bibinfo{person}{Xiaotao Feng}, \bibinfo{person}{Ruoxi Sun}, \bibinfo{person}{Xiaogang Zhu}, \bibinfo{person}{Minhui Xue}, \bibinfo{person}{Sheng Wen}, \bibinfo{person}{Dongxi Liu}, \bibinfo{person}{Surya Nepal}, {and} \bibinfo{person}{Yang Xiang}.} \bibinfo{year}{2021}\natexlab{}.
\newblock \showarticletitle{Snipuzz: Black-Box Fuzzing of IoT Firmware via Message Snippet Inference}. In \bibinfo{booktitle}{\emph{2021 ACM SIGSAC Conference on Computer and Communications Security (CCS)}}. \bibinfo{pages}{337–350}.
\newblock


\bibitem[Fiterau-Brostean et~al\mbox{.}(2020)]%
        {sec20-dtls}
\bibfield{author}{\bibinfo{person}{Paul Fiterau-Brostean}, \bibinfo{person}{Bengt Jonsson}, \bibinfo{person}{Robert Merget}, \bibinfo{person}{Joeri de Ruiter}, \bibinfo{person}{Konstantinos Sagonas}, {and} \bibinfo{person}{Juraj Somorovsky}.} \bibinfo{year}{2020}\natexlab{}.
\newblock \showarticletitle{Analysis of {DTLS} Implementations Using Protocol State Fuzzing}. In \bibinfo{booktitle}{\emph{29th USENIX Security Symposium (USENIX Security)}}. \bibinfo{pages}{2523--2540}.
\newblock


\bibitem[Fiterau-Brostean et~al\mbox{.}(2022)]%
        {icst22-dtlsfuzzer}
\bibfield{author}{\bibinfo{person}{P. Fiterau-Brostean}, \bibinfo{person}{B. Jonsson}, \bibinfo{person}{K. Sagonas}, {and} \bibinfo{person}{F. Taquist}.} \bibinfo{year}{2022}\natexlab{}.
\newblock \showarticletitle{DTLS-Fuzzer: A DTLS Protocol State Fuzzer}. In \bibinfo{booktitle}{\emph{2022 IEEE Conference on Software Testing, Verification and Validation (ICST)}}. \bibinfo{pages}{456--458}.
\newblock


\bibitem[Fiterau-Brostean et~al\mbox{.}(2023)]%
        {fiterau2023automata}
\bibfield{author}{\bibinfo{person}{Paul Fiterau-Brostean}, \bibinfo{person}{Bengt Jonsson}, \bibinfo{person}{Konstantinos Sagonas}, {and} \bibinfo{person}{Fredrik T{\aa}quist}.} \bibinfo{year}{2023}\natexlab{}.
\newblock \showarticletitle{Automata-Based Automated Detection of State Machine Bugs in Protocol Implementations.}. In \bibinfo{booktitle}{\emph{30rd Annual Network and Distributed System Security Symposium (NDSS)}}.
\newblock


\bibitem[Foundation(2015)]%
        {openflowdoc}
\bibfield{author}{\bibinfo{person}{Open~Networking Foundation}.} \bibinfo{year}{2015}\natexlab{}.
\newblock \bibinfo{title}{OpenFlow Switch Specification}.
\newblock
\newblock


\bibitem[Fu et~al\mbox{.}(2023)]%
        {fu2023framework}
\bibfield{author}{\bibinfo{person}{Junsong Fu}, \bibinfo{person}{Shuai Xiong}, \bibinfo{person}{Na Wang}, \bibinfo{person}{Ruiping Ren}, \bibinfo{person}{Ang Zhou}, {and} \bibinfo{person}{Bharat~K Bhargava}.} \bibinfo{year}{2023}\natexlab{}.
\newblock \showarticletitle{A Framework of High-Speed Network Protocol Fuzzing Based on Shared Memory}.
\newblock \bibinfo{journal}{\emph{IEEE Transactions on Dependable and Secure Computing}} (\bibinfo{year}{2023}).
\newblock


\bibitem[Gao et~al\mbox{.}(2022)]%
        {gao2022fw}
\bibfield{author}{\bibinfo{person}{Zicong Gao}, \bibinfo{person}{Weiyu Dong}, \bibinfo{person}{Rui Chang}, {and} \bibinfo{person}{Yisen Wang}.} \bibinfo{year}{2022}\natexlab{}.
\newblock \showarticletitle{Fw-fuzz: A code coverage-guided fuzzing framework for network protocols on firmware}.
\newblock \bibinfo{journal}{\emph{Concurrency and Computation: Practice and Experience}} \bibinfo{volume}{34}, \bibinfo{number}{16} (\bibinfo{year}{2022}), \bibinfo{pages}{e5756}.
\newblock


\bibitem[Garbelini et~al\mbox{.}(2022a)]%
        {sec22-braktooth}
\bibfield{author}{\bibinfo{person}{Matheus~E. Garbelini}, \bibinfo{person}{Vaibhav Bedi}, \bibinfo{person}{Sudipta Chattopadhyay}, \bibinfo{person}{Sumei Sun}, {and} \bibinfo{person}{Ernest Kurniawan}.} \bibinfo{year}{2022}\natexlab{a}.
\newblock \showarticletitle{{BrakTooth}: Causing Havoc on Bluetooth Link Manager via Directed Fuzzing}. In \bibinfo{booktitle}{\emph{31st USENIX Security Symposium (USENIX Security 22)}}. \bibinfo{pages}{1025--1042}.
\newblock
\showISBNx{978-1-939133-31-1}


\bibitem[Garbelini et~al\mbox{.}(2022b)]%
        {10001673}
\bibfield{author}{\bibinfo{person}{Matheus~E. Garbelini}, \bibinfo{person}{Zewen Shang}, \bibinfo{person}{Sudipta Chattopadhyay}, \bibinfo{person}{Sumei Sun}, {and} \bibinfo{person}{Ernest Kurniawan}.} \bibinfo{year}{2022}\natexlab{b}.
\newblock \showarticletitle{Towards Automated Fuzzing of 4G/5G Protocol Implementations Over the Air}. In \bibinfo{booktitle}{\emph{GLOBECOM 2022 - 2022 IEEE Global Communications Conference}}. \bibinfo{pages}{86--92}.
\newblock


\bibitem[Garbelini et~al\mbox{.}(2022c)]%
        {tdsc22-greyhound}
\bibfield{author}{\bibinfo{person}{Matheus~E. Garbelini}, \bibinfo{person}{Chundong Wang}, {and} \bibinfo{person}{Sudipta Chattopadhyay}.} \bibinfo{year}{2022}\natexlab{c}.
\newblock \showarticletitle{Greyhound: Directed Greybox Wi-Fi Fuzzing}.
\newblock \bibinfo{journal}{\emph{IEEE Transactions on Dependable and Secure Computing}} \bibinfo{volume}{19}, \bibinfo{number}{2} (\bibinfo{year}{2022}), \bibinfo{pages}{817--834}.
\newblock


\bibitem[Garbelini et~al\mbox{.}(2020)]%
        {atc20-sweyntooth}
\bibfield{author}{\bibinfo{person}{Matheus~E. Garbelini}, \bibinfo{person}{Chundong Wang}, \bibinfo{person}{Sudipta Chattopadhyay}, \bibinfo{person}{Sun Sumei}, {and} \bibinfo{person}{Ernest Kurniawan}.} \bibinfo{year}{2020}\natexlab{}.
\newblock \showarticletitle{{SweynTooth}: Unleashing Mayhem over Bluetooth Low Energy}. In \bibinfo{booktitle}{\emph{2020 USENIX Annual Technical Conference (USENIX ATC 20)}}. \bibinfo{pages}{911--925}.
\newblock


\bibitem[Gascon et~al\mbox{.}(2015)]%
        {securecomm15-pulsar}
\bibfield{author}{\bibinfo{person}{Hugo Gascon}, \bibinfo{person}{Christian Wressnegger}, \bibinfo{person}{Fabian Yamaguchi}, \bibinfo{person}{Daniel Arp}, {and} \bibinfo{person}{Konrad Rieck}.} \bibinfo{year}{2015}\natexlab{}.
\newblock \showarticletitle{Pulsar: Stateful Black-Box Fuzzing of Proprietary Network Protocols}. In \bibinfo{booktitle}{\emph{Security and Privacy in Communication Networks - 11th International Conference (SecureComm)}}, Vol.~\bibinfo{volume}{164}. \bibinfo{pages}{330--347}.
\newblock


\bibitem[Gorenc and Molinyawe(2014)]%
        {defcon22-sms}
\bibfield{author}{\bibinfo{person}{Brian Gorenc} {and} \bibinfo{person}{Matt Molinyawe}.} \bibinfo{year}{2014}\natexlab{}.
\newblock \bibinfo{title}{Blowing up the Celly: Building Your Own SMS/MMS Fuzzer}.
\newblock
\newblock
\urldef\tempurl%
\url{https://media.defcon.org/DEF\%20CON\%2022/DEF\%20CON\%2022\%20presentations/DEF\%20CON\%2022\%20-\%20Brian-Gorenc-Matt-Molinyawe-Blowing-Up-The-Celly.pdf}
\showURL{%
\tempurl}


\bibitem[Goubault-Larrecq and Parrennes(2005)]%
        {10.1007/978-3-540-30579-8_24}
\bibfield{author}{\bibinfo{person}{Jean Goubault-Larrecq} {and} \bibinfo{person}{Fabrice Parrennes}.} \bibinfo{year}{2005}\natexlab{}.
\newblock \showarticletitle{Cryptographic Protocol Analysis on Real C Code}. In \bibinfo{booktitle}{\emph{Verification, Model Checking, and Abstract Interpretation}}. \bibinfo{pages}{363--379}.
\newblock
\showISBNx{978-3-540-30579-8}


\bibitem[Group(2018)]%
        {ssodoc}
\bibfield{author}{\bibinfo{person}{The~Open Group}.} \bibinfo{year}{2018}\natexlab{}.
\newblock \bibinfo{title}{Single Sign-On}.
\newblock
\newblock
\urldef\tempurl%
\url{http://www.opengroup.org/security/sso/}
\showURL{%
\tempurl}


\bibitem[Gupta et~al\mbox{.}(2015)]%
        {gupta2015survey}
\bibfield{author}{\bibinfo{person}{Lav Gupta}, \bibinfo{person}{Raj Jain}, {and} \bibinfo{person}{Gabor Vaszkun}.} \bibinfo{year}{2015}\natexlab{}.
\newblock \showarticletitle{Survey of important issues in UAV communication networks}.
\newblock \bibinfo{journal}{\emph{IEEE communications surveys \& tutorials}} \bibinfo{volume}{18}, \bibinfo{number}{2} (\bibinfo{year}{2015}), \bibinfo{pages}{1123--1152}.
\newblock


\bibitem[Hawkes(2022)]%
        {googlevulsheet}
\bibfield{author}{\bibinfo{person}{Ben Hawkes}.} \bibinfo{year}{2022}\natexlab{}.
\newblock \bibinfo{title}{0day In the Wild}.
\newblock
\newblock
\urldef\tempurl%
\url{https://googleprojectzero.blogspot.com/p/0day.html}
\showURL{%
\tempurl}


\bibitem[He et~al\mbox{.}(2022)]%
        {9868872}
\bibfield{author}{\bibinfo{person}{Fengjiao He}, \bibinfo{person}{Wenchuan Yang}, \bibinfo{person}{Baojiang Cui}, {and} \bibinfo{person}{Jia Cui}.} \bibinfo{year}{2022}\natexlab{}.
\newblock \showarticletitle{Intelligent Fuzzing Algorithm for 5G NAS Protocol Based on Predefined Rules}. In \bibinfo{booktitle}{\emph{2022 International Conference on Computer Communications and Networks (ICCCN)}}. \bibinfo{pages}{1--7}.
\newblock


\bibitem[Heinze et~al\mbox{.}({[n.\,d.]})]%
        {woot20-ToothPicker}
\bibfield{author}{\bibinfo{person}{Dennis Heinze}, \bibinfo{person}{Jiska Classen}, {and} \bibinfo{person}{Matthias Hollick}.} \bibinfo{year}{[n.\,d.]}\natexlab{}.
\newblock \showarticletitle{{ToothPicker}: Apple Picking in the {iOS} Bluetooth Stack}. In \bibinfo{booktitle}{\emph{14th USENIX Workshop on Offensive Technologies (WOOT)}}.
\newblock


\bibitem[Hoque et~al\mbox{.}(2017)]%
        {dsn17-modelchecking}
\bibfield{author}{\bibinfo{person}{Endadul Hoque}, \bibinfo{person}{Omar Chowdhury}, \bibinfo{person}{Sze~Yiu Chau}, \bibinfo{person}{Cristina Nita-Rotaru}, {and} \bibinfo{person}{Ninghui Li}.} \bibinfo{year}{2017}\natexlab{}.
\newblock \showarticletitle{Analyzing Operational Behavior of Stateful Protocol Implementations for Detecting Semantic Bugs}. In \bibinfo{booktitle}{\emph{47th IEEE/IFIP International Conference on Dependable Systems and Networks (DSN)}}. \bibinfo{pages}{627--638}.
\newblock


\bibitem[Hussain et~al\mbox{.}(2021)]%
        {hussain2021noncompliance}
\bibfield{author}{\bibinfo{person}{Syed~Rafiul Hussain}, \bibinfo{person}{Imtiaz Karim}, \bibinfo{person}{Abdullah~Al Ishtiaq}, \bibinfo{person}{Omar Chowdhury}, {and} \bibinfo{person}{Elisa Bertino}.} \bibinfo{year}{2021}\natexlab{}.
\newblock \showarticletitle{Noncompliance as deviant behavior: An automated black-box noncompliance checker for 4g lte cellular devices}. In \bibinfo{booktitle}{\emph{2021 ACM SIGSAC Conference on Computer and Communications Security (CCS)}}. \bibinfo{pages}{1082--1099}.
\newblock


\bibitem[Iyengar and Thomson(2021)]%
        {quicdoc}
\bibfield{author}{\bibinfo{person}{Jana Iyengar} {and} \bibinfo{person}{Martin Thomson}.} \bibinfo{year}{2021}\natexlab{}.
\newblock \bibinfo{title}{{QUIC: A UDP-Based Multiplexed and Secure Transport}}.
\newblock \bibinfo{howpublished}{RFC 9000}.
\newblock
\urldef\tempurl%
\url{https://www.rfc-editor.org/info/rfc9000}
\showURL{%
\tempurl}


\bibitem[Jelassi et~al\mbox{.}(2012)]%
        {6129367}
\bibfield{author}{\bibinfo{person}{Sofiene Jelassi}, \bibinfo{person}{Gerardo Rubino}, \bibinfo{person}{Hugh Melvin}, \bibinfo{person}{Habib Youssef}, {and} \bibinfo{person}{Guy Pujolle}.} \bibinfo{year}{2012}\natexlab{}.
\newblock \showarticletitle{Quality of Experience of VoIP Service: A Survey of Assessment Approaches and Open Issues}.
\newblock \bibinfo{journal}{\emph{IEEE Communications Surveys \& Tutorials}} \bibinfo{volume}{14}, \bibinfo{number}{2} (\bibinfo{year}{2012}), \bibinfo{pages}{491--513}.
\newblock


\bibitem[Jero et~al\mbox{.}(2018)]%
        {ndss18-tcpcongestion}
\bibfield{author}{\bibinfo{person}{Samuel Jero}, \bibinfo{person}{Md.~Endadul Hoque}, \bibinfo{person}{David~R. Choffnes}, \bibinfo{person}{Alan Mislove}, {and} \bibinfo{person}{Cristina Nita{-}Rotaru}.} \bibinfo{year}{2018}\natexlab{}.
\newblock \showarticletitle{Automated Attack Discovery in {TCP} Congestion Control Using a Model-guided Approach}. In \bibinfo{booktitle}{\emph{25th Annual Network and Distributed System Security Symposium (NDSS)}}.
\newblock


\bibitem[Jero et~al\mbox{.}(2015)]%
        {dsn15-snake}
\bibfield{author}{\bibinfo{person}{Samuel Jero}, \bibinfo{person}{Hyojeong Lee}, {and} \bibinfo{person}{Cristina Nita-Rotaru}.} \bibinfo{year}{2015}\natexlab{}.
\newblock \showarticletitle{Leveraging State Information for Automated Attack Discovery in Transport Protocol Implementations}. In \bibinfo{booktitle}{\emph{2015 45th Annual IEEE/IFIP International Conference on Dependable Systems and Networks}}. \bibinfo{pages}{1--12}.
\newblock


\bibitem[Jero et~al\mbox{.}(2019)]%
        {aaai19-specification-learning}
\bibfield{author}{\bibinfo{person}{Samuel Jero}, \bibinfo{person}{Maria~Leonor Pacheco}, \bibinfo{person}{Dan Goldwasser}, {and} \bibinfo{person}{Cristina Nita-Rotaru}.} \bibinfo{year}{2019}\natexlab{}.
\newblock \showarticletitle{Leveraging Textual Specifications for Grammar-Based Fuzzing of Network Protocols}.
\newblock \bibinfo{journal}{\emph{AAAI Conference on Artificial Intelligence}} \bibinfo{volume}{33}, \bibinfo{number}{01} (\bibinfo{year}{2019}), \bibinfo{pages}{9478--9483}.
\newblock
\urldef\tempurl%
\url{https://ojs.aaai.org/index.php/AAAI/article/view/5002}
\showURL{%
\tempurl}


\bibitem[Ji and Xu(2023)]%
        {ji2023finding}
\bibfield{author}{\bibinfo{person}{Ru Ji} {and} \bibinfo{person}{Meng Xu}.} \bibinfo{year}{2023}\natexlab{}.
\newblock \showarticletitle{Finding Specification Blind Spots via Fuzz Testing}. In \bibinfo{booktitle}{\emph{2023 IEEE Symposium on Security and Privacy (SP)}}. \bibinfo{pages}{2708--2725}.
\newblock


\bibitem[Jung et~al\mbox{.}(2001)]%
        {jung2001dns}
\bibfield{author}{\bibinfo{person}{Jaeyeon Jung}, \bibinfo{person}{Emil Sit}, \bibinfo{person}{Hari Balakrishnan}, {and} \bibinfo{person}{Robert Morris}.} \bibinfo{year}{2001}\natexlab{}.
\newblock \showarticletitle{DNS performance and the effectiveness of caching}. In \bibinfo{booktitle}{\emph{1st ACM SIGCOMM Workshop on Internet Measurement}}. \bibinfo{pages}{153--167}.
\newblock


\bibitem[Jurjens(2006)]%
        {ase06-javaprovers}
\bibfield{author}{\bibinfo{person}{Jan Jurjens}.} \bibinfo{year}{2006}\natexlab{}.
\newblock \showarticletitle{Security Analysis of Crypto-based Java Programs using Automated Theorem Provers}. In \bibinfo{booktitle}{\emph{21st IEEE/ACM International Conference on Automated Software Engineering (ASE'06)}}. \bibinfo{pages}{167--176}.
\newblock


\bibitem[Kakarla and Beckett(2023)]%
        {kakarla2023oracle}
\bibfield{author}{\bibinfo{person}{Siva Kesava~Reddy Kakarla} {and} \bibinfo{person}{Ryan Beckett}.} \bibinfo{year}{2023}\natexlab{}.
\newblock \showarticletitle{Oracle-based Protocol Testing with Eywa}.
\newblock \bibinfo{journal}{\emph{arXiv:2312.06875}} (\bibinfo{year}{2023}).
\newblock


\bibitem[Katz and Shin(2006)]%
        {katz2006parallel}
\bibfield{author}{\bibinfo{person}{Jonathan Katz} {and} \bibinfo{person}{Ji~Sun Shin}.} \bibinfo{year}{2006}\natexlab{}.
\newblock \showarticletitle{Parallel and concurrent security of the HB and HB+ protocols}. In \bibinfo{booktitle}{\emph{25th International Cryptology Conference (EUROCRYPT)}}. \bibinfo{pages}{73--87}.
\newblock


\bibitem[Kim et~al\mbox{.}(2023)]%
        {kim2023fuzz}
\bibfield{author}{\bibinfo{person}{Kyungtae Kim}, \bibinfo{person}{Sungwoo Kim}, \bibinfo{person}{Kevin~RB Butler}, \bibinfo{person}{Antonio Bianchi}, \bibinfo{person}{Rick Kennell}, {and} \bibinfo{person}{Dave~Jing Tian}.} \bibinfo{year}{2023}\natexlab{}.
\newblock \showarticletitle{Fuzz The Power: Dual-role State Guided Black-box Fuzzing for USB Power Delivery}. In \bibinfo{booktitle}{\emph{32nd USENIX Security Symposium (USENIX Security 23)}}. \bibinfo{pages}{5845--5861}.
\newblock


\bibitem[Kim et~al\mbox{.}(2017)]%
        {sp17-iotcube}
\bibfield{author}{\bibinfo{person}{Seulbae Kim}, \bibinfo{person}{Seunghoon Woo}, \bibinfo{person}{Heejo Lee}, {and} \bibinfo{person}{Hakjoo Oh}.} \bibinfo{year}{2017}\natexlab{}.
\newblock \showarticletitle{Poster: Iotcube: an automated analysis platform for finding security vulnerabilities}. In \bibinfo{booktitle}{\emph{38th IEEE Symposium on Poster presented at Security and Privacy}}.
\newblock


\bibitem[Krupp et~al\mbox{.}(2022)]%
        {sec22-ampfuzz}
\bibfield{author}{\bibinfo{person}{Johannes Krupp}, \bibinfo{person}{Ilya Grishchenko}, {and} \bibinfo{person}{Christian Rossow}.} \bibinfo{year}{2022}\natexlab{}.
\newblock \showarticletitle{{AmpFuzz}: Fuzzing for Amplification {DDoS} Vulnerabilities}. In \bibinfo{booktitle}{\emph{31st USENIX Security Symposium (USENIX Security 22)}}. \bibinfo{pages}{1043--1060}.
\newblock
\showISBNx{978-1-939133-31-1}


\bibitem[Lee et~al\mbox{.}(2018)]%
        {blackhat-us-18-delta}
\bibfield{author}{\bibinfo{person}{Seungsoo Lee}, \bibinfo{person}{Jinwoo Kim}, \bibinfo{person}{Seungwon Woo}, {and} \bibinfo{person}{Seungwon Shin}.} \bibinfo{year}{2018}\natexlab{}.
\newblock \showarticletitle{The Finest Penetration Testing Framework for Software-Defined Networks}. In \bibinfo{booktitle}{\emph{Blackhat US 2018}}.
\newblock


\bibitem[Li et~al\mbox{.}(2022b)]%
        {arxiv22-SPIDER}
\bibfield{author}{\bibinfo{person}{Ao Li}, \bibinfo{person}{Rohan Padhye}, {and} \bibinfo{person}{Vyas Sekar}.} \bibinfo{year}{2022}\natexlab{b}.
\newblock \bibinfo{title}{SPIDER: A Practical Fuzzing Framework to Uncover Stateful Performance Issues in SDN Controllers}.
\newblock
\newblock
\urldef\tempurl%
\url{https://arxiv.org/abs/2209.04026}
\showURL{%
\tempurl}


\bibitem[Li et~al\mbox{.}(2022a)]%
        {arxiv2022-snpsfuzzer}
\bibfield{author}{\bibinfo{person}{Junqiang Li}, \bibinfo{person}{Senyi Li}, \bibinfo{person}{Gang Sun}, \bibinfo{person}{Ting Chen}, {and} \bibinfo{person}{Hongfang Yu}.} \bibinfo{year}{2022}\natexlab{a}.
\newblock \showarticletitle{SNPSFuzzer: {A} Fast Greybox Fuzzer for Stateful Network Protocols using Snapshots}.
\newblock  (\bibinfo{year}{2022}).
\newblock
\showeprint[arXiv]{2202.03643}


\bibitem[Liang et~al\mbox{.}(2018)]%
        {tr18-fuzzing-survey}
\bibfield{author}{\bibinfo{person}{Hongliang Liang}, \bibinfo{person}{Xiaoxiao Pei}, \bibinfo{person}{Xiaodong Jia}, \bibinfo{person}{Wuwei Shen}, {and} \bibinfo{person}{Jian Zhang}.} \bibinfo{year}{2018}\natexlab{}.
\newblock \showarticletitle{Fuzzing: State of the Art}.
\newblock \bibinfo{journal}{\emph{IEEE Transactions on Reliability}} \bibinfo{volume}{67}, \bibinfo{number}{3} (\bibinfo{year}{2018}), \bibinfo{pages}{1199--1218}.
\newblock


\bibitem[Liu et~al\mbox{.}(2020)]%
        {ase20-legion}
\bibfield{author}{\bibinfo{person}{Dongge Liu}, \bibinfo{person}{Gidon Ernst}, \bibinfo{person}{Toby Murray}, {and} \bibinfo{person}{Benjamin I.~P. Rubinstein}.} \bibinfo{year}{2020}\natexlab{}.
\newblock \showarticletitle{Legion: Best-First Concolic Testing}. In \bibinfo{booktitle}{\emph{35th IEEE/ACM International Conference on Automated Software Engineering (ASE)}}. \bibinfo{pages}{54–65}.
\newblock


\bibitem[Liu et~al\mbox{.}(2022a)]%
        {dongge-state-selection}
\bibfield{author}{\bibinfo{person}{D. Liu}, \bibinfo{person}{V. Pham}, \bibinfo{person}{G. Ernst}, \bibinfo{person}{T. Murray}, {and} \bibinfo{person}{B.~P. Rubinstein}.} \bibinfo{year}{2022}\natexlab{a}.
\newblock \showarticletitle{State Selection Algorithms and Their Impact on The Performance of Stateful Network Protocol Fuzzing}. In \bibinfo{booktitle}{\emph{2022 IEEE International Conference on Software Analysis, Evolution and Reengineering (SANER)}}. \bibinfo{pages}{720--730}.
\newblock


\bibitem[Liu et~al\mbox{.}(2022b)]%
        {liu2022fuzzing}
\bibfield{author}{\bibinfo{person}{Puzhuo Liu}, \bibinfo{person}{Yaowen Zheng}, \bibinfo{person}{Zhanwei Song}, \bibinfo{person}{Dongliang Fang}, \bibinfo{person}{Shichao Lv}, {and} \bibinfo{person}{Limin Sun}.} \bibinfo{year}{2022}\natexlab{b}.
\newblock \showarticletitle{Fuzzing proprietary protocols of programmable controllers to find vulnerabilities that affect physical control}.
\newblock \bibinfo{journal}{\emph{Journal of Systems Architecture}}  \bibinfo{volume}{127} (\bibinfo{year}{2022}), \bibinfo{pages}{102483}.
\newblock


\bibitem[Liu et~al\mbox{.}(2021)]%
        {9678653}
\bibfield{author}{\bibinfo{person}{Qiang Liu}, \bibinfo{person}{Cen Zhang}, \bibinfo{person}{Lin Ma}, \bibinfo{person}{Muhui Jiang}, \bibinfo{person}{Yajin Zhou}, \bibinfo{person}{Lei Wu}, \bibinfo{person}{Wenbo Shen}, \bibinfo{person}{Xiapu Luo}, \bibinfo{person}{Yang Liu}, {and} \bibinfo{person}{Kui Ren}.} \bibinfo{year}{2021}\natexlab{}.
\newblock \showarticletitle{FirmGuide: Boosting the Capability of Rehosting Embedded Linux Kernels through Model-Guided Kernel Execution}. In \bibinfo{booktitle}{\emph{36th IEEE/ACM International Conference on Automated Software Engineering (ASE)}}. \bibinfo{pages}{792--804}.
\newblock


\bibitem[Luo et~al\mbox{.}(2024a)]%
        {ndss24-dynpre}
\bibfield{author}{\bibinfo{person}{Zhengxiong Luo}, \bibinfo{person}{Kai Liang}, \bibinfo{person}{Yanyang Zhao}, \bibinfo{person}{Feifan Wu}, \bibinfo{person}{Junze Yu}, \bibinfo{person}{Heyuan Shi}, {and} \bibinfo{person}{Yu Jiang}.} \bibinfo{year}{2024}\natexlab{a}.
\newblock \showarticletitle{DYNPRE: Protocol Reverse Engineering via Dynamic Inference}. In \bibinfo{booktitle}{\emph{31th Annual Network and Distributed System Security Symposium (NDSS)}}. \bibinfo{pages}{1--18}.
\newblock


\bibitem[Luo et~al\mbox{.}(2024b)]%
        {emsoft24-mpfuzz}
\bibfield{author}{\bibinfo{person}{Zhengxiong Luo}, \bibinfo{person}{Junze Yu}, \bibinfo{person}{Qingpeng Du}, \bibinfo{person}{Yanyang Zhao}, \bibinfo{person}{Feifan Wu}, {and} \bibinfo{person}{Heyuan Shi}.} \bibinfo{year}{2024}\natexlab{b}.
\newblock \showarticletitle{Parallel Fuzzing of IoT Messaging Protocols through Collaborative Packet Generation}. In \bibinfo{booktitle}{\emph{ACM SIGBED International Conference on Embedded Software (EMSOFT)}}. \bibinfo{pages}{1--12}.
\newblock


\bibitem[Luo et~al\mbox{.}(2023)]%
        {sec23-bleem}
\bibfield{author}{\bibinfo{person}{Zhengxiong Luo}, \bibinfo{person}{Junze Yu}, \bibinfo{person}{Feilong Zuo}, \bibinfo{person}{Jianzhong Liu}, \bibinfo{person}{Yu Jiang}, \bibinfo{person}{Ting Chen}, \bibinfo{person}{Abhik Roychoudhury}, {and} \bibinfo{person}{Jiaguang Sun}.} \bibinfo{year}{2023}\natexlab{}.
\newblock \showarticletitle{Bleem: Packet Sequence Oriented Fuzzing for Protocol Implementations}. In \bibinfo{booktitle}{\emph{32nd USENIX Security Symposium (USENIX Security 23)}}. \bibinfo{pages}{4481--4498}.
\newblock


\bibitem[Luo et~al\mbox{.}(2019)]%
        {tecs19-polar}
\bibfield{author}{\bibinfo{person}{Zhengxiong Luo}, \bibinfo{person}{Feilong Zuo}, \bibinfo{person}{Yu Jiang}, \bibinfo{person}{Jian Gao}, \bibinfo{person}{Xun Jiao}, {and} \bibinfo{person}{Jiaguang Sun}.} \bibinfo{year}{2019}\natexlab{}.
\newblock \showarticletitle{Polar: Function Code Aware Fuzz Testing of {ICS} Protocol}.
\newblock \bibinfo{journal}{\emph{{ACM} Trans. Embed. Comput. Syst.}} \bibinfo{volume}{18}, \bibinfo{number}{5s} (\bibinfo{year}{2019}), \bibinfo{pages}{93:1--93:22}.
\newblock


\bibitem[Luo et~al\mbox{.}(2020)]%
        {dac20-icsprotocol}
\bibfield{author}{\bibinfo{person}{Zhengxiong Luo}, \bibinfo{person}{Feilong Zuo}, \bibinfo{person}{Yuheng Shen}, \bibinfo{person}{Xun Jiao}, \bibinfo{person}{Wanli Chang}, {and} \bibinfo{person}{Yu Jiang}.} \bibinfo{year}{2020}\natexlab{}.
\newblock \showarticletitle{{ICS} Protocol Fuzzing: Coverage Guided Packet Crack and Generation}. In \bibinfo{booktitle}{\emph{57th {ACM/IEEE} Design Automation Conference (DAC)}}. \bibinfo{pages}{1--6}.
\newblock


\bibitem[Ma et~al\mbox{.}(2023a)]%
        {ma2023loki}
\bibfield{author}{\bibinfo{person}{Fuchen Ma}, \bibinfo{person}{Yuanliang Chen}, \bibinfo{person}{Meng Ren}, \bibinfo{person}{Yuanhang Zhou}, \bibinfo{person}{Yu Jiang}, \bibinfo{person}{Ting Chen}, \bibinfo{person}{Huizhong Li}, {and} \bibinfo{person}{Jiaguang Sun}.} \bibinfo{year}{2023}\natexlab{a}.
\newblock \showarticletitle{LOKI: State-Aware Fuzzing Framework for the Implementation of Blockchain Consensus Protocols}. In \bibinfo{booktitle}{\emph{Proceedings 2023 Network and Distributed System Security Symposium}}.
\newblock


\bibitem[Ma et~al\mbox{.}(2024)]%
        {sec24-mgptfuzz}
\bibfield{author}{\bibinfo{person}{Xiaoyue Ma}, \bibinfo{person}{Lannan Luo}, {and} \bibinfo{person}{Qiang Zeng}.} \bibinfo{year}{2024}\natexlab{}.
\newblock \showarticletitle{From One Thousand Pages of Specification to Unveiling Hidden Bugs: Large Language Model Assisted Fuzzing of Matter {IoT} Devices}. In \bibinfo{booktitle}{\emph{33rd USENIX Security Symposium (USENIX Security 24)}}. \bibinfo{pages}{4783--4800}.
\newblock


\bibitem[Ma et~al\mbox{.}(2023b)]%
        {ma2023no}
\bibfield{author}{\bibinfo{person}{Xiaoyue Ma}, \bibinfo{person}{Qiang Zeng}, \bibinfo{person}{Haotian Chi}, {and} \bibinfo{person}{Lannan Luo}.} \bibinfo{year}{2023}\natexlab{b}.
\newblock \showarticletitle{No more companion apps hacking but one dongle: Hub-based blackbox fuzzing of iot firmware}. In \bibinfo{booktitle}{\emph{21st Annual International Conference on Mobile Systems, Applications and Services (Mobisys)}}. \bibinfo{pages}{205--218}.
\newblock


\bibitem[Maier et~al\mbox{.}(2020)]%
        {wisec20-BaseSAFE}
\bibfield{author}{\bibinfo{person}{Dominik Maier}, \bibinfo{person}{Lukas Seidel}, {and} \bibinfo{person}{Shinjo Park}.} \bibinfo{year}{2020}\natexlab{}.
\newblock \showarticletitle{BaseSAFE: Baseband SAnitized Fuzzing through Emulation}.
\newblock  (\bibinfo{year}{2020}).
\newblock
\showeprint[arXiv]{2005.07797}


\bibitem[Manes et~al\mbox{.}(2021)]%
        {tse19-fuzzing-survey}
\bibfield{author}{\bibinfo{person}{V.~M. Manes}, \bibinfo{person}{H. Han}, \bibinfo{person}{C. Han}, \bibinfo{person}{S. Cha}, \bibinfo{person}{M. Egele}, \bibinfo{person}{E.~J. Schwartz}, {and} \bibinfo{person}{M. Woo}.} \bibinfo{year}{2021}\natexlab{}.
\newblock \showarticletitle{The Art, Science, and Engineering of Fuzzing: A Survey}.
\newblock \bibinfo{journal}{\emph{IEEE Transactions on Software Engineering}} \bibinfo{volume}{47}, \bibinfo{number}{11} (\bibinfo{year}{2021}), \bibinfo{pages}{2312--2331}.
\newblock


\bibitem[Marcussen(2018)]%
        {doona}
\bibfield{author}{\bibinfo{person}{Eldar Marcussen}.} \bibinfo{year}{2018}\natexlab{}.
\newblock \bibinfo{title}{Doona - Network fuzzing tool}.
\newblock
\newblock
\urldef\tempurl%
\url{Available: https://github.com/wireghoul/doona}
\showURL{%
\tempurl}


\bibitem[Marsden(1986)]%
        {marsden1986communication}
\bibfield{author}{\bibinfo{person}{B.W. Marsden}.} \bibinfo{year}{1986}\natexlab{}.
\newblock \bibinfo{booktitle}{\emph{Communication Network Protocols}}.
\newblock
\showISBNx{9780862381066}
\showLCCN{lc89188307}


\bibitem[Mathis et~al\mbox{.}(2019)]%
        {pldi19-parser}
\bibfield{author}{\bibinfo{person}{Bj\"{o}rn Mathis}, \bibinfo{person}{Rahul Gopinath}, \bibinfo{person}{Micha\"{e}l Mera}, \bibinfo{person}{Alexander Kampmann}, \bibinfo{person}{Matthias H\"{o}schele}, {and} \bibinfo{person}{Andreas Zeller}.} \bibinfo{year}{2019}\natexlab{}.
\newblock \showarticletitle{Parser-Directed Fuzzing}. In \bibinfo{booktitle}{\emph{40th ACM SIGPLAN Conference on Programming Language Design and Implementation (PLDI)}}. \bibinfo{pages}{548–560}.
\newblock


\bibitem[McMahon~Stone et~al\mbox{.}(2018)]%
        {esorics18-extendstatefuzzing}
\bibfield{author}{\bibinfo{person}{Chris McMahon~Stone}, \bibinfo{person}{Tom Chothia}, {and} \bibinfo{person}{Joeri de Ruiter}.} \bibinfo{year}{2018}\natexlab{}.
\newblock \showarticletitle{Extending Automated Protocol State Learning for the 802.11 4-Way Handshake}. In \bibinfo{booktitle}{\emph{Computer Security}}. \bibinfo{pages}{325--345}.
\newblock
\showISBNx{978-3-319-99073-6}


\bibitem[McMahon~Stone et~al\mbox{.}(2022)]%
        {sec22-stateinspector}
\bibfield{author}{\bibinfo{person}{Chris McMahon~Stone}, \bibinfo{person}{Sam~L. Thomas}, \bibinfo{person}{Mathy Vanhoef}, \bibinfo{person}{James Henderson}, \bibinfo{person}{Nicolas Bailluet}, {and} \bibinfo{person}{Tom Chothia}.} \bibinfo{year}{2022}\natexlab{}.
\newblock \showarticletitle{The Closer You Look, The More You Learn: A Grey-box Approach to Protocol State Machine Learning}. In \bibinfo{booktitle}{\emph{2022 ACM SIGSAC Conference on Computer and Communications Security (CCS)}}. \bibinfo{pages}{2265–2278}.
\newblock


\bibitem[Meng et~al\mbox{.}(2021)]%
        {icse22-ltlprop}
\bibfield{author}{\bibinfo{person}{Ruijie Meng}, \bibinfo{person}{Zhen Dong}, \bibinfo{person}{Jialin Li}, \bibinfo{person}{Ivan Beschastnikh}, {and} \bibinfo{person}{Abhik Roychoudhury}.} \bibinfo{year}{2021}\natexlab{}.
\newblock \showarticletitle{Finding Counterexamples of Temporal Logic properties in Software Implementations via Greybox Fuzzing}.
\newblock \bibinfo{journal}{\emph{CoRR}}  \bibinfo{volume}{abs/2109.02312} (\bibinfo{year}{2021}).
\newblock
\showeprint[arXiv]{2109.02312}


\bibitem[Meng et~al\mbox{.}(2024)]%
        {ndss24-chatafl}
\bibfield{author}{\bibinfo{person}{Ruijie Meng}, \bibinfo{person}{Martin Mirchev}, \bibinfo{person}{Marcel B{\"{o}}hme}, {and} \bibinfo{person}{Abhik Roychoudhury}.} \bibinfo{year}{2024}\natexlab{}.
\newblock \showarticletitle{Large Language Model guided Protocol Fuzzing}. In \bibinfo{booktitle}{\emph{31st Annual Network and Distributed System Security Symposium (NDSS)}}. \bibinfo{pages}{1--17}.
\newblock


\bibitem[Meng et~al\mbox{.}(2023)]%
        {ccs23-Mallory}
\bibfield{author}{\bibinfo{person}{Ruijie Meng}, \bibinfo{person}{George P\^{\i}rlea}, \bibinfo{person}{Abhik Roychoudhury}, {and} \bibinfo{person}{Ilya Sergey}.} \bibinfo{year}{2023}\natexlab{}.
\newblock \showarticletitle{Greybox Fuzzing of Distributed Systems}. In \bibinfo{booktitle}{\emph{2023 ACM SIGSAC Conference on Computer and Communications Security (CCS)}}. \bibinfo{pages}{1615–1629}.
\newblock


\bibitem[Microsoft(2007)]%
        {rdpdoc}
\bibfield{author}{\bibinfo{person}{Microsoft}.} \bibinfo{year}{2007}\natexlab{}.
\newblock \bibinfo{title}{Remote Desktop Protocol: Basic Connectivity and Graphics Remoting}.
\newblock
\newblock
\urldef\tempurl%
\url{https://learn.microsoft.com/en-us/openspecs/windows_protocols/ms-rdpbcgr/5073f4ed-1e93-45e1-b039-6e30c385867c}
\showURL{%
\tempurl}


\bibitem[Miller et~al\mbox{.}(1990)]%
        {miller1990empirical}
\bibfield{author}{\bibinfo{person}{Barton~P Miller}, \bibinfo{person}{Lars Fredriksen}, {and} \bibinfo{person}{Bryan So}.} \bibinfo{year}{1990}\natexlab{}.
\newblock \showarticletitle{An empirical study of the reliability of UNIX utilities}.
\newblock \bibinfo{journal}{\emph{Commun. ACM}} \bibinfo{volume}{33}, \bibinfo{number}{12} (\bibinfo{year}{1990}), \bibinfo{pages}{32--44}.
\newblock


\bibitem[Musuvathi and Engler(2004)]%
        {nsdi04}
\bibfield{author}{\bibinfo{person}{Madanlal Musuvathi} {and} \bibinfo{person}{Dawson~R. Engler}.} \bibinfo{year}{2004}\natexlab{}.
\newblock \showarticletitle{Model Checking Large Network Protocol Implementations}. In \bibinfo{booktitle}{\emph{First Symposium on Networked Systems Design and Implementation (NSDI 04)}}. \bibinfo{pages}{12}.
\newblock


\bibitem[Mutton(2014)]%
        {heartbleedblog}
\bibfield{author}{\bibinfo{person}{Paul Mutton}.} \bibinfo{year}{2014}\natexlab{}.
\newblock \bibinfo{title}{Half a million widely trusted websites vulnerable to Heartbleed bug}.
\newblock
\newblock
\urldef\tempurl%
\url{https://news.netcraft.com/archives/2014/04/08/half-a-million-widely-trusted-websites-vulnerable-to-heartbleed-bug.html}
\showURL{%
\tempurl}


\bibitem[Natella(2022)]%
        {arxiv21-stateafl}
\bibfield{author}{\bibinfo{person}{Roberto Natella}.} \bibinfo{year}{2022}\natexlab{}.
\newblock \showarticletitle{Stateafl: Greybox fuzzing for stateful network servers}.
\newblock \bibinfo{journal}{\emph{Empirical Software Engineering}} \bibinfo{volume}{27}, \bibinfo{number}{7} (\bibinfo{year}{2022}), \bibinfo{pages}{191}.
\newblock


\bibitem[Nikbakht et~al\mbox{.}(2024)]%
        {nikbakht2024tspec}
\bibfield{author}{\bibinfo{person}{Rasoul Nikbakht}, \bibinfo{person}{Mohamed Benzaghta}, {and} \bibinfo{person}{Giovanni Geraci}.} \bibinfo{year}{2024}\natexlab{}.
\newblock \showarticletitle{TSpec-LLM: An Open-source Dataset for LLM Understanding of 3GPP Specifications}.
\newblock \bibinfo{journal}{\emph{arXiv:2406.01768}} (\bibinfo{year}{2024}).
\newblock


\bibitem[NOCHVAY(2019)]%
        {codesysdoc}
\bibfield{author}{\bibinfo{person}{NOCHVAY}.} \bibinfo{year}{2019}\natexlab{}.
\newblock \bibinfo{title}{Security research: CODESYS Runtime, a PLC control framework}.
\newblock
\newblock
\urldef\tempurl%
\url{https://ics-cert.kaspersky.com/publications/reports/2019/09/18/security-research-codesys-runtime-a-plc-control-framework-part-1/}
\showURL{%
\tempurl}


\bibitem[OASIS(2019)]%
        {mqttdoc}
\bibfield{author}{\bibinfo{person}{OASIS}.} \bibinfo{year}{2019}\natexlab{}.
\newblock \bibinfo{title}{MQTT Version 5.0}.
\newblock
\newblock
\urldef\tempurl%
\url{https://docs.oasis-open.org/mqtt/mqtt/v5.0/mqtt-v5.0.html}
\showURL{%
\tempurl}


\bibitem[Ohkawa et~al\mbox{.}(2019)]%
        {ohkawa2019high}
\bibfield{author}{\bibinfo{person}{Takeshi Ohkawa}, \bibinfo{person}{Yuhei Sugata}, \bibinfo{person}{Harumi Watanabe}, \bibinfo{person}{Nobuhiko Ogura}, \bibinfo{person}{Kanemitsu Ootsu}, {and} \bibinfo{person}{Takashi Yokota}.} \bibinfo{year}{2019}\natexlab{}.
\newblock \showarticletitle{High level synthesis of ROS protocol interpretation and communication circuit for FPGA}. In \bibinfo{booktitle}{\emph{2nd International Workshop on Robotics Software Engineering (RoSE)}}. \bibinfo{pages}{33--36}.
\newblock


\bibitem[OMG(2018)]%
        {rtpsdoc}
\bibfield{author}{\bibinfo{person}{OMG}.} \bibinfo{year}{2018}\natexlab{}.
\newblock \bibinfo{title}{The Real-time Publish-Subscribe Protocol (RTPS) DDS Interoperability Wire Protocol Specification}.
\newblock
\newblock
\urldef\tempurl%
\url{https://www.omg.org/spec/DDSI-RTPS/2.3/Beta1/PDF}
\showURL{%
\tempurl}


\bibitem[openvpn(2014)]%
        {openvpndoc}
\bibfield{author}{\bibinfo{person}{openvpn}.} \bibinfo{year}{2014}\natexlab{}.
\newblock \bibinfo{title}{OpenVPN: OpenVPN source code documentation}.
\newblock
\newblock
\urldef\tempurl%
\url{https://build.openvpn.net/doxygen/}
\showURL{%
\tempurl}


\bibitem[Ozavci(2013)]%
        {defcon21-Ozavci}
\bibfield{author}{\bibinfo{person}{Fatih Ozavci}.} \bibinfo{year}{2013}\natexlab{}.
\newblock \showarticletitle{VoIP Wars : Return of the SIP}. In \bibinfo{booktitle}{\emph{DEFCON 21}}.
\newblock


\bibitem[Pacheco et~al\mbox{.}(2022)]%
        {sp22-auto-fsm-extraction}
\bibfield{author}{\bibinfo{person}{Maria~Leonor Pacheco}, \bibinfo{person}{Max von Hippel}, \bibinfo{person}{Ben Weintraub}, \bibinfo{person}{Dan Goldwasser}, {and} \bibinfo{person}{Cristina Nita{-}Rotaru}.} \bibinfo{year}{2022}\natexlab{}.
\newblock \showarticletitle{Automated Attack Synthesis by Extracting Finite State Machines from Protocol Specification Documents}.
\newblock  (\bibinfo{year}{2022}).
\newblock
\showeprint[arXiv]{2202.09470}


\bibitem[Park et~al\mbox{.}(2019)]%
        {blackhat-eu-19-park}
\bibfield{author}{\bibinfo{person}{Chun~Sung Park}, \bibinfo{person}{Yeongjin Jang}, \bibinfo{person}{Seungjoo Kim}, {and} \bibinfo{person}{Ki~Taek Lee}.} \bibinfo{year}{2019}\natexlab{}.
\newblock \bibinfo{title}{Fuzzing and Exploiting Virtual Channels in Microsoft Remote Desktop Protocol for Fun and Profit}.
\newblock
\newblock
\urldef\tempurl%
\url{https://www.blackhat.com/eu-19/briefings/schedule/#fuzzing-and-exploiting-virtual-channels-in-microsoft-remote-desktop-protocol-for-fun-and-profit-17789}
\showURL{%
\tempurl}


\bibitem[Park et~al\mbox{.}(2022)]%
        {dsn22-l2fuzz}
\bibfield{author}{\bibinfo{person}{H. Park}, \bibinfo{person}{C. Nkuba}, \bibinfo{person}{S. Woo}, {and} \bibinfo{person}{H. Lee}.} \bibinfo{year}{2022}\natexlab{}.
\newblock \showarticletitle{L2Fuzz: Discovering Bluetooth L2CAP Vulnerabilities Using Stateful Fuzz Testing}. In \bibinfo{booktitle}{\emph{2022 52nd Annual IEEE/IFIP International Conference on Dependable Systems and Networks (DSN)}}. \bibinfo{pages}{343--354}.
\newblock


\bibitem[Pearson et~al\mbox{.}(2022)]%
        {infocom22-fume}
\bibfield{author}{\bibinfo{person}{Bryan Pearson}, \bibinfo{person}{Yue Zhang}, \bibinfo{person}{Cliff Zou}, {and} \bibinfo{person}{Xinwen Fu}.} \bibinfo{year}{2022}\natexlab{}.
\newblock \showarticletitle{FUME: Fuzzing Message Queuing Telemetry Transport Brokers}. In \bibinfo{booktitle}{\emph{2022 IEEE Conference on Computer Communications (INFOCOM)}}. \bibinfo{pages}{1699--1708}.
\newblock


\bibitem[Peterson et~al\mbox{.}(2020)]%
        {raid20-abbrate}
\bibfield{author}{\bibinfo{person}{Anthony Peterson}, \bibinfo{person}{Samuel Jero}, \bibinfo{person}{Endadul Hoque}, \bibinfo{person}{David Choffnes}, {and} \bibinfo{person}{Cristina Nita-Rotaru}.} \bibinfo{year}{2020}\natexlab{}.
\newblock \showarticletitle{{aBBRate}: Automating {BBR} Attack Exploration Using a {Model-Based} Approach}. In \bibinfo{booktitle}{\emph{23rd International Symposium on Research in Attacks, Intrusions and Defenses (RAID)}}. \bibinfo{pages}{225--240}.
\newblock


\bibitem[Pham et~al\mbox{.}(2020)]%
        {icst20-aflnet}
\bibfield{author}{\bibinfo{person}{Van{-}Thuan Pham}, \bibinfo{person}{Marcel B{\"{o}}hme}, {and} \bibinfo{person}{Abhik Roychoudhury}.} \bibinfo{year}{2020}\natexlab{}.
\newblock \showarticletitle{{AFLNET:} {A} Greybox Fuzzer for Network Protocols}. In \bibinfo{booktitle}{\emph{13th {IEEE} International Conference on Software Testing, Validation and Verification (ICST)}}. \bibinfo{pages}{460--465}.
\newblock


\bibitem[Pham et~al\mbox{.}(2019)]%
        {pham2019smart}
\bibfield{author}{\bibinfo{person}{Van-Thuan Pham}, \bibinfo{person}{Marcel B{\"o}hme}, \bibinfo{person}{Andrew~E Santosa}, \bibinfo{person}{Alexandru~R{\u{a}}zvan C{\u{a}}ciulescu}, {and} \bibinfo{person}{Abhik Roychoudhury}.} \bibinfo{year}{2019}\natexlab{}.
\newblock \showarticletitle{Smart greybox fuzzing}.
\newblock \bibinfo{journal}{\emph{IEEE Transactions on Software Engineering}} \bibinfo{volume}{47}, \bibinfo{number}{9} (\bibinfo{year}{2019}), \bibinfo{pages}{1980--1997}.
\newblock


\bibitem[Poncelet et~al\mbox{.}(2022)]%
        {poncelet2022so}
\bibfield{author}{\bibinfo{person}{Cl{\'e}ment Poncelet}, \bibinfo{person}{Konstantinos Sagonas}, {and} \bibinfo{person}{Nicolas Tsiftes}.} \bibinfo{year}{2022}\natexlab{}.
\newblock \showarticletitle{So Many Fuzzers, So Little Time\*: Experience from Evaluating Fuzzers on the Contiki-NG Network (Hay) Stack}. In \bibinfo{booktitle}{\emph{37th IEEE/ACM International Conference on Automated Software Engineering}}. \bibinfo{pages}{1--12}.
\newblock


\bibitem[Project(2022)]%
        {openssl}
\bibfield{author}{\bibinfo{person}{OpenSSL Project}.} \bibinfo{year}{2022}\natexlab{}.
\newblock \bibinfo{title}{OpenSSL}.
\newblock
\newblock
\urldef\tempurl%
\url{Available: https://github.com/openssl/openssl}
\showURL{%
\tempurl}


\bibitem[Qin et~al\mbox{.}(2023)]%
        {qin2023nsfuzz}
\bibfield{author}{\bibinfo{person}{Shisong Qin}, \bibinfo{person}{Fan Hu}, \bibinfo{person}{Zheyu Ma}, \bibinfo{person}{Bodong Zhao}, \bibinfo{person}{Tingting Yin}, {and} \bibinfo{person}{Chao Zhang}.} \bibinfo{year}{2023}\natexlab{}.
\newblock \showarticletitle{NSFuzz: Towards Efficient and State-Aware Network Service Fuzzing}.
\newblock \bibinfo{journal}{\emph{ACM Transactions on Software Engineering and Methodology}} (\bibinfo{year}{2023}).
\newblock


\bibitem[Qu et~al\mbox{.}(2021)]%
        {blackhat-eu-21-badmesher}
\bibfield{author}{\bibinfo{person}{Lewei Qu}, \bibinfo{person}{Dongxiang Ke}, \bibinfo{person}{Ye Zhang}, {and} \bibinfo{person}{Ying Wang}.} \bibinfo{year}{2021}\natexlab{}.
\newblock \showarticletitle{BadMesher: New Attack Surfaces of Wi-Fi Mesh Network}. In \bibinfo{booktitle}{\emph{Blackhat EU 2021}}.
\newblock


\bibitem[Quynh and Vu(2015)]%
        {quynh2015unicorn}
\bibfield{author}{\bibinfo{person}{NGUYEN~Anh Quynh} {and} \bibinfo{person}{DANG~Hoang Vu}.} \bibinfo{year}{2015}\natexlab{}.
\newblock \showarticletitle{Unicorn: Next generation cpu emulator framework}.
\newblock \bibinfo{journal}{\emph{BlackHat USA}}  \bibinfo{volume}{476} (\bibinfo{year}{2015}).
\newblock


\bibitem[Rangnau et~al\mbox{.}(2020)]%
        {9233212}
\bibfield{author}{\bibinfo{person}{Thorsten Rangnau}, \bibinfo{person}{Remco~v. Buijtenen}, \bibinfo{person}{Frank Fransen}, {and} \bibinfo{person}{Fatih Turkmen}.} \bibinfo{year}{2020}\natexlab{}.
\newblock \showarticletitle{Continuous Security Testing: A Case Study on Integrating Dynamic Security Testing Tools in CI/CD Pipelines}. In \bibinfo{booktitle}{\emph{24th International Enterprise Distributed Object Computing Conference (EDOC)}}. \bibinfo{pages}{145--154}.
\newblock


\bibitem[Reen and Rossow(2020)]%
        {acsac20-dpifuzz}
\bibfield{author}{\bibinfo{person}{Gaganjeet~Singh Reen} {and} \bibinfo{person}{Christian Rossow}.} \bibinfo{year}{2020}\natexlab{}.
\newblock \showarticletitle{DPIFuzz: A Differential Fuzzing Framework to Detect DPI Elusion Strategies for QUIC}. In \bibinfo{booktitle}{\emph{Annual Computer Security Applications Conference (ACSAC)}}. \bibinfo{pages}{332–344}.
\newblock


\bibitem[Ren et~al\mbox{.}(2021)]%
        {wisec21-zfuzzer}
\bibfield{author}{\bibinfo{person}{Mengfei Ren}, \bibinfo{person}{Xiaolei Ren}, \bibinfo{person}{Huadong Feng}, \bibinfo{person}{Jiang Ming}, {and} \bibinfo{person}{Yu Lei}.} \bibinfo{year}{2021}\natexlab{}.
\newblock \showarticletitle{Z-Fuzzer: device-agnostic fuzzing of Zigbee protocol implementation}. In \bibinfo{booktitle}{\emph{14th {ACM} Conference on Security and Privacy in Wireless and Mobile Networks (WiSec)}}. \bibinfo{pages}{347--358}.
\newblock


\bibitem[Ren et~al\mbox{.}(2023)]%
        {esorics23-taintbfuzz}
\bibfield{author}{\bibinfo{person}{Mengfei Ren}, \bibinfo{person}{Haotian Zhang}, \bibinfo{person}{Xiaolei Ren}, \bibinfo{person}{Jiang Ming}, {and} \bibinfo{person}{Yu Lei}.} \bibinfo{year}{2023}\natexlab{}.
\newblock \showarticletitle{Intelligent Zigbee Protocol Fuzzing via Constraint-Field Dependency Inference}. In \bibinfo{booktitle}{\emph{European Symposium on Research in Computer Security (ESORICS)}}. \bibinfo{pages}{467--486}.
\newblock


\bibitem[Rescorla(2012)]%
        {dtlsdoc}
\bibfield{author}{\bibinfo{person}{E. Rescorla}.} \bibinfo{year}{2012}\natexlab{}.
\newblock \bibinfo{title}{RFC6347: Datagram Transport Layer Security Version 1.2}.
\newblock
\newblock
\urldef\tempurl%
\url{https://datatracker.ietf.org/doc/html/rfc6347}
\showURL{%
\tempurl}


\bibitem[Roitburd et~al\mbox{.}(2021)]%
        {cns21-Roitburd}
\bibfield{author}{\bibinfo{person}{Gerbert Roitburd}, \bibinfo{person}{Matthias Ortmann}, \bibinfo{person}{Matthias Hollick}, {and} \bibinfo{person}{Jiska Classen}.} \bibinfo{year}{2021}\natexlab{}.
\newblock \showarticletitle{Very Pwnable Network: Cisco AnyConnect Security Analysis}. In \bibinfo{booktitle}{\emph{2021 IEEE Conference on Communications and Network Security (CNS)}}. \bibinfo{pages}{56--64}.
\newblock


\bibitem[Romero and Rivas(2019)]%
        {defcon27-fuzzowski}
\bibfield{author}{\bibinfo{person}{Daniel Romero} {and} \bibinfo{person}{Mario Rivas}.} \bibinfo{year}{2019}\natexlab{}.
\newblock \showarticletitle{Why you should fear your 'mundane' office equipment}. In \bibinfo{booktitle}{\emph{DEFCON 27}}.
\newblock


\bibitem[Rossow(2014)]%
        {rossow2014amplification}
\bibfield{author}{\bibinfo{person}{Christian Rossow}.} \bibinfo{year}{2014}\natexlab{}.
\newblock \showarticletitle{Amplification Hell: Revisiting Network Protocols for DDoS Abuse}. In \bibinfo{booktitle}{\emph{21st Annual Network and Distributed System Security Symposium (NDSS)}}. \bibinfo{pages}{1--15}.
\newblock


\bibitem[Ruge et~al\mbox{.}({[n.\,d.]})]%
        {sec20-frankenstein}
\bibfield{author}{\bibinfo{person}{Jan Ruge}, \bibinfo{person}{Jiska Classen}, \bibinfo{person}{Francesco Gringoli}, {and} \bibinfo{person}{Matthias Hollick}.} \bibinfo{year}{[n.\,d.]}\natexlab{}.
\newblock \showarticletitle{Frankenstein: Advanced Wireless Fuzzing to Exploit New Bluetooth Escalation Targets}. In \bibinfo{booktitle}{\emph{29th USENIX Security Symposium (USENIX Security)}}. \bibinfo{pages}{19--36}.
\newblock


\bibitem[Sagonas and Typaldos(2023)]%
        {sagonas2023edhoc}
\bibfield{author}{\bibinfo{person}{Konstantinos Sagonas} {and} \bibinfo{person}{Thanasis Typaldos}.} \bibinfo{year}{2023}\natexlab{}.
\newblock \showarticletitle{EDHOC-Fuzzer: An EDHOC Protocol State Fuzzer}. In \bibinfo{booktitle}{\emph{32nd ACM SIGSOFT International Symposium on Software Testing and Analysis (ISSTA)}}. \bibinfo{pages}{1495--1498}.
\newblock


\bibitem[Sasnauskas et~al\mbox{.}(2008)]%
        {sensys08-kleenet}
\bibfield{author}{\bibinfo{person}{Raimondas Sasnauskas}, \bibinfo{person}{J\'{o} \'{A}gila~Bitsch Link}, \bibinfo{person}{Muhammad~Hamad Alizai}, {and} \bibinfo{person}{Klaus Wehrle}.} \bibinfo{year}{2008}\natexlab{}.
\newblock \showarticletitle{KleeNet: Automatic Bug Hunting in Sensor Network Applications}. In \bibinfo{booktitle}{\emph{6th ACM Conference on Embedded Network Sensor Systems (SenSys)}}. \bibinfo{pages}{425–426}.
\newblock


\bibitem[Schepers et~al\mbox{.}(2021)]%
        {wisec21-schepers}
\bibfield{author}{\bibinfo{person}{Domien Schepers}, \bibinfo{person}{Mathy Vanhoef}, {and} \bibinfo{person}{Aanjhan Ranganathan}.} \bibinfo{year}{2021}\natexlab{}.
\newblock \showarticletitle{A Framework to Test and Fuzz Wi-Fi Devices}. In \bibinfo{booktitle}{\emph{14th ACM Conference on Security and Privacy in Wireless and Mobile Networks (WiSec)}}. \bibinfo{pages}{368–370}.
\newblock


\bibitem[Schumilo et~al\mbox{.}(2022)]%
        {eurosys22-nyxnet}
\bibfield{author}{\bibinfo{person}{Sergej Schumilo}, \bibinfo{person}{Cornelius Aschermann}, \bibinfo{person}{Andrea Jemmett}, \bibinfo{person}{Ali Abbasi}, {and} \bibinfo{person}{Thorsten Holz}.} \bibinfo{year}{2022}\natexlab{}.
\newblock \showarticletitle{Nyx-Net: Network Fuzzing with Incremental Snapshots}. In \bibinfo{booktitle}{\emph{17th European Conference on Computer Systems (EuroSys)}}. \bibinfo{pages}{166–180}.
\newblock


\bibitem[Serebryany et~al\mbox{.}(2012)]%
        {serebryany2012addresssanitizer}
\bibfield{author}{\bibinfo{person}{Konstantin Serebryany}, \bibinfo{person}{Derek Bruening}, \bibinfo{person}{Alexander Potapenko}, {and} \bibinfo{person}{Dmitriy Vyukov}.} \bibinfo{year}{2012}\natexlab{}.
\newblock \showarticletitle{AddressSanitizer: A Fast Address Sanity Checker}. In \bibinfo{booktitle}{\emph{2012 USENIX Annual Technical Conference (USENIX ATC 12)}}. \bibinfo{pages}{309--318}.
\newblock


\bibitem[Serebryany and Iskhodzhanov(2009)]%
        {serebryany2009threadsanitizer}
\bibfield{author}{\bibinfo{person}{Konstantin Serebryany} {and} \bibinfo{person}{Timur Iskhodzhanov}.} \bibinfo{year}{2009}\natexlab{}.
\newblock \showarticletitle{ThreadSanitizer: data race detection in practice}. In \bibinfo{booktitle}{\emph{workshop on binary instrumentation and applications}}. \bibinfo{pages}{62--71}.
\newblock


\bibitem[Sesterhenn and Muench(2013)]%
        {bed}
\bibfield{author}{\bibinfo{person}{Eric Sesterhenn} {and} \bibinfo{person}{Martin~J. Muench}.} \bibinfo{year}{2013}\natexlab{}.
\newblock \bibinfo{title}{Bruteforce Exploit Detector}.
\newblock
\newblock
\urldef\tempurl%
\url{Available: https://gitlab.com/kalilinux/packages/bed}
\showURL{%
\tempurl}


\bibitem[Shelby(2014)]%
        {coapdoc}
\bibfield{author}{\bibinfo{person}{Z. Shelby}.} \bibinfo{year}{2014}\natexlab{}.
\newblock \bibinfo{title}{RFC7252: The Constrained Application Protocol (CoAP)}.
\newblock
\newblock
\urldef\tempurl%
\url{https://www.rfc-editor.org/rfc/rfc7252}
\showURL{%
\tempurl}


\bibitem[Shi et~al\mbox{.}(2023a)]%
        {ccs23-netlifter}
\bibfield{author}{\bibinfo{person}{Qingkai Shi}, \bibinfo{person}{Junyang Shao}, \bibinfo{person}{Yapeng Ye}, \bibinfo{person}{Mingwei Zheng}, {and} \bibinfo{person}{Xiangyu Zhang}.} \bibinfo{year}{2023}\natexlab{a}.
\newblock \showarticletitle{Lifting Network Protocol Implementation to Precise Format Specification with Security Applications}. In \bibinfo{booktitle}{\emph{2023 ACM SIGSAC Conference on Computer and Communications Security (CCS)}}. \bibinfo{pages}{1287–1301}.
\newblock


\bibitem[Shi et~al\mbox{.}(2023b)]%
        {sec23-statelifter}
\bibfield{author}{\bibinfo{person}{Qingkai Shi}, \bibinfo{person}{Xiangzhe Xu}, {and} \bibinfo{person}{Xiangyu Zhang}.} \bibinfo{year}{2023}\natexlab{b}.
\newblock \showarticletitle{Extracting protocol format as state machine via controlled static loop analysis}. In \bibinfo{booktitle}{\emph{32nd USENIX Security Symposium (USENIX Security 23)}}. \bibinfo{pages}{7019--7036}.
\newblock


\bibitem[Shi et~al\mbox{.}(2019)]%
        {asiaccs19-mossot}
\bibfield{author}{\bibinfo{person}{Shangcheng Shi}, \bibinfo{person}{Xianbo Wang}, {and} \bibinfo{person}{Wing~Cheong Lau}.} \bibinfo{year}{2019}\natexlab{}.
\newblock \showarticletitle{MoSSOT: An Automated Blackbox Tester for Single Sign-On Vulnerabilities in Mobile Applications}. In \bibinfo{booktitle}{\emph{2019 ACM Asia Conference on Computer and Communications Security (AsiaCCS)}}. \bibinfo{pages}{269–282}.
\newblock


\bibitem[SIG(2016)]%
        {blespec}
\bibfield{author}{\bibinfo{person}{Bluetooth SIG}.} \bibinfo{year}{2016}\natexlab{}.
\newblock \bibinfo{title}{Bluetooth Core Specifications}.
\newblock
\newblock
\urldef\tempurl%
\url{https://www.bluetooth.com/specifications/ bluetooth-core-specification}
\showURL{%
\tempurl}


\bibitem[Sommer et~al\mbox{.}(2017)]%
        {blackhat-us-17-flowfuzz}
\bibfield{author}{\bibinfo{person}{Manuel Sommer}, \bibinfo{person}{Nicholas Gray}, \bibinfo{person}{Phuoc Tran-Gia}, {and} \bibinfo{person}{Thomas Zinner}.} \bibinfo{year}{2017}\natexlab{}.
\newblock \showarticletitle{FlowFuzz: A Framework for Fuzzing OpenFlow-enabled Software and Hardware Switches}. In \bibinfo{booktitle}{\emph{Blackhat US 2017}}.
\newblock


\bibitem[Sommer et~al\mbox{.}(2018)]%
        {defcon26-RFfuzzer}
\bibfield{author}{\bibinfo{person}{Manuel Sommer}, \bibinfo{person}{Nicholas Gray}, \bibinfo{person}{Phuoc Tran-Gia}, {and} \bibinfo{person}{Thomas Zinner}.} \bibinfo{year}{2018}\natexlab{}.
\newblock \showarticletitle{Designing and Applying Extensible RF Fuzzing Tools to Expose PHY Layer Vulnerabilities}. In \bibinfo{booktitle}{\emph{DEFCON 26}}.
\newblock


\bibitem[Somorovsky(2016)]%
        {ccs16-tlsattacker}
\bibfield{author}{\bibinfo{person}{Juraj Somorovsky}.} \bibinfo{year}{2016}\natexlab{}.
\newblock \showarticletitle{Systematic Fuzzing and Testing of TLS Libraries}. In \bibinfo{booktitle}{\emph{2016 ACM SIGSAC Conference on Computer and Communications Security (CCS)}}. \bibinfo{pages}{1492–1504}.
\newblock


\bibitem[Song et~al\mbox{.}(2019)]%
        {song2019sok}
\bibfield{author}{\bibinfo{person}{Dokyung Song}, \bibinfo{person}{Julian Lettner}, \bibinfo{person}{Prabhu Rajasekaran}, \bibinfo{person}{Yeoul Na}, \bibinfo{person}{Stijn Volckaert}, \bibinfo{person}{Per Larsen}, {and} \bibinfo{person}{Michael Franz}.} \bibinfo{year}{2019}\natexlab{}.
\newblock \showarticletitle{SoK: Sanitizing for security}. In \bibinfo{booktitle}{\emph{2019 IEEE Symposium on Security and Privacy (SP)}}. \bibinfo{pages}{1275--1295}.
\newblock


\bibitem[Song et~al\mbox{.}(2014)]%
        {tse14-SymbexNet}
\bibfield{author}{\bibinfo{person}{JaeSeung Song}, \bibinfo{person}{Cristian Cadar}, {and} \bibinfo{person}{Peter Pietzuch}.} \bibinfo{year}{2014}\natexlab{}.
\newblock \showarticletitle{SymbexNet: Testing Network Protocol Implementations with Symbolic Execution and Rule-Based Specifications}.
\newblock \bibinfo{journal}{\emph{IEEE Transactions on Software Engineering}} \bibinfo{volume}{40}, \bibinfo{number}{7} (\bibinfo{year}{2014}), \bibinfo{pages}{695--709}.
\newblock


\bibitem[Song et~al\mbox{.}(2022)]%
        {defcon30-someip}
\bibfield{author}{\bibinfo{person}{Jonghyuk Song}, \bibinfo{person}{Soohwan Oh}, {and} \bibinfo{person}{Woongjo Choi}.} \bibinfo{year}{2022}\natexlab{}.
\newblock \showarticletitle{Automotive Ethernet Fuzzing: From Purchasing ECU to SOME/IP Fuzzing}. In \bibinfo{booktitle}{\emph{DEFCON 30}}.
\newblock


\bibitem[Stepanov and Serebryany(2015)]%
        {stepanov2015memorysanitizer}
\bibfield{author}{\bibinfo{person}{Evgeniy Stepanov} {and} \bibinfo{person}{Konstantin Serebryany}.} \bibinfo{year}{2015}\natexlab{}.
\newblock \showarticletitle{MemorySanitizer: fast detector of uninitialized memory use in C++}. In \bibinfo{booktitle}{\emph{2015 IEEE/ACM International Symposium on Code Generation and Optimization (CGO)}}. \bibinfo{pages}{46--55}.
\newblock


\bibitem[Stephens et~al\mbox{.}(2016)]%
        {ndss16-driller}
\bibfield{author}{\bibinfo{person}{Nick Stephens}, \bibinfo{person}{John Grosen}, \bibinfo{person}{Christopher Salls}, \bibinfo{person}{Andrew Dutcher}, \bibinfo{person}{Ruoyu Wang}, \bibinfo{person}{Jacopo Corbetta}, \bibinfo{person}{Yan Shoshitaishvili}, \bibinfo{person}{Christopher Kruegel}, {and} \bibinfo{person}{Giovanni Vigna}.} \bibinfo{year}{2016}\natexlab{}.
\newblock \showarticletitle{Driller: Augmenting Fuzzing Through Selective Symbolic Execution}. In \bibinfo{booktitle}{\emph{23rd Annual Network and Distributed System Security Symposium (NDSS)}}.
\newblock


\bibitem[Sudhodanan et~al\mbox{.}(2016)]%
        {sudhodanan2016attack}
\bibfield{author}{\bibinfo{person}{Avinash Sudhodanan}, \bibinfo{person}{Alessandro Armando}, \bibinfo{person}{Roberto Carbone}, \bibinfo{person}{Luca Compagna}, {et~al\mbox{.}}} \bibinfo{year}{2016}\natexlab{}.
\newblock \showarticletitle{Attack Patterns for Black-Box Security Testing of Multi-Party Web Applications.}. In \bibinfo{booktitle}{\emph{23st Annual Network and Distributed System Security Symposium (NDSS)}}.
\newblock


\bibitem[Sun et~al\mbox{.}(2017)]%
        {issta17-symbolic}
\bibfield{author}{\bibinfo{person}{Wei Sun}, \bibinfo{person}{Lisong Xu}, {and} \bibinfo{person}{Sebastian Elbaum}.} \bibinfo{year}{2017}\natexlab{}.
\newblock \showarticletitle{Improving the Cost-Effectiveness of Symbolic Testing Techniques for Transport Protocol Implementations under Packet Dynamics}. In \bibinfo{booktitle}{\emph{26th ACM SIGSOFT International Symposium on Software Testing and Analysis (ISSTA)}}. \bibinfo{pages}{79–89}.
\newblock


\bibitem[Sun et~al\mbox{.}(2018)]%
        {icc18-symbolic}
\bibfield{author}{\bibinfo{person}{Wei Sun}, \bibinfo{person}{Lisong Xu}, {and} \bibinfo{person}{Sebastian Elbaum}.} \bibinfo{year}{2018}\natexlab{}.
\newblock \showarticletitle{Scalably Testing Congestion Control Algorithms of Real-World TCP Implementations}. In \bibinfo{booktitle}{\emph{2018 IEEE International Conference on Communications (ICC)}}. \bibinfo{pages}{1--7}.
\newblock


\bibitem[Sun et~al\mbox{.}(2019)]%
        {nsdi19-act}
\bibfield{author}{\bibinfo{person}{Wei Sun}, \bibinfo{person}{Lisong Xu}, \bibinfo{person}{Sebastian Elbaum}, {and} \bibinfo{person}{Di Zhao}.} \bibinfo{year}{2019}\natexlab{}.
\newblock \showarticletitle{{Model-Agnostic} and Efficient Exploration of Numerical State Space of {Real-World} {TCP} Congestion Control Implementations}. In \bibinfo{booktitle}{\emph{16th USENIX Symposium on Networked Systems Design and Implementation (NSDI 19)}}. \bibinfo{pages}{719--734}.
\newblock


\bibitem[Sun et~al\mbox{.}(2023)]%
        {tdsc23-spenny}
\bibfield{author}{\bibinfo{person}{Yue Sun}, \bibinfo{person}{Zhi Li}, \bibinfo{person}{Shichao Lv}, {and} \bibinfo{person}{Limin Sun}.} \bibinfo{year}{2023}\natexlab{}.
\newblock \showarticletitle{Spenny: Extensive ICS Protocol Reverse Analysis via Field Guided Symbolic Execution}.
\newblock \bibinfo{journal}{\emph{IEEE Transactions on Dependable and Secure Computing}} \bibinfo{volume}{20}, \bibinfo{number}{6} (\bibinfo{year}{2023}), \bibinfo{pages}{4502--4518}.
\newblock


\bibitem[Sun(2014)]%
        {sun2005satellite}
\bibfield{author}{\bibinfo{person}{Zhili Sun}.} \bibinfo{year}{2014}\natexlab{}.
\newblock \bibinfo{booktitle}{\emph{Satellite Networking: Principles and Protocols} (\bibinfo{edition}{2nd} ed.)}.
\newblock \bibinfo{publisher}{Wiley Publishing}.
\newblock
\showISBNx{1118351606}


\bibitem[Sutskever et~al\mbox{.}(2014)]%
        {sutskever2014sequence}
\bibfield{author}{\bibinfo{person}{Ilya Sutskever}, \bibinfo{person}{Oriol Vinyals}, {and} \bibinfo{person}{Quoc~V Le}.} \bibinfo{year}{2014}\natexlab{}.
\newblock \showarticletitle{Sequence to sequence learning with neural networks}.
\newblock \bibinfo{journal}{\emph{Advances in neural information processing systems}}  \bibinfo{volume}{27} (\bibinfo{year}{2014}).
\newblock


\bibitem[Synopsys(2014)]%
        {heartbleed}
\bibfield{author}{\bibinfo{person}{Inc. Synopsys}.} \bibinfo{year}{2014}\natexlab{}.
\newblock \bibinfo{title}{Heartbleed Vulnerability}.
\newblock
\newblock
\urldef\tempurl%
\url{Available: https://heartbleed.com/}
\showURL{%
\tempurl}


\bibitem[Trimberger and Moore(2014)]%
        {trimberger2014fpga}
\bibfield{author}{\bibinfo{person}{Stephen~M Trimberger} {and} \bibinfo{person}{Jason~J Moore}.} \bibinfo{year}{2014}\natexlab{}.
\newblock \showarticletitle{FPGA security: Motivations, features, and applications}.
\newblock \bibinfo{journal}{\emph{IEEE}} \bibinfo{volume}{102}, \bibinfo{number}{8} (\bibinfo{year}{2014}), \bibinfo{pages}{1248--1265}.
\newblock


\bibitem[Tsankov et~al\mbox{.}(2013)]%
        {issta13-semi-valid}
\bibfield{author}{\bibinfo{person}{Petar Tsankov}, \bibinfo{person}{Mohammad~Torabi Dashti}, {and} \bibinfo{person}{David Basin}.} \bibinfo{year}{2013}\natexlab{}.
\newblock \showarticletitle{Semi-Valid Input Coverage for Fuzz Testing}. In \bibinfo{booktitle}{\emph{2013 International Symposium on Software Testing and Analysis (ISSTA)}}. \bibinfo{pages}{56–66}.
\newblock


\bibitem[Vanhoef(2017)]%
        {blackhat-us-17-wifuzz}
\bibfield{author}{\bibinfo{person}{Mathy Vanhoef}.} \bibinfo{year}{2017}\natexlab{}.
\newblock \showarticletitle{WiFuzz: Detecting and Exploiting Logical Flaws in the Wi-Fi Cryptographic Handshake}. In \bibinfo{booktitle}{\emph{Blackhat US 2017}}.
\newblock


\bibitem[Walz and Sikora(2020)]%
        {tdsc17-differencialTLS}
\bibfield{author}{\bibinfo{person}{Andreas Walz} {and} \bibinfo{person}{Axel Sikora}.} \bibinfo{year}{2020}\natexlab{}.
\newblock \showarticletitle{Exploiting Dissent: Towards Fuzzing-Based Differential Black-Box Testing of TLS Implementations}.
\newblock \bibinfo{journal}{\emph{IEEE Transactions on Dependable and Secure Computing}} \bibinfo{volume}{17}, \bibinfo{number}{2} (\bibinfo{year}{2020}), \bibinfo{pages}{278--291}.
\newblock


\bibitem[Wang et~al\mbox{.}(2017)]%
        {wang2017skyfire}
\bibfield{author}{\bibinfo{person}{Junjie Wang}, \bibinfo{person}{Bihuan Chen}, \bibinfo{person}{Lei Wei}, {and} \bibinfo{person}{Yang Liu}.} \bibinfo{year}{2017}\natexlab{}.
\newblock \showarticletitle{Skyfire: Data-driven seed generation for fuzzing}. In \bibinfo{booktitle}{\emph{2017 IEEE Symposium on Security and Privacy (SP)}}. \bibinfo{pages}{579--594}.
\newblock


\bibitem[Wang et~al\mbox{.}(2024)]%
        {sp24-llmif}
\bibfield{author}{\bibinfo{person}{J. Wang}, \bibinfo{person}{L. Yu}, {and} \bibinfo{person}{X. Luo}.} \bibinfo{year}{2024}\natexlab{}.
\newblock \showarticletitle{LLMIF: Augmented Large Language Model for Fuzzing IoT Devices}. In \bibinfo{booktitle}{\emph{2024 IEEE Symposium on Security and Privacy (SP)}}. \bibinfo{pages}{196--196}.
\newblock


\bibitem[Wang et~al\mbox{.}(2021)]%
        {sec21-mpinspector}
\bibfield{author}{\bibinfo{person}{Qinying Wang}, \bibinfo{person}{Shouling Ji}, \bibinfo{person}{Yuan Tian}, \bibinfo{person}{Xuhong Zhang}, \bibinfo{person}{Binbin Zhao}, \bibinfo{person}{Yuhong Kan}, \bibinfo{person}{Zhaowei Lin}, \bibinfo{person}{Changting Lin}, \bibinfo{person}{Shuiguang Deng}, \bibinfo{person}{Alex~X. Liu}, {and} \bibinfo{person}{Raheem Beyah}.} \bibinfo{year}{2021}\natexlab{}.
\newblock \showarticletitle{MPInspector: {A} Systematic and Automatic Approach for Evaluating the Security of IoT Messaging Protocols}. In \bibinfo{booktitle}{\emph{30th {USENIX} Security Symposium (USENIX Security)}}. \bibinfo{pages}{4205--4222}.
\newblock


\bibitem[Wang and Wang(2023)]%
        {wang2023nlp}
\bibfield{author}{\bibinfo{person}{Zhuzhu Wang} {and} \bibinfo{person}{Ying Wang}.} \bibinfo{year}{2023}\natexlab{}.
\newblock \showarticletitle{NLP-based Cross-Layer 5G Vulnerabilities Detection via Fuzzing Generated Run-Time Profiling}.
\newblock \bibinfo{journal}{\emph{arXiv:2305.08226}} (\bibinfo{year}{2023}).
\newblock


\bibitem[Wu et~al\mbox{.}(2024)]%
        {issta24-logos}
\bibfield{author}{\bibinfo{person}{Feifan Wu}, \bibinfo{person}{Zhengxiong Luo}, \bibinfo{person}{Yanyang Zhao}, \bibinfo{person}{Qingpeng Du}, \bibinfo{person}{Junze Yu}, \bibinfo{person}{Ruikang Peng}, \bibinfo{person}{Heyuan Shi}, {and} \bibinfo{person}{Yu Jiang}.} \bibinfo{year}{2024}\natexlab{}.
\newblock \showarticletitle{Logos: Log Guided Fuzzing for Protocol Implementations}. In \bibinfo{booktitle}{\emph{33rd ACM SIGSOFT International Symposium on Software Testing and Analysis (ISSTA)}}. \bibinfo{pages}{1--13}.
\newblock


\bibitem[Wu and Li(2021)]%
        {blackhat-as-21-wu}
\bibfield{author}{\bibinfo{person}{Huiyu Wu} {and} \bibinfo{person}{Yuxiang Li}.} \bibinfo{year}{2021}\natexlab{}.
\newblock \showarticletitle{X-in-the-Middle: Attacking Fast Charging Piles and Electric Vehicles}. In \bibinfo{booktitle}{\emph{Blackhat Asia 2021}}.
\newblock


\bibitem[Wu et~al\mbox{.}(2021)]%
        {sec21-LIGHTBLUE}
\bibfield{author}{\bibinfo{person}{Jianliang Wu}, \bibinfo{person}{Ruoyu Wu}, \bibinfo{person}{Daniele Antonioli}, \bibinfo{person}{Mathias Payer}, \bibinfo{person}{Nils~Ole Tippenhauer}, \bibinfo{person}{Dongyan Xu}, \bibinfo{person}{Dave~(Jing) Tian}, {and} \bibinfo{person}{Antonio Bianchi}.} \bibinfo{year}{2021}\natexlab{}.
\newblock \showarticletitle{{LIGHTBLUE}: Automatic {Profile-Aware} Debloating of Bluetooth Stacks}. In \bibinfo{booktitle}{\emph{30th USENIX Security Symposium (USENIX Security 21)}}. \bibinfo{pages}{339--356}.
\newblock
\showISBNx{978-1-939133-24-3}


\bibitem[Xie et~al\mbox{.}(2020)]%
        {blackhat-as-20-xie}
\bibfield{author}{\bibinfo{person}{Haikuo Xie}, \bibinfo{person}{Ying Wang}, {and} \bibinfo{person}{Ye Zhang}.} \bibinfo{year}{2020}\natexlab{}.
\newblock \showarticletitle{WIFI-Important Remote Attack Surface: Threat is Expanding}. In \bibinfo{booktitle}{\emph{Blackhat Asia 2020}}.
\newblock


\bibitem[Yan et~al\mbox{.}(2022)]%
        {blackhat-us-22-brokenmesh}
\bibfield{author}{\bibinfo{person}{Han Yan}, \bibinfo{person}{Lewei Qu}, {and} \bibinfo{person}{Dongxiang Ke}.} \bibinfo{year}{2022}\natexlab{}.
\newblock \showarticletitle{BrokenMesh: New Attack Surfaces of Bluetooth Mesh}. In \bibinfo{booktitle}{\emph{Blackhat US 2022}}.
\newblock


\bibitem[Yang et~al\mbox{.}(2021)]%
        {osdi21-consensus}
\bibfield{author}{\bibinfo{person}{Youngseok Yang}, \bibinfo{person}{Taesoo Kim}, {and} \bibinfo{person}{Byung-Gon Chun}.} \bibinfo{year}{2021}\natexlab{}.
\newblock \showarticletitle{Finding Consensus Bugs in Ethereum via Multi-transaction Differential Fuzzing}. In \bibinfo{booktitle}{\emph{15th USENIX Symposium on Operating Systems Design and Implementation (OSDI)}}. \bibinfo{pages}{349--365}.
\newblock


\bibitem[Yasin et~al\mbox{.}(2020)]%
        {greyliterature}
\bibfield{author}{\bibinfo{person}{Affan Yasin}, \bibinfo{person}{Rubia Fatima}, \bibinfo{person}{Lijie Wen}, \bibinfo{person}{Wasif Afzal}, \bibinfo{person}{Muhammad Azhar}, {and} \bibinfo{person}{Richard Torkar}.} \bibinfo{year}{2020}\natexlab{}.
\newblock \showarticletitle{On Using Grey Literature and Google Scholar in Systematic Literature Reviews in Software Engineering}.
\newblock \bibinfo{journal}{\emph{IEEE Access}}  \bibinfo{volume}{8} (\bibinfo{year}{2020}), \bibinfo{pages}{36226--36243}.
\newblock


\bibitem[Ye et~al\mbox{.}(2023)]%
        {ye2023detecting}
\bibfield{author}{\bibinfo{person}{Mingxi Ye}, \bibinfo{person}{Yuhong Nan}, \bibinfo{person}{Zibin Zheng}, \bibinfo{person}{Dongpeng Wu}, {and} \bibinfo{person}{Huizhong Li}.} \bibinfo{year}{2023}\natexlab{}.
\newblock \showarticletitle{Detecting state inconsistency bugs in dapps via on-chain transaction replay and fuzzing}. In \bibinfo{booktitle}{\emph{32nd ACM SIGSOFT International Symposium on Software Testing and Analysis (ISSTA)}}. \bibinfo{pages}{298--309}.
\newblock


\bibitem[Yen et~al\mbox{.}(2021)]%
        {blackhat-eu-21-dds}
\bibfield{author}{\bibinfo{person}{Ta-Lun Yen}, \bibinfo{person}{Federico Maggi}, \bibinfo{person}{Erik Boasson}, \bibinfo{person}{Victor Mayoral-Vilches}, \bibinfo{person}{Mars Cheng}, \bibinfo{person}{Patrick Kuo}, {and} \bibinfo{person}{Chizuru Toyama}.} \bibinfo{year}{2021}\natexlab{}.
\newblock \showarticletitle{The Data Distribution Service (DDS) Protocol is Critical Let's Use it Securely!}. In \bibinfo{booktitle}{\emph{Blackhat EU 2021}}.
\newblock


\bibitem[Yu et~al\mbox{.}(2019)]%
        {ccs19-iothunter}
\bibfield{author}{\bibinfo{person}{Bo Yu}, \bibinfo{person}{Pengfei Wang}, \bibinfo{person}{Tai Yue}, {and} \bibinfo{person}{Yong Tang}.} \bibinfo{year}{2019}\natexlab{}.
\newblock \showarticletitle{Poster: Fuzzing IoT Firmware via Multi-Stage Message Generation}. In \bibinfo{booktitle}{\emph{2019 ACM SIGSAC Conference on Computer and Communications Security (CCS)}}. \bibinfo{pages}{2525–2527}.
\newblock


\bibitem[Yu et~al\mbox{.}(2023)]%
        {icdcs23-dp-reverser}
\bibfield{author}{\bibinfo{person}{Le Yu}, \bibinfo{person}{Rui Yao}, {and} \bibinfo{person}{Zhanlei Zhang}.} \bibinfo{year}{2023}\natexlab{}.
\newblock \showarticletitle{Poster: DP-Reverser: Automatically Reverse Engineering Vehicle Diagnostic Protocols}. In \bibinfo{booktitle}{\emph{2023 IEEE 43rd International Conference on Distributed Computing Systems (ICDCS)}}. \bibinfo{pages}{1053--1054}.
\newblock


\bibitem[Yu et~al\mbox{.}(2022)]%
        {iot22-cgfuzzer}
\bibfield{author}{\bibinfo{person}{Zhenhua Yu}, \bibinfo{person}{Haolu Wang}, \bibinfo{person}{Dan Wang}, \bibinfo{person}{Zhiwu Li}, {and} \bibinfo{person}{Houbing Song}.} \bibinfo{year}{2022}\natexlab{}.
\newblock \showarticletitle{CGFuzzer: A Fuzzing Approach Based on Coverage-Guided Generative Adversarial Networks for Industrial IoT Protocols}.
\newblock \bibinfo{journal}{\emph{IEEE Internet Things J.}} \bibinfo{volume}{9}, \bibinfo{number}{21} (\bibinfo{year}{2022}), \bibinfo{pages}{21607--21619}.
\newblock


\bibitem[Yun et~al\mbox{.}(2018)]%
        {yun2018qsym}
\bibfield{author}{\bibinfo{person}{Insu Yun}, \bibinfo{person}{Sangho Lee}, \bibinfo{person}{Meng Xu}, \bibinfo{person}{Yeongjin Jang}, {and} \bibinfo{person}{Taesoo Kim}.} \bibinfo{year}{2018}\natexlab{}.
\newblock \showarticletitle{QSYM: A practical concolic execution engine tailored for hybrid fuzzing}. In \bibinfo{booktitle}{\emph{27th USENIX Security Symposium (USENIX Security 18)}}. \bibinfo{pages}{745--761}.
\newblock


\bibitem[Zalewski(2015)]%
        {afl}
\bibfield{author}{\bibinfo{person}{Michal Zalewski}.} \bibinfo{year}{2015}\natexlab{}.
\newblock \bibinfo{title}{American fuzzy lop}.
\newblock
\newblock
\urldef\tempurl%
\url{https://github.com/google/AFL}
\showURL{%
\tempurl}


\bibitem[zardus(2019)]%
        {preeny}
\bibfield{author}{\bibinfo{person}{zardus}.} \bibinfo{year}{2019}\natexlab{}.
\newblock \bibinfo{title}{Preeny: Some helpful preload libraries for pwning stuff.}
\newblock
\newblock
\urldef\tempurl%
\url{https://github.com/zardus/preeny}
\showURL{%
\tempurl}


\bibitem[Zhang et~al\mbox{.}(2023b)]%
        {zhang2023automata}
\bibfield{author}{\bibinfo{person}{Cen Zhang}, \bibinfo{person}{Yuekang Li}, \bibinfo{person}{Hao Zhou}, \bibinfo{person}{Xiaohan Zhang}, \bibinfo{person}{Yaowen Zheng}, \bibinfo{person}{Xian Zhan}, \bibinfo{person}{Xiaofei Xie}, \bibinfo{person}{Xiapu Luo}, \bibinfo{person}{Xinghua Li}, \bibinfo{person}{Yang Liu}, {et~al\mbox{.}}} \bibinfo{year}{2023}\natexlab{b}.
\newblock \showarticletitle{Automata-Guided Control-Flow-Sensitive Fuzz Driver Generation.}. In \bibinfo{booktitle}{\emph{32nd USENIX Security Symposium}}. \bibinfo{pages}{2867--2884}.
\newblock


\bibitem[Zhang et~al\mbox{.}(2021)]%
        {zhang2021apicraft}
\bibfield{author}{\bibinfo{person}{Cen Zhang}, \bibinfo{person}{Xingwei Lin}, \bibinfo{person}{Yuekang Li}, \bibinfo{person}{Yinxing Xue}, \bibinfo{person}{Jundong Xie}, \bibinfo{person}{Hongxu Chen}, \bibinfo{person}{Xinlei Ying}, \bibinfo{person}{Jiashui Wang}, {and} \bibinfo{person}{Yang Liu}.} \bibinfo{year}{2021}\natexlab{}.
\newblock \showarticletitle{$\{$APICraft$\}$: Fuzz driver generation for closed-source $\{$SDK$\}$ libraries}. In \bibinfo{booktitle}{\emph{30th USENIX Security Symposium (USENIX Security 21)}}. \bibinfo{pages}{2811--2828}.
\newblock


\bibitem[Zhang et~al\mbox{.}(2024)]%
        {issta24-llmdriver}
\bibfield{author}{\bibinfo{person}{Cen Zhang}, \bibinfo{person}{Yaowen Zheng}, \bibinfo{person}{Mingqiang Bai}, \bibinfo{person}{Yeting Li}, \bibinfo{person}{Wei Ma}, \bibinfo{person}{Xiaofei Xie}, \bibinfo{person}{Yuekang Li}, \bibinfo{person}{Limin Sun}, {and} \bibinfo{person}{Yang Liu}.} \bibinfo{year}{2024}\natexlab{}.
\newblock \showarticletitle{How Effective Are They? Exploring Large Language Model Based Fuzz Driver Generation}. In \bibinfo{booktitle}{\emph{33rd ACM SIGSOFT International Symposium on Software Testing and Analysis (ISSTA)}}. \bibinfo{pages}{1--13}.
\newblock


\bibitem[Zhang et~al\mbox{.}(2023a)]%
        {10.1145/3580597}
\bibfield{author}{\bibinfo{person}{Zenong Zhang}, \bibinfo{person}{George Klees}, \bibinfo{person}{Eric Wang}, \bibinfo{person}{Michael Hicks}, {and} \bibinfo{person}{Shiyi Wei}.} \bibinfo{year}{2023}\natexlab{a}.
\newblock \showarticletitle{Fuzzing Configurations of Program Options}.
\newblock \bibinfo{journal}{\emph{ACM Trans. Softw. Eng. Methodol.}} \bibinfo{volume}{32}, \bibinfo{number}{2} (\bibinfo{year}{2023}), \bibinfo{pages}{1--21}.
\newblock


\bibitem[Zhao et~al\mbox{.}(2019)]%
        {icst19-seqfuzzer}
\bibfield{author}{\bibinfo{person}{Hui Zhao}, \bibinfo{person}{Zhihui Li}, \bibinfo{person}{Hansheng Wei}, \bibinfo{person}{Jianqi Shi}, {and} \bibinfo{person}{Yanhong Huang}.} \bibinfo{year}{2019}\natexlab{}.
\newblock \showarticletitle{SeqFuzzer: An Industrial Protocol Fuzzing Framework from a Deep Learning Perspective}. In \bibinfo{booktitle}{\emph{12th {IEEE} Conference on Software Testing, Validation and Verification ({ICST})}}. \bibinfo{pages}{59--67}.
\newblock


\bibitem[Zheng et~al\mbox{.}(2024)]%
        {zheng2024pardiff}
\bibfield{author}{\bibinfo{person}{Mingwei Zheng}, \bibinfo{person}{Qingkai Shi}, \bibinfo{person}{Xuwei Liu}, \bibinfo{person}{Xiangzhe Xu}, \bibinfo{person}{Le Yu}, \bibinfo{person}{Congyu Liu}, \bibinfo{person}{Guannan Wei}, {and} \bibinfo{person}{Xiangyu Zhang}.} \bibinfo{year}{2024}\natexlab{}.
\newblock \showarticletitle{ParDiff: Practical Static Differential Analysis of Network Protocol Parsers}.
\newblock \bibinfo{journal}{\emph{Proceedings of the ACM on Programming Languages (OOPSLA)}}  \bibinfo{volume}{8} (\bibinfo{year}{2024}), \bibinfo{pages}{1208--1234}.
\newblock


\bibitem[Zheng et~al\mbox{.}(2022)]%
        {zheng2022efficient}
\bibfield{author}{\bibinfo{person}{Yaowen Zheng}, \bibinfo{person}{Yuekang Li}, \bibinfo{person}{Cen Zhang}, \bibinfo{person}{Hongsong Zhu}, \bibinfo{person}{Yang Liu}, {and} \bibinfo{person}{Limin Sun}.} \bibinfo{year}{2022}\natexlab{}.
\newblock \showarticletitle{Efficient greybox fuzzing of applications in Linux-based IoT devices via enhanced user-mode emulation}. In \bibinfo{booktitle}{\emph{31st ACM SIGSOFT International Symposium on Software Testing and Analysis}}. \bibinfo{pages}{417--428}.
\newblock


\bibitem[Zheng and Sun(2022)]%
        {zheng2022ipspex}
\bibfield{author}{\bibinfo{person}{Yaowen Zheng} {and} \bibinfo{person}{Limin Sun}.} \bibinfo{year}{2022}\natexlab{}.
\newblock \showarticletitle{IPSpex: Enabling Efficient Fuzzing via Specification Extraction on ICS Protocol}. In \bibinfo{booktitle}{\emph{Applied Cryptography and Network Security: 20th International Conference (ACNS)}}, Vol.~\bibinfo{volume}{13269}. \bibinfo{pages}{356}.
\newblock


\bibitem[Zhu et~al\mbox{.}(2022)]%
        {fuzzingsurvey-roadmap}
\bibfield{author}{\bibinfo{person}{Xiaogang Zhu}, \bibinfo{person}{Sheng Wen}, \bibinfo{person}{Seyit Camtepe}, {and} \bibinfo{person}{Yang Xiang}.} \bibinfo{year}{2022}\natexlab{}.
\newblock \showarticletitle{Fuzzing: A Survey for Roadmap}.
\newblock \bibinfo{journal}{\emph{Comput. Surveys}} \bibinfo{volume}{54}, \bibinfo{number}{11s} (\bibinfo{year}{2022}).
\newblock


\bibitem[Zou et~al\mbox{.}(2020)]%
        {arxiv20-zou}
\bibfield{author}{\bibinfo{person}{Qingtian Zou}, \bibinfo{person}{Anoop Singhal}, \bibinfo{person}{Xiaoyan Sun}, {and} \bibinfo{person}{Peng Liu}.} \bibinfo{year}{2020}\natexlab{}.
\newblock \showarticletitle{Generating Comprehensive Data with Protocol Fuzzing for Applying Deep Learning to Detect Network Attacks}.
\newblock  (\bibinfo{year}{2020}).
\newblock
\showeprint[arXiv]{2012.12743}


\bibitem[Zou et~al\mbox{.}({[n.\,d.]})]%
        {atc21-tcpfuzz}
\bibfield{author}{\bibinfo{person}{Yong-Hao Zou}, \bibinfo{person}{Jia-Ju Bai}, \bibinfo{person}{Jielong Zhou}, \bibinfo{person}{Jianfeng Tan}, \bibinfo{person}{Chenggang Qin}, {and} \bibinfo{person}{Shi-Min Hu}.} \bibinfo{year}{[n.\,d.]}\natexlab{}.
\newblock \showarticletitle{{TCP-Fuzz}: Detecting Memory and Semantic Bugs in {TCP} Stacks with Fuzzing}. In \bibinfo{booktitle}{\emph{2021 USENIX Annual Technical Conference (USENIX ATC)}}. \bibinfo{pages}{489--502}.
\newblock


\bibitem[Zuo et~al\mbox{.}(2022)]%
        {tcad22-charon}
\bibfield{author}{\bibinfo{person}{Feilong Zuo}, \bibinfo{person}{Zhengxiong Luo}, \bibinfo{person}{Junze Yu}, \bibinfo{person}{Ting Chen}, \bibinfo{person}{Zichen Xu}, \bibinfo{person}{Aiguo Cui}, {and} \bibinfo{person}{Yu Jiang}.} \bibinfo{year}{2022}\natexlab{}.
\newblock \showarticletitle{Vulnerability Detection of ICS Protocols via Cross-State Fuzzing}.
\newblock \bibinfo{journal}{\emph{IEEE Transactions on Computer-Aided Design of Integrated Circuits and Systems}} \bibinfo{volume}{41}, \bibinfo{number}{11} (\bibinfo{year}{2022}), \bibinfo{pages}{4457--4468}.
\newblock


\bibitem[Zuo et~al\mbox{.}(2021)]%
        {dac21-PAVFuzz}
\bibfield{author}{\bibinfo{person}{Feilong Zuo}, \bibinfo{person}{Zhengxiong Luo}, \bibinfo{person}{Junze Yu}, \bibinfo{person}{Zhe Liu}, {and} \bibinfo{person}{Yu Jiang}.} \bibinfo{year}{2021}\natexlab{}.
\newblock \showarticletitle{PAVFuzz: State-Sensitive Fuzz Testing of Protocols in Autonomous Vehicles}.
\newblock \bibinfo{journal}{\emph{58th ACM/IEEE Design Automation Conference (DAC)}} (\bibinfo{year}{2021}), \bibinfo{pages}{823--828}.
\newblock


\end{thebibliography}

\appendix

\end{document}